\pdfoutput=1

\documentclass[11pt,twoside,a4paper,cmspaper,final,collab]{cms-tdr}
\def\svnVersion{494048P}\def\cmsCernNoTag{CERN-EP-2018-322}\def\cmsCernDate{\today}\def\cmsMessage{Published in the Journal of High Energy Physics as \href{http://dx.doi.org/10.1007/JHEP04(2019)122}{\doi{10.1007/JHEP04(2019)122.}}}

\begin{document}\cmsNoteHeader{SMP-18-002}

\hyphenation{had-ron-i-za-tion}
\hyphenation{cal-or-i-me-ter}
\hyphenation{de-vices}
\RCS$HeadURL: svn+ssh://svn.cern.ch/reps/tdr2/papers/SMP-18-002/trunk/SMP-18-002.tex $
\RCS$Id: SMP-18-002.tex 492928 2019-03-28 08:56:28Z vischia $
\newlength\cmsFigWidth
\ifthenelse{\boolean{cms@external}}{\setlength\cmsFigWidth{0.49\textwidth}}{\setlength\cmsFigWidth{0.65\textwidth}}
\ifthenelse{\boolean{cms@external}}{\providecommand{\cmsLeft}{upper\xspace}}{\providecommand{\cmsLeft}{left\xspace}}
\ifthenelse{\boolean{cms@external}}{\providecommand{\cmsRight}{lower\xspace}}{\providecommand{\cmsRight}{right\xspace}}

\newcommand{\fulllumi}{\ensuremath{35.9\fbinv}\xspace}

\newcommand{\WZ}{\ensuremath{\PW\cPZ}\xspace}
\newcommand{\mwz}{\ensuremath{M(\WZ)}\xspace}
\newcommand{\ttZ}{\ensuremath{\ttbar\cPZ}\xspace}
\newcommand{\ZZ}{\ensuremath{\PZ\cPZ}\xspace}
\newcommand{\tZq}{\ensuremath{\cPqt\cPZ\cPq}\xspace}
\newcommand{\sigmatot}{\ensuremath{\sigma_{\text{tot}}}}
\newcommand{\wzprod}{\ensuremath{\Pp\Pp\to\PW\cPZ}\xspace}
\newcommand{\eee}{\Pe\Pe\Pe\xspace}
\newcommand{\eem}{\Pe\Pe\Pgm\xspace}
\newcommand{\emm}{\Pe\Pgm\Pgm\xspace}
\newcommand{\mmm}{\Pgm\Pgm\Pgm\xspace}
\newcommand{\mZ}{\ensuremath{m_{\PZ}}\xspace}
\newcommand{\mW}{\ensuremath{m_{\PW}}\xspace}
\newcommand{\xgamma}{\ensuremath{\mathrm{X+}\gamma}\xspace}
\newcommand{\PV}{\ensuremath{\cmsSymbolFace{V}}\xspace}
\newcommand{\vvv}{\ensuremath{\cmsSymbolFace{VVV}}\xspace}
\newcommand{\awz}{\ensuremath{A^{+-}_{\WZ}}\xspace}
\newcommand{\cw}{\ensuremath{c_{\PW}}\xspace}
\newcommand{\cwww}{\ensuremath{c_{\PW\PW\PW}}\xspace}
\newcommand{\cb}{\ensuremath{c_{\cPqb}}\xspace}
\newcommand{\nbkg}{\ensuremath{N_{\text{bkg}}}\xspace}
\newcommand{\nobs}{\ensuremath{N_{\text{obs}}}\xspace}
\newcommand{\ptzone}{\ensuremath{\pt(\ell_{\PZ 1})}\xspace}
\newcommand{\ptztwo}{\ensuremath{\pt(\ell_{\PZ 2})}\xspace}
\newcommand{\ptw}{\ensuremath{\pt(\ell_{\PW})}\xspace}

\newcommand{\sqrts}{\ensuremath{\sqrt{s}}\xspace}
\providecommand{\NA}{\ensuremath{\text{---}}}
\providecommand{\cmsTable}[1]{\resizebox{\textwidth}{!}{#1}}
\newlength\cmsTabSkip\setlength{\cmsTabSkip}{1ex}

\cmsNoteHeader{SMP-18-002}
\title{Measurements of the \wzprod inclusive and differential production cross sections and constraints on charged anomalous triple gauge couplings at $\sqrts = 13\TeV$}

\date{\today}

\abstract{
The \WZ production cross section is measured in proton-proton collisions at a centre-of-mass energy $\sqrts = 13\TeV$ using data collected with the CMS detector, corresponding to an integrated luminosity of \fulllumi. The inclusive cross section is measured to be $\sigmatot(\wzprod) = 48.09 \mathrm{ }^{+1.00}_{-0.96}\stat \mathrm{ }^{+0.44}_{-0.37}\thy \mathrm{ }^{+2.39}_{-2.17}\syst  \pm 1.39\lum$\unit{pb}, resulting in a total uncertainty of $-2.78$/$+2.98$\unit{pb}. Fiducial cross section and ratios of charge-dependent cross section measurements are provided. Differential cross section measurements are also presented with respect to three variables: the \cPZ{} boson transverse momentum \pt, the leading jet \pt, and the \mwz variable, defined as the invariant mass of the system composed of the three leptons and the missing transverse momentum. Differential measurements with respect to the \PW\ boson \pt, separated by charge, are also shown. Results are consistent with standard model predictions, favouring next-to-next-to-leading-order predictions over those at next-to-leading order. Constraints on anomalous triple gauge couplings are derived via a binned maximum likelihood fit to the \mwz variable.}

\hypersetup{%
pdfauthor={CMS Collaboration},%
pdftitle={Measurements of the pp to WZ inclusive and differential production cross section and constraints on charged anomalous triple gauge couplings at sqrt(s) = 13 TeV},%
pdfsubject={CMS},%
pdfkeywords={CMS, physics, aTGC, WZ production}}

\maketitle

\section{Introduction}

The measurement of the diboson production cross section is sensitive to the self-interaction between gauge bosons via triple gauge couplings (TGCs) and is therefore an important test of the standard model (SM). Such couplings directly result from the nonabelian $\mathrm{SU}(2){\times}\mathrm{U}(1)$ gauge symmetry of the SM.
In the SM, the values of the couplings are fully determined by the structure of the Lagrangian; any deviation of the observed
diboson production strength from the SM prediction, typically manifested as a change in the cross section, would indicate new physics.
The expected change would lead to an overall increase of the cross section, although in some portions of phase space there will be a negative interference between the SM and new physics beyond the SM (BSM).

Associated \WZ\ production is particularly interesting, as it is the only process directly sensitive to the \PW\WZ\ coupling with a \cPZ\ boson in the final state.
Furthermore, \WZ\ production is a major background to searches for new physics in multilepton final states; a precise determination of its cross section is crucial to improve the sensitivity of these searches.
In addition, initial state radiation can be used as a probe of the boost of the \WZ\ system through a differential study of the leading jet transverse momentum, since an initial state particle can radiate a jet and this jet will recoil against the \WZ\ system.

\begin{figure}[!hbtp]
  \centering
  \includegraphics[width=0.32\linewidth]{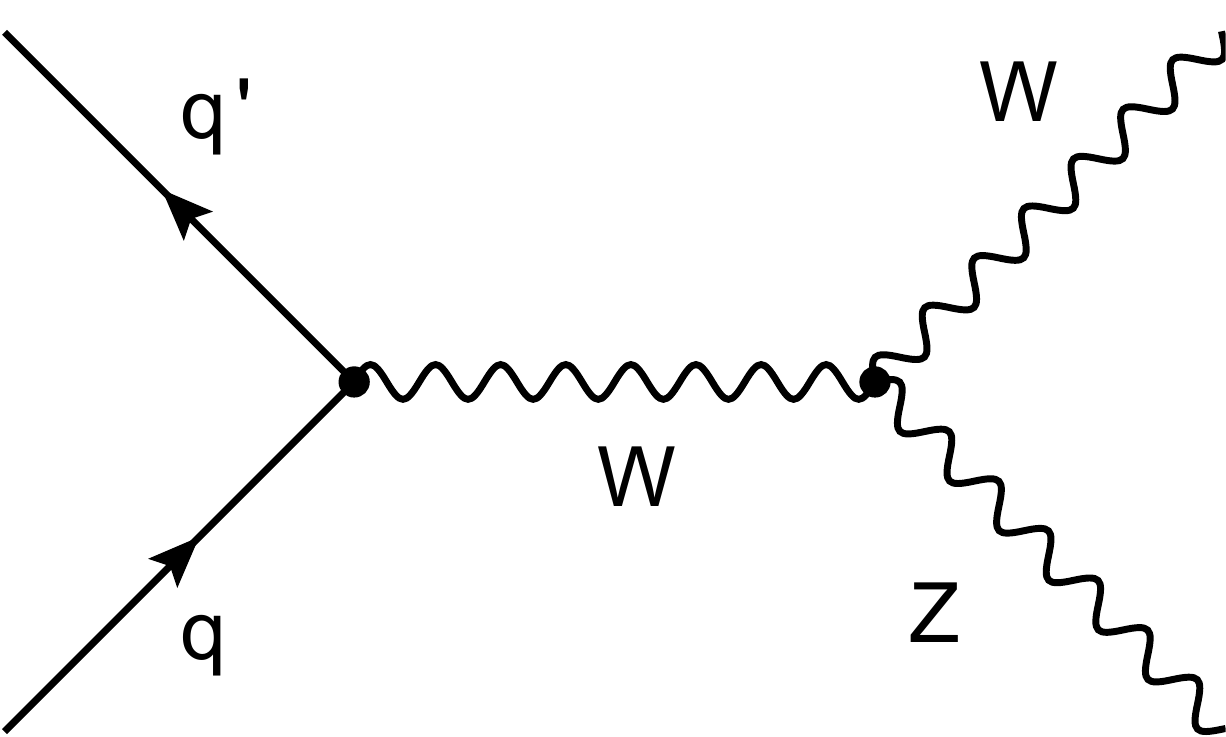}
  \includegraphics[width=0.32\linewidth]{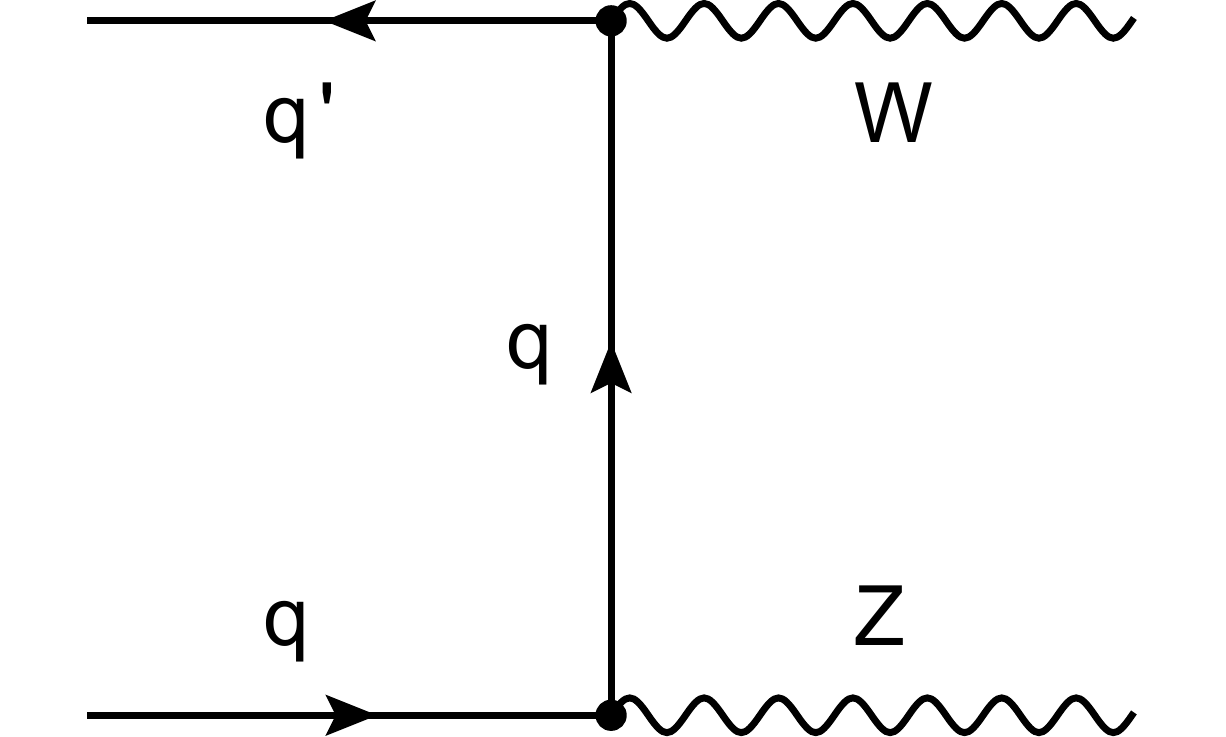}
  \includegraphics[width=0.32\linewidth]{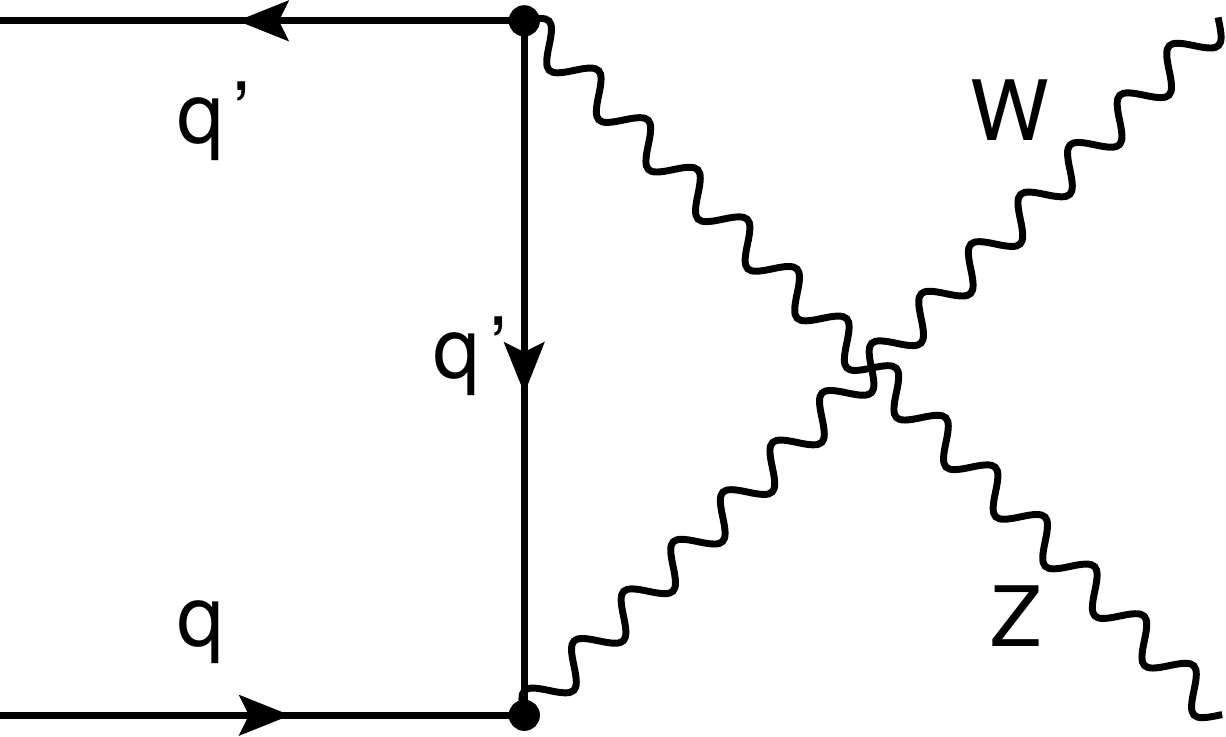}
  \caption{Feynman diagrams for \WZ\ production at leading order in perturbative QCD in proton-proton collisions for the $s$-channel (left), $t$-channel (middle), and $u$-channel (right). The contribution from $s$-channel proceeds through TGC.}
  \label{fig:feynmanwz}
\end{figure}

In the SM at leading order (LO) in perturbative quantum chromodynamics (QCD), \WZ\ production in proton-proton (\Pp\Pp) collisions proceeds via quark-antiquark interactions in the $s$-, $t$-, and $u$-channels. Figure~\ref{fig:feynmanwz} shows the tree-level production diagrams for each channel. The $s$-channel, which proceeds through the \PW\WZ\ TGC, is the only channel sensitive to anomalous values of this coupling.

After a first inconclusive observation of candidate events for \WZ\ production at UA1~\cite{1987389}, standard model \WZ\ production has been studied at $\sqrts = 1.96\TeV$ at the Fermilab Tevatron~\cite{wzdzero,wzcdf} and also in \Pp\Pp~collisions at the CERN LHC by the ATLAS~\cite{Aad:2011cx,Aad:2012twa,Aad:2014mda,Aad:2016ett,Aaboud:2016yus,Aaboud:2016uuk,Aaboud:2017cgf} and CMS~\cite{Chatrchyan:2012jra,Chatrchyan:2012bd,Chatrchyan:2014aqa,Khachatryan:2016tgp,Khachatryan:2016poo,Sirunyan:2017bey} Collaborations.
The most relevant of these results to this paper is the CMS analysis reporting the \wzprod\ production cross section at $\sqrts = 7$ and $8\TeV$ as well as a search for anomalous TGCs (aTGCs) at $\sqrts = 8\TeV$ in the multilepton final state using the full 2011 and 2012 data sets\ \cite{Khachatryan:2016poo}.
The ATLAS Collaboration has similarly analyzed the full 8\TeV data set~\cite{Aad:2016ett}, measured the inclusive and differential cross section, and set limits on aTGCs.
The inclusive \wzprod\ production cross section at $\sqrts = 13\TeV$ was measured in the multilepton final state by the ATLAS~\cite{Aaboud:2016yus} and CMS~\cite{Khachatryan:2016tgp} Collaborations, using the full 2015 data set.

This paper presents a new analysis of \wzprod\ production at $\sqrts = 13\TeV$ using multilepton final states in which the \PZ boson decays into a pair of electrons or muons, and the \PW\ boson decays into a neutrino and either an electron or a muon.
Compared to the previous results, the inclusive and differential cross sections are measured with increased precision (the overall uncertainty in the inclusive cross section is reduced by half), and more stringent confidence intervals on aTGCs are set, yielding the current best limits in two of the parameters.

This paper is organized as follows: the detector is described in Section~\ref{sec:cms}; the data and Monte Carlo (MC) simulated samples are described in Section~\ref{sec:samples}; the object definition and the event selection are described in Section~\ref{sec:objects} and Section~\ref{sec:selection}, respectively; the background estimation is described in Section~\ref{sec:background}, and the systematic uncertainties affecting the analysis are described in Section~\ref{sec:systematics}. Finally, the inclusive cross section measurement is presented in Section~\ref{sec:inclusive}, the differential cross section measurement is presented in Section~\ref{sec:differential}, and the confidence regions for aTGCs are presented in Section~\ref{sec:ano}. A summary of the results is shown in Section~\ref{sec:conclusions}.

\section{The CMS Detector}
\label{sec:cms}

The central feature of the CMS apparatus is a superconducting solenoid of 6\unit{m} internal diameter, providing a magnetic field of 3.8\unit{T}. Within the solenoid volume are a silicon pixel and strip tracker, a lead tungstate crystal electromagnetic calorimeter (ECAL), and a brass and scintillator hadron calorimeter (HCAL), each composed of a barrel and two endcap sections. Forward calorimeters extend the pseudorapidity coverage provided by the barrel and endcap detectors. Muons are detected in gas-ionization chambers embedded in the steel flux-return yoke outside the solenoid.

Events of interest are selected using a two-tiered trigger system~\cite{Khachatryan:2016bia}. The first level (L1), composed of custom hardware processors, uses information from the calorimeters and muon detectors to select events at a rate of around 100\unit{kHz} within a time interval of less than 4\mus. The second level, known as the high-level trigger (HLT), consists of a farm of processors running a version of the full event reconstruction software optimized for fast processing, and reduces the event rate to around 1\unit{kHz} before data storage.

A more detailed description of the CMS detector, together with a definition of the coordinate system used and the relevant kinematic variables, can be found in Ref.~\cite{Chatrchyan:2008zzk}.

\section{Data and simulated samples}\label{sec:samples}

This study is performed using proton-proton (\Pp\Pp) collisions at a centre-of-mass energy of 13\TeV at the LHC. Data taken in 2016 with the CMS detector are analyzed, corresponding to a total integrated luminosity of \fulllumi. The data are filtered to remove detector noise and unphysical events.

Event generators based on the MC method are used to simulate the behaviour of signal and background processes. The \POWHEG\ v2.0~\cite{Melia:2011tj,Nason:2013ydw} software is used to generate both the \WZ\ signal and the \ZZ\ background samples without additional partons besides the ones included in the matrix element calculations at next-to-leading order (NLO) in perturbative QCD. The rest of the SM background samples are produced with the \MGvATNLO\ v2.3.3 generator~\cite{MADGRAPH5} at LO or NLO accuracy, including up to one or two additional partons in the matrix element calculations. The procedure for accounting correctly for parton multiplicities larger than one is referred to as the merging scheme; where applicable, the FxFx merging scheme~\cite{Frederix:2012ps} is used for the NLO samples, and the MLM merging scheme~\cite{Alwall:2007fs} is used for the LO samples.
The modelling of the aTGCs is done by applying the matrix element reweighting method~\cite{MElReweight} at LO accuracy to a signal sample generated with the \MGvATNLO\ v2.3.3 generator~\cite{MADGRAPH5} at NLO accuracy. The procedure is applied to a set of samples produced for different ranges of the \PZ\ boson transverse momentum ($\pt^\PZ$) such that the statistical power of the MC at higher energies, where anomalous couplings are expected to dominate, is enhanced.
The NNPDF3.0LO (NNPDF3.0NLO)~\cite{Ball:2014uwa}  parton distribution functions (PDFs) are used for the simulated samples generated at LO (NLO). The computations are interfaced with the \PYTHIA\ v8.205 generator~\cite{Sjostrand:2014zea}\ to include the effects of parton showering and hadronization using the CUETP8M1 tune~\cite{Skands:2014pea,CMS-PAS-GEN-14-001}.

The effect of additional interactions in the same or adjacent bunch crossing (referred to as pileup) is accounted for by simulating additional minimum bias interactions for each hard scattering event. Simulated events are then reweighted so that the pileup distribution matches that observed in data, which is characterized by an average of 23 collisions per bunch crossing. The generated events are interfaced with a model of the CMS detector response implemented using the \GEANTfour\ package~\cite{Geant} and reconstructed using the same software as the real data.

\section{Event reconstruction and object selection}
\label{sec:objects}

\subsection{Event Reconstruction}
\label{sec:pflow}
Events are reconstructed using the particle-flow (PF) algorithm~\cite{CMS-PRF-14-001} by matching information from all CMS subdetectors to obtain a global description of the event. The resulting objects are classified into mutually exclusive categories: charged hadrons, neutral hadrons, photons, electrons, and muons.

Interaction vertices are identified by grouping tracks consistent with originating from the same location in the beam interaction region. The reconstructed vertex with the largest value of summed physics-object $\pt^2$ is taken to be the primary \Pp\Pp\ interaction vertex. The aforementioned physics objects are the jets, clustered using the jet finding algorithm~\cite{Cacciari:2008gp,Cacciari:2011ma} with the tracks assigned to the vertex as inputs, and the associated missing transverse momentum, taken as the negative vector sum of the \pt of those jets. More details are given in Section 9.4.1 of Ref.~\cite{CMS-TDR-15-02}.

Photons are identified as ECAL energy clusters not linked to the extrapolation of any charged particle trajectory to the ECAL, but are not used in this analysis. Electrons are identified as a primary charged particle accompanied by potentially many ECAL energy clusters~\cite{Khachatryan:2015hwa}; such clusters are matched to the extrapolation of this track to the ECAL and to possible bremsstrahlung photons emitted along the way through the tracker material. Muons are identified as a track in the central tracker consistent with either a track or several hits in the muon system, in association with an energy deficit in the calorimeters. Charged hadrons are identified as charged particle tracks neither identified as electrons, nor as muons. Finally, neutral hadrons are identified as HCAL energy clusters not linked to any charged-hadron trajectory, or as ECAL and HCAL energy excesses with respect to the expected charged-hadron energy deposit.
The energy of photons is directly obtained from the ECAL measurement, corrected for zero-suppression effects. The energy of electrons is determined from a combination of the track momentum at the main interaction vertex, the corresponding ECAL cluster energy, and the energy sum of all bremsstrahlung photons attached to the track. The energy of muons is obtained from the corresponding track momentum.
The energy of charged hadrons is determined from a combination of the track momentum and the corresponding ECAL and HCAL energy, corrected for zero-suppression effects and for the response function of the calorimeters to hadronic showers. Finally, the energy of neutral hadrons is obtained from the corresponding corrected ECAL and HCAL energy.

\subsection{Electrons and muons}

In this analysis, leptons~\cite{PhysRevLett.19.1264} coming from the primary vertex play a prominent role amongst all reconstructed event objects because of the very distinct trilepton ($3\ell$) signature of the signal process. \emph{Prompt signal leptons} are defined as the light leptons (electrons or muons) from the decays of particles in the signal processes, such as those coming from\ \PW\ and \PZ boson and \Pgt\ lepton decays. Leptons originating from hadrons, primarily \cPqb\ hadron decays, are referred to as \emph{nonprompt leptons}.

Electrons are reconstructed as described in Section~\ref{sec:pflow};
candidates are further required to have $\abs{\eta} < 2.5$, to be within the tracking acceptance, and to have $\pt > 7\GeV$.
The identification is performed using a multivariate discriminant with inputs related to the shower shape and to the tracking and track-cluster matching.
Additional identification criteria are applied for electrons with $\pt > 30\GeV$ to mimic the identification applied at trigger level described in Section~\ref{sec:selection}; this ensures consistency between the \emph{measurement} region and \emph{application} region of the misidentification rate estimate.

Muon candidates are reconstructed as described in Section~\ref{sec:pflow} by combining the information from both the silicon tracker and the muon spectrometer in a global fit~\cite{1748-0221-13-06-P06015}.
Candidates are identified by checking the quality of the geometrical matching between the tracker and the muon system measurements.
Only candidates within the muon system acceptance $\abs{\eta} < 2.4$ and with $\pt > 5\GeV$ are considered.

The energy scale of the leptons is corrected to account for mismeasurements in the tracker and muon systems, and in the ECAL. For both electrons and muons, the average difference between the corrected and uncorrected energies is zero; however, a spread is induced in the lepton \pt of about 1\% that is assigned as a systematic uncertainty in the energy of each lepton.

In order to improve the rejection of pileup and misreconstructed tracks and, more importantly, to reject background leptons from \cPqb\ hadron decays, loose selections are applied to variables related to the track impact parameter, as described in Refs.~\cite{Sirunyan:2017lae,Sirunyan:2018shy}.

The charged leptons produced in decays of heavy particles, such as\ \PW\ and\ \cPZ\ bosons,
are typically spatially isolated from the hadronic activity in the event,
whereas the leptons produced in the decays of hadrons or misidentified leptons are usually spatially embedded in jets.
For high-energy \PW\ and \PZ\ bosons the decay products tend to be collimated (a boosted system) and this distinction based on a simple definition of isolation is not effective anymore.

Therefore, the PF-based isolation definition used in the $\sqrts = 8\TeV$ analysis~\cite{Khachatryan:2016poo}, which included all the photons and the neutral and charged hadrons in a cone of {\tolerance=3000 $\Delta R = \sqrt{\smash[b]{(\eta^\ell - \eta^i)^{2} + (\phi^\ell- \phi^i)^{2}}} < 0.3$
(where $i$ indicates the hadrons and $\ell$ the lepton) around the leptons, is improved~\cite{Rehermann:2010vq,Khachatryan:2016kod} by using a \pt-dependent cone size given by the formula:\par}
\begin{equation}
  \Delta R(\pt(\ell)) = \frac{10\GeV}{\min\big[\max(\pt(\ell),50\GeV), 200\GeV\big]}.
\end{equation}
The discrimination between prompt leptons and nonprompt leptons is improved by exploiting the differences in isolation-related variables and in impact-parameter-related variables between the two categories of leptons (prompt and nonprompt). An identification algorithm, based on a multivariate analysis (MVA) using boosted decision trees (BDTs), is trained to discriminate signal leptons (from \PW\ and\ \PZ\ decays) from background leptons (mostly \cPqb\ hadron decays).
The resulting classifier is referred to as the \emph{lepton MVA discriminator}, and was trained using a sample of \ttZ\ events: signal leptons originate from leptonic \ttZ\ decays, and background leptons originate mostly from \cPqb\ hadron decays. Further details on the lepton MVA discriminator can be found in Refs.~\cite{Sirunyan:2017lae,Sirunyan:2018shy}. The efficiencies have a high dependence on the lepton \pt\ and $\eta$; typical values for electrons are 3--7\% misidentification efficiency and 20/40/80/90\% identification efficiency for low \pt\ electrons in the endcap, low \pt\ electrons in the barrel, high \pt\ electrons in the endcap, and high \pt\ electrons in the barrel, respectively. For muons, typical values are 2--10\% misidentification efficiency and 80--100\% identification efficiencies where higher values correspond to higher \pt\ muons.

Throughout the analysis, leptons passing a high threshold on the lepton MVA discriminator are referred to as \emph{tight} leptons.

\subsection{Jets}
Jets are reconstructed by clustering PF candidates using the anti-\kt\ algorithm~\cite{Cacciari:2008gp, Cacciari:2011ma} with a distance parameter of 0.4. The jet momentum is determined as the vector sum of all particle momenta in the jet, and is estimated from simulation to be within 5--10\% of the true momentum over the whole \pt spectrum and detector acceptance~\cite{CMS-DP-2016-020}. Charged hadrons not originating from the primary vertex are subtracted from the PF candidates considered in the clustering; this procedure is referred to as charged-hadron subtraction.
Jet energy corrections are derived from simulation to bring the measured response of jets to that of particle level jets on average, and are applied to the energy of the jet as a function of the jet \pt and $\eta$. In situ measurements of the momentum balance in dijet, photon+jet, \PZ+jet, and multijet events are used to correct for any residual differences in jet energy scale between data and simulation~\cite{Khachatryan:2016kdb}. The jet energy resolution amounts typically to 15\% at 10\GeV, 8\% at 100\GeV, and 4\% at 1\TeV. Additional selection criteria are applied to each jet to remove jets potentially dominated by anomalous contributions from various subdetector components or reconstruction failures~\cite{CMS-PAS-JME-10-003}.
Jets with a minimum $\pt > 30\GeV$ are required to be separated from any lepton candidate passing the minimal lepton selection by selecting $\Delta R = \sqrt{\smash[b]{(\eta^{\ell} - \eta^{jet})^{2} + (\phi^{\ell} - \phi^{jet})^{2}}} > 0.4$.

The combined secondary vertex (CSV) \cPqb\ tagging algorithm~\cite{Sirunyan:2017ezt} is used to identify jets that are likely to originate from the hadronization of bottom quarks (referred to as \cPqb\ jets).  This algorithm combines both secondary vertex information and track impact parameter information together in a likelihood discriminant with output values ranging from zero to one. A jet is tagged as a \cPqb\ jet if the CSV discriminator output exceeds a threshold value, referred to as the \emph{working point}. This analysis uses the medium working point corresponding to requiring that $\mathrm{CSV} > 0.8$. This working point gives approximately 70\% efficiency for tagging \cPqb\ jets and 1.5\% efficiency for mistakenly tagging jets coming from light quarks or gluons~\cite{Sirunyan:2017ezt}. Jets that pass the medium CSV working point and have a minimum $\pt > 30\GeV$ are defined in this analysis as \cPqb\ jets.

Corrections accounting for differences in the \cPqb\ tagging performance between data and simulation are derived by applying weights dependent on the jet \pt, $\eta$, \cPqb\ tagging discriminator, and flavour to each simulated jet~\cite{Sirunyan:2017ezt}. The jet flavour is defined as the flavour of the object originating the jet, which is known in simulation. The weights are derived from \ttbar\ and Z+jets events. The per-event weight is defined as the product of the per-jet weights, including the weights of the jets overlapping with leptons. Uncertainties in the weights are propagated throughout the analysis as systematic uncertainties.

\subsection{Missing transverse momentum}
The missing transverse momentum vector is computed as the negative vector \pt sum of all PF objects identified in the event. The magnitude of this vector is referred to as \ptmiss. The jet energy corrections, introduced previously, are propagated to the estimation of \ptmiss.

\section{Event selection}
\label{sec:selection}

Events with contributions from beam halo processes or calorimeter noise are rejected using dedicated filters~\cite{JME-13-003, CMS-PAS-JME-16-004}. The remaining events are required to pass one of several triggers involving either a single loosely isolated light lepton or a pair of them with any flavour composition. For the single-lepton cases, the \pt threshold is 27 (24) \GeV for electrons (muons). The lower \pt thresholds for the same-flavour dilepton triggers are 23 (17)\GeV for the leading and 12 (8)\GeV for the subleading electron (muon). The cross-flavour triggers require a leading lepton \pt of $23\GeV$ and a subleading lepton \pt of $8\GeV$.

The baseline selection is defined by the presence of at least three tight leptons with at least one opposite-sign same-flavour (OSSF) pair. To exploit the specific kinematic properties of the process, each of the three leading leptons is tentatively assigned to its most likely parent boson.
The first stage of the algorithm assigns the OSSF pair of leptons with an invariant mass closest to the \PZ\ boson mass~\cite{PDG2016}, \mZ, to the \PZ\ boson.
These two leptons are denoted as $\ell_{\PZ 1}$ (leading) and $\ell_{\PZ 2}$ (subleading), ranked by \pt. The remaining lepton is assigned to the \PW\ boson and is denoted as $\ell_{\PW}$. The performance of the algorithm is studied by using simulated events and comparing the assigned parent particle with MC generator level truth; this algorithm properly assigns the leptons in about 95\% of cases.

The baseline selection includes additional requirements on the \pt of each lepton: $\ptzone > 25\GeV$, $\ptztwo > 10\GeV$, and $\pt(\ell_{\PW}) > 25\GeV$. The total efficiency of the set of triggers used to record the data is measured to be close to 99\% with respect to this baseline selection; this is because all the trigger paths can be triggered by more than one object, yielding a very high efficiency for the combined set of triggers even if the efficiency of the individual triggers is lower. The baseline selection is split, on the basis of the flavour composition of the leptonic triplet, into four categories denoted as: \eee, \eem, \emm, and \mmm.

\begin{figure}[]
\centering
\includegraphics[width=0.48\textwidth]{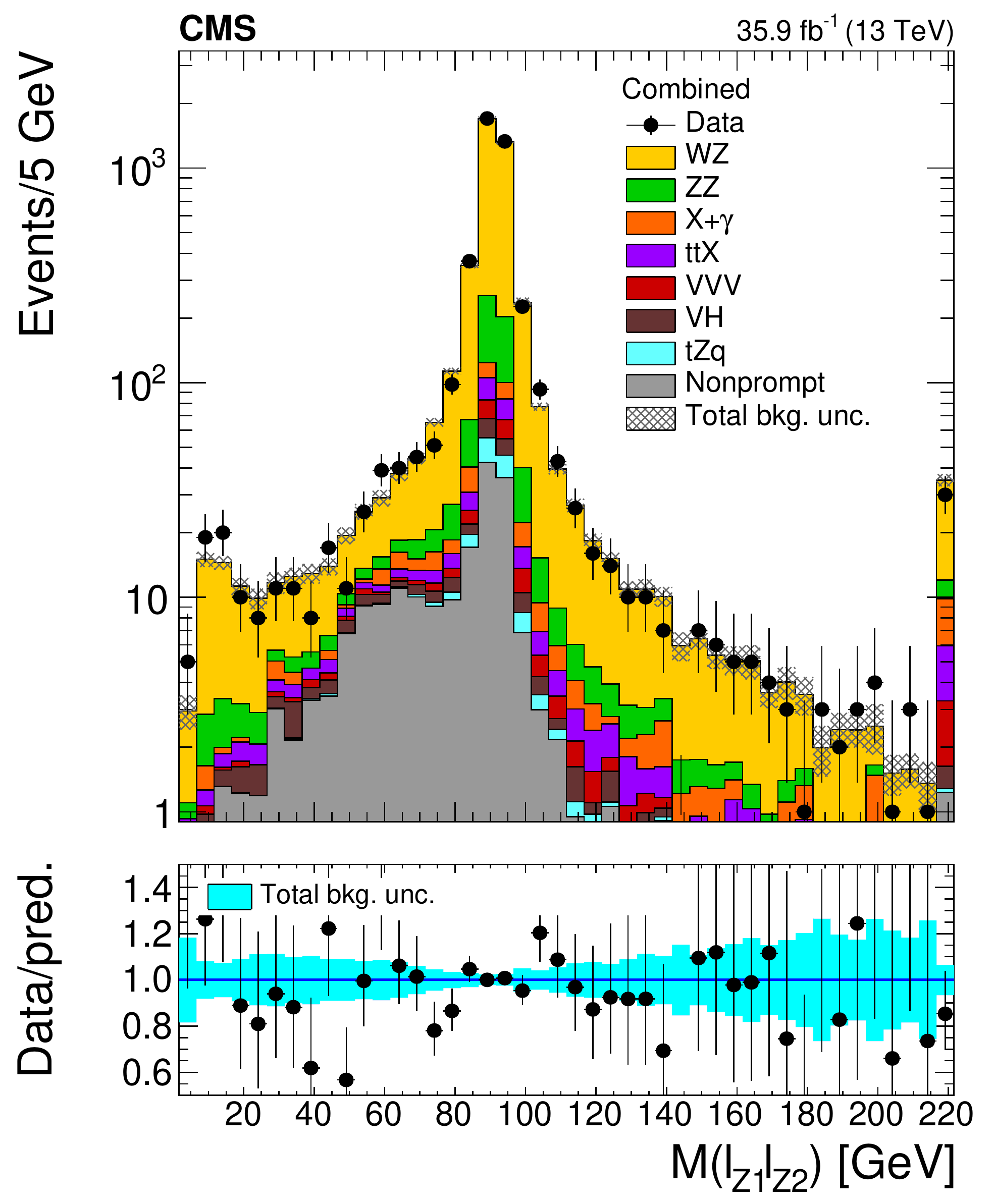}
\includegraphics[width=0.48\textwidth]{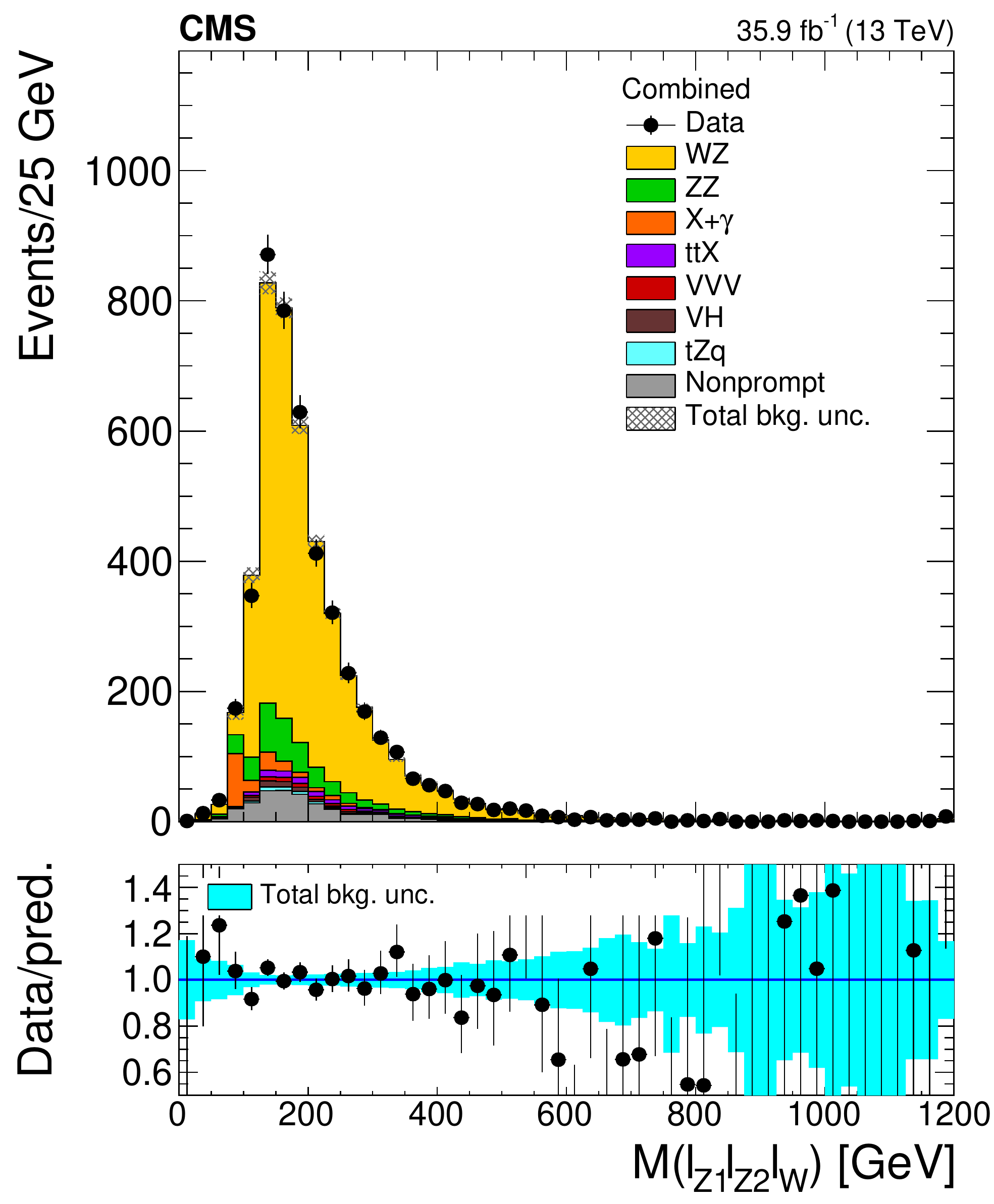}\\
\includegraphics[width=0.48\textwidth]{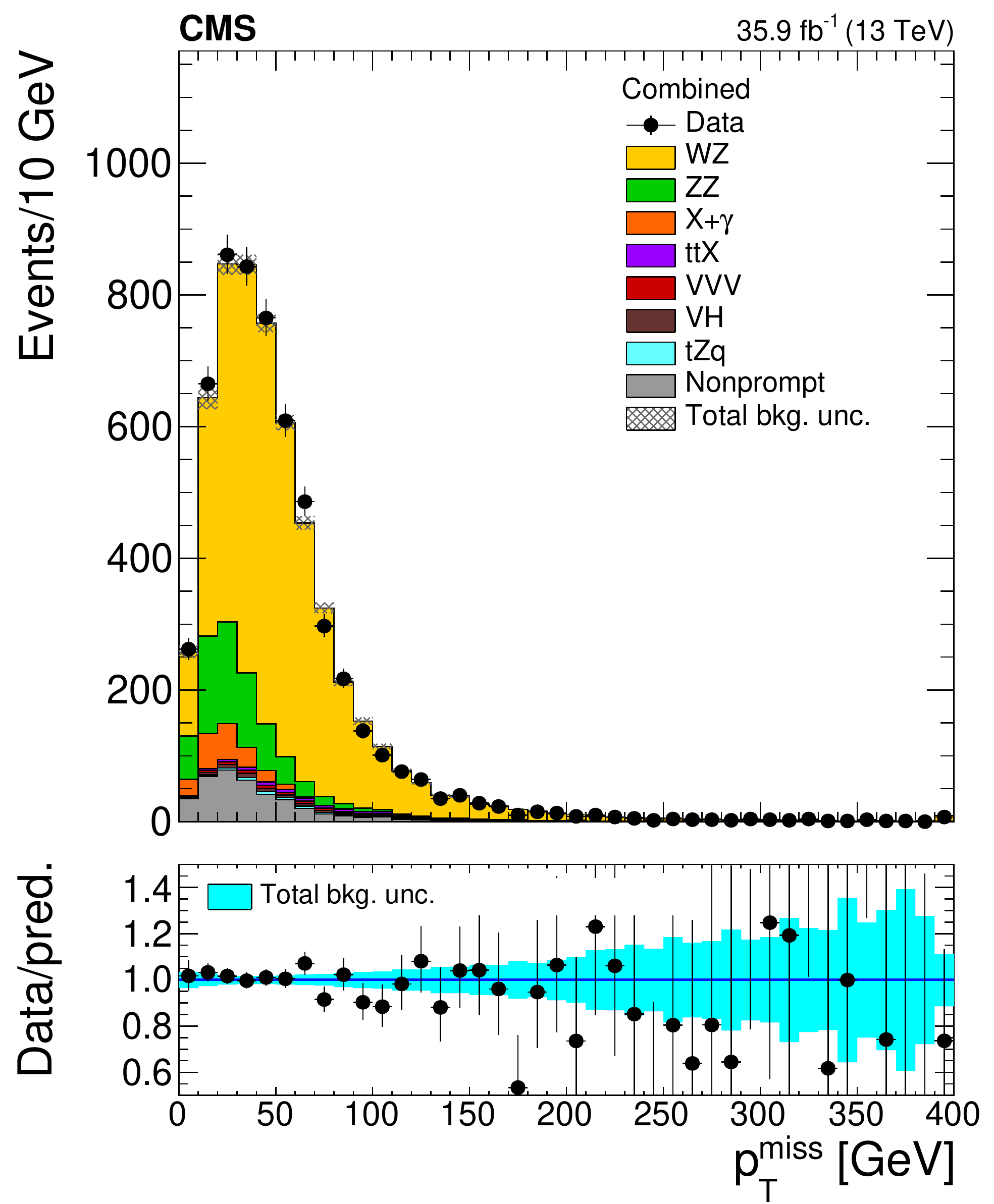}
\includegraphics[width=0.48\textwidth]{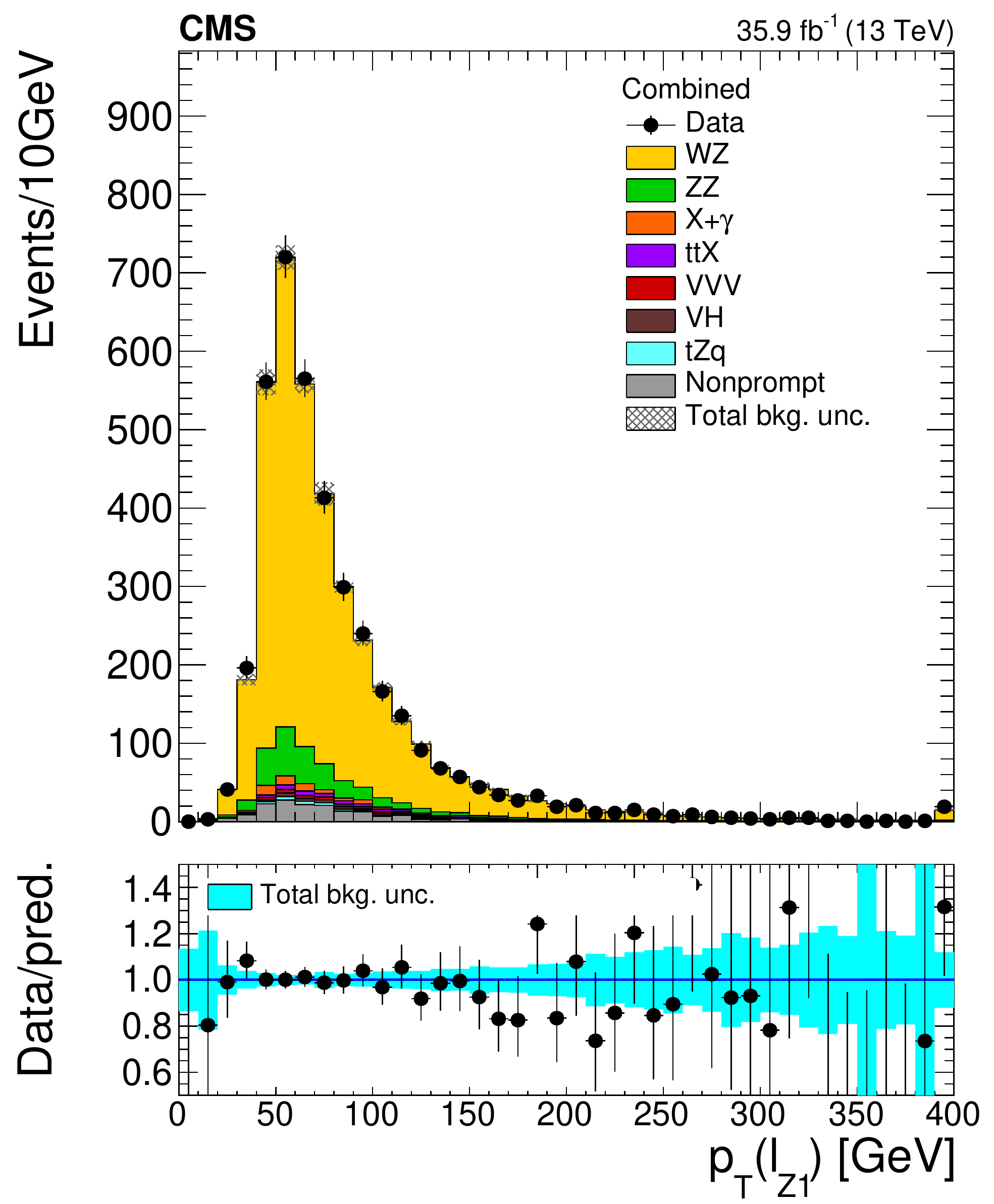}
\caption{Distribution of key observables in the signal region after the signal extraction fit: invariant mass of the lepton pair assigned to the \PZ boson (top left), invariant mass of the three-lepton system (top right), missing transverse momentum (bottom left), and transverse momentum of the leading lepton assigned to the \PZ boson. For each distribution all the signal region requirements are applied except the requirement relating to the particular observable so that the effect of the requirement on that observable can be easily seen. The last bin contains the overflow. Vertical bars on the data points include the statistical uncertainty and shaded bands over the prediction include the contributions of the different sources of uncertainty at their values after the signal extraction fit.}
\label{fig:selec}
\end{figure}

The signal region (SR) is defined by applying to the baseline selection additional requirements that are designed to increase the purity of the region by reducing specific background contributions.
Consistency with the \cPZ\ boson mass peak is enforced by requiring the invariant mass of the two leptons assigned to the \cPZ\ boson to be close to \mZ, $\abs{M(\ell_{\PZ 1}\ell_{\PZ 2}) - \mZ} < 15\GeV$. This requirement greatly reduces the contribution from nonresonant multilepton production processes such as \ttbar\ production.
A requirement that $\ptmiss > 30\GeV$ is found to greatly reduce the \cPZ+jets background contribution; in the following, residual events are included in the contribution labelled \emph{nonprompt}. A large reduction in the number of events that include both a \ttbar\ pair and a \PZ\ boson is obtained by rejecting events that contain one or more \cPqb-tagged jets. The invariant mass of the trilepton system $M(\ell_{\PZ 1}\ell_{\PZ 2}\ell_{\PW})$ is required to exceed 100\GeV.
Finally, contributions from tetraleptonic decays in \ZZ\ events are reduced by rejecting any event with an additional fourth lepton that passes a looser lepton selection.
Generator-level requirements for the signal, designed to avoid infrared divergences, are matched at reconstruction level by vetoing events that do not contain a lepton pair passing a minimum invariant mass requirement of $M(\ell\ell') > 4\GeV$. This requirement also has the desirable effect of reducing the contribution from low-mass resonance processes.
The distribution of the key observables used in the definition of the signal region after the signal extraction fit is displayed in Fig.~\ref{fig:selec}.

Multiple control regions (CRs) are defined to cross-check or estimate the different background processes. Each of them follows the same selection as the signal region, except that individual specific selection criteria are inverted in order to increase the fraction of the targeted background process in the region. A summary of the three orthogonal control regions used in the analysis is presented in Table~\ref{tab:SRCR}. The control regions are labelled according to the expected dominant background process in each region. A detailed explanation of their use is given in the next section.

\begin{table}[h!]
\centering
\topcaption{\label{tab:SRCR} Requirements for the definition of the signal region of the analysis and the three different regions designed to estimate the main background sources.}
\cmsTable{
\begin{tabular}{ccccccccc}
\hline
Region    & $N_{\ell}$ & $\pt\{\ell_{\PZ 1},\ell_{\PZ 2},\ell_{\PW},\ell_{4}\}$               & $N_{\mathrm{OSSF}}$ & $\abs{M(\ell_{\PZ 1}\ell_{\PZ 2}) - \mZ}$ & \ptmiss & $N_{\cPqb\,\mathrm{tag}}$ &  $\min(M(\ell\ell'))$ & $M(\ell_{\PZ 1}\ell_{\PZ 2}\ell_{\PW} )$  \\
          &            &    [\GeVns{}]                      &            & [\GeVns{}]      & [\GeVns{}]                  &      &    [\GeVns{}]             &    [\GeVns{}]      \\\hline
SR        & $=$3   & ${>}\{25,10,25\}$    & $\geq$1   & $<$15 & $>$30 & $=$0 &  $>$4 & $>$100  \\
CR-top    & $=$3   & ${>}\{25,10,25\}$    & $\geq$1   & $>$5 &  $>$30 & $>$0 &  $>$4 & $>$100  \\
CR-$\Z\Z$ & $=$4   & ${>}\{25,10,25,10\}$ & $\geq$1   & $<$15 & $>$30 & $=$0 &  $>$4 & $>$100  \\
CR-Conv   & $=$3   & ${>}\{25,10,25\}$    & $\geq$1   & $>$15 & $\leq$30 & $=$0 & $>$4 & $<$100 \\\hline
\end{tabular}
}
\end{table}

\section{Background estimation}\label{sec:background}

The background contributions fall in two categories, depending on the origin of the final-state leptons.
\emph{Prompt} background sources consist of the SM processes where the leptons originate in the decay of an SM boson or \Pgt\ lepton;
the \emph{nonprompt} backgrounds consist of SM processes where the leptons originate in the decay of \cPqb\ hadrons.

The nonprompt background contributions are heavily dominated by \cPZ+jets production, with additional contributions from dileptonic \ttbar\ decays.
The total contribution of these processes to the signal region is estimated using the \emph{tight-to-loose} method described in detail in Ref.~\cite{Khachatryan:2016kod}.
The probability for a loose lepton to pass the tight criteria is measured in a single-lepton+jets signal region enriched in nonprompt leptons.
For each specific selection, an application region is defined starting from the same requirements and additionally requesting that at least one of the leptons passes the loose selection but fails the tight selection.
Depending on the \pt, $\abs{\eta}$, and multiplicity of the failing leptons, the extrapolation from the control region to the application region is derived for each event as a transfer factor, based on the previously measured probability. The contamination of the application region due to the prompt contribution is estimated from simulation and its effect is subtracted from the total nonprompt estimation in the selection, using the same transfer factors.
Uncertainties in the determination of the nonprompt contribution are estimated with simulated events by comparing the prediction of the method and the one derived directly from simulation; they are found to be dominated by the statistical uncertainty due to the limited amount of simulated events, and estimated to be about 30\%. An additional source of systematic uncertainty is estimated to range between 5 and 30\% from the differences observed amongst a number of methods used for the subtraction of the prompt background processes in the signal region.

\begin{figure}[]
\centering
\includegraphics[width=0.48\textwidth]{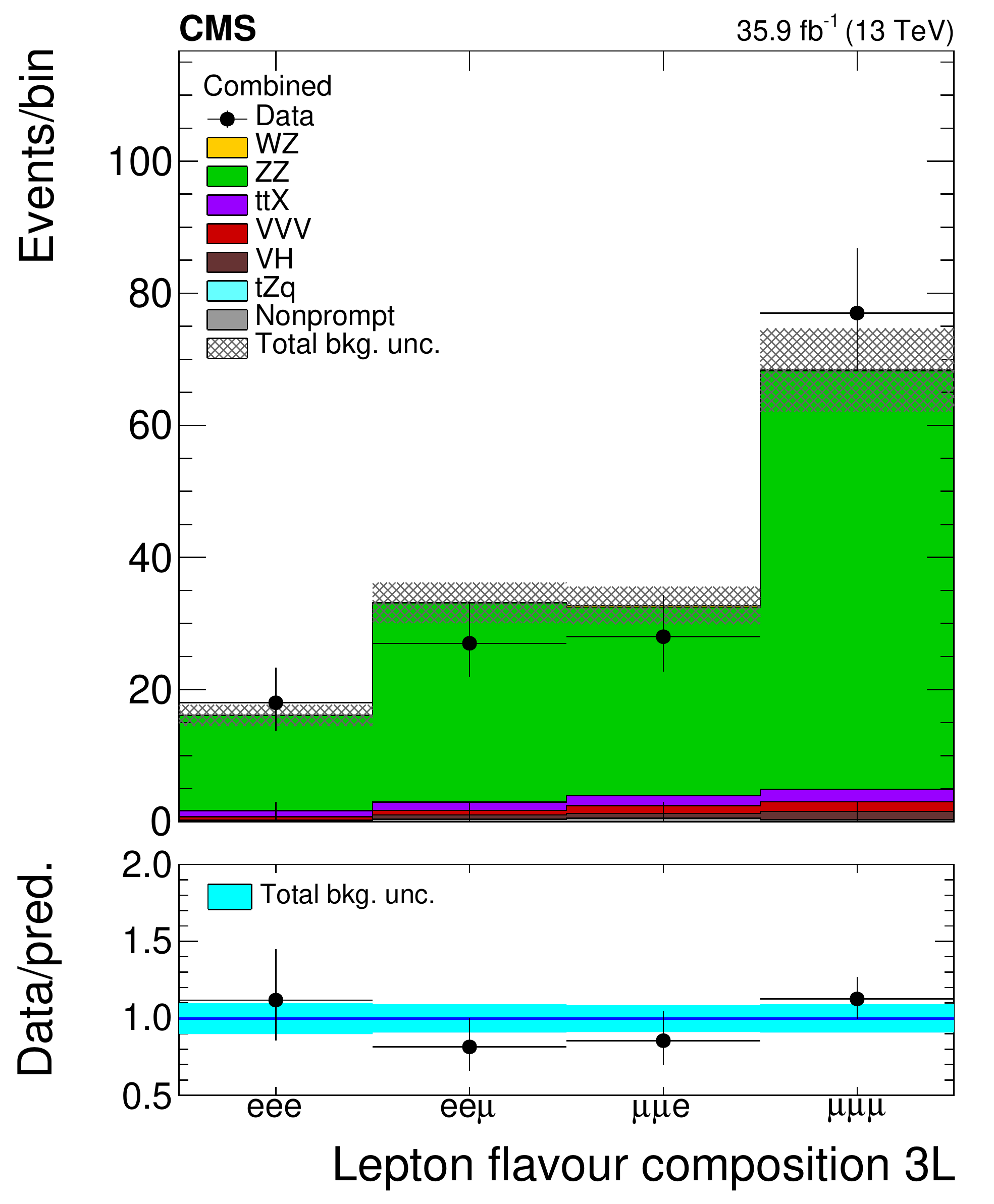}
\includegraphics[width=0.48\textwidth]{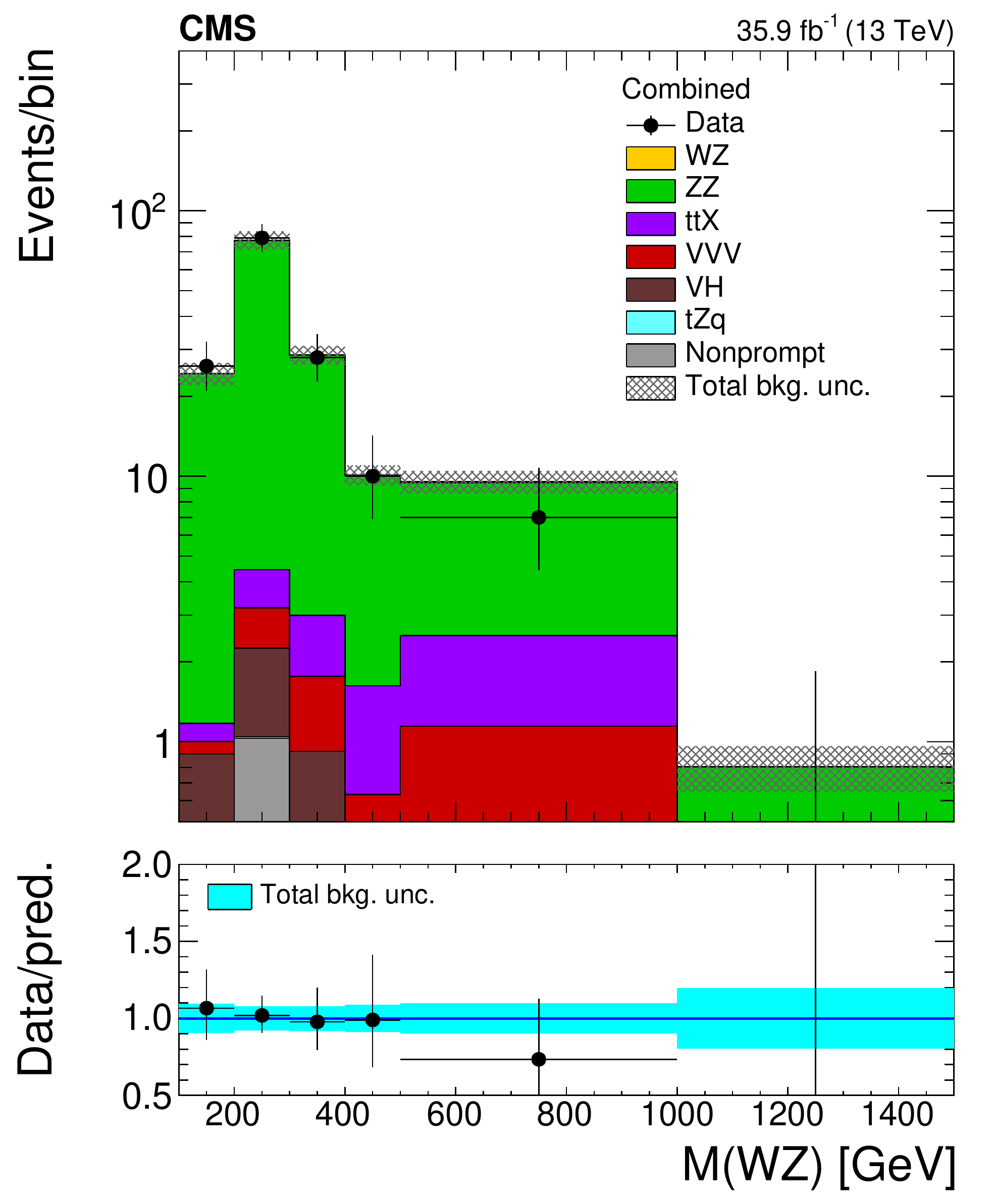}\\
\includegraphics[width=0.48\textwidth]{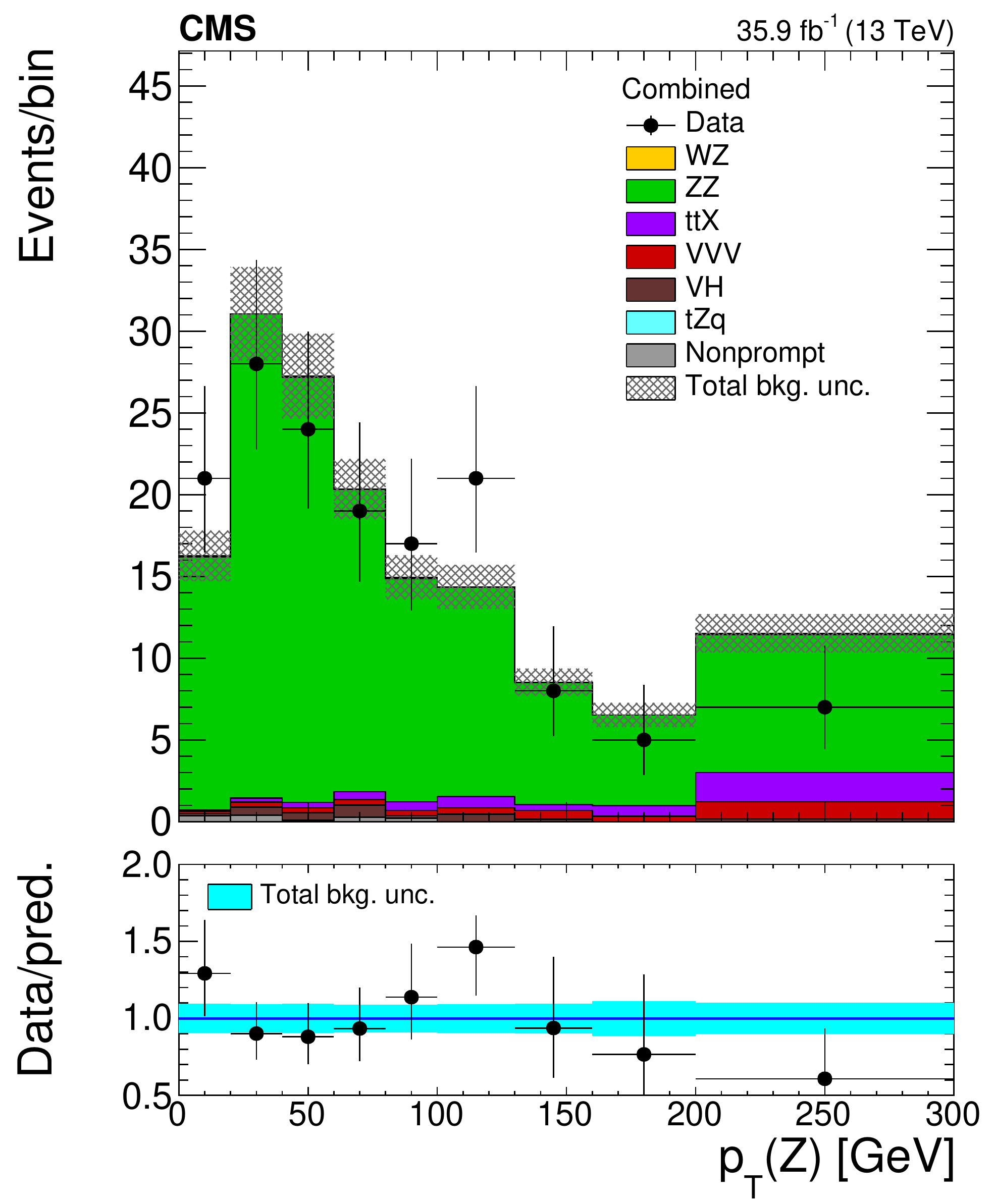}
\includegraphics[width=0.48\textwidth]{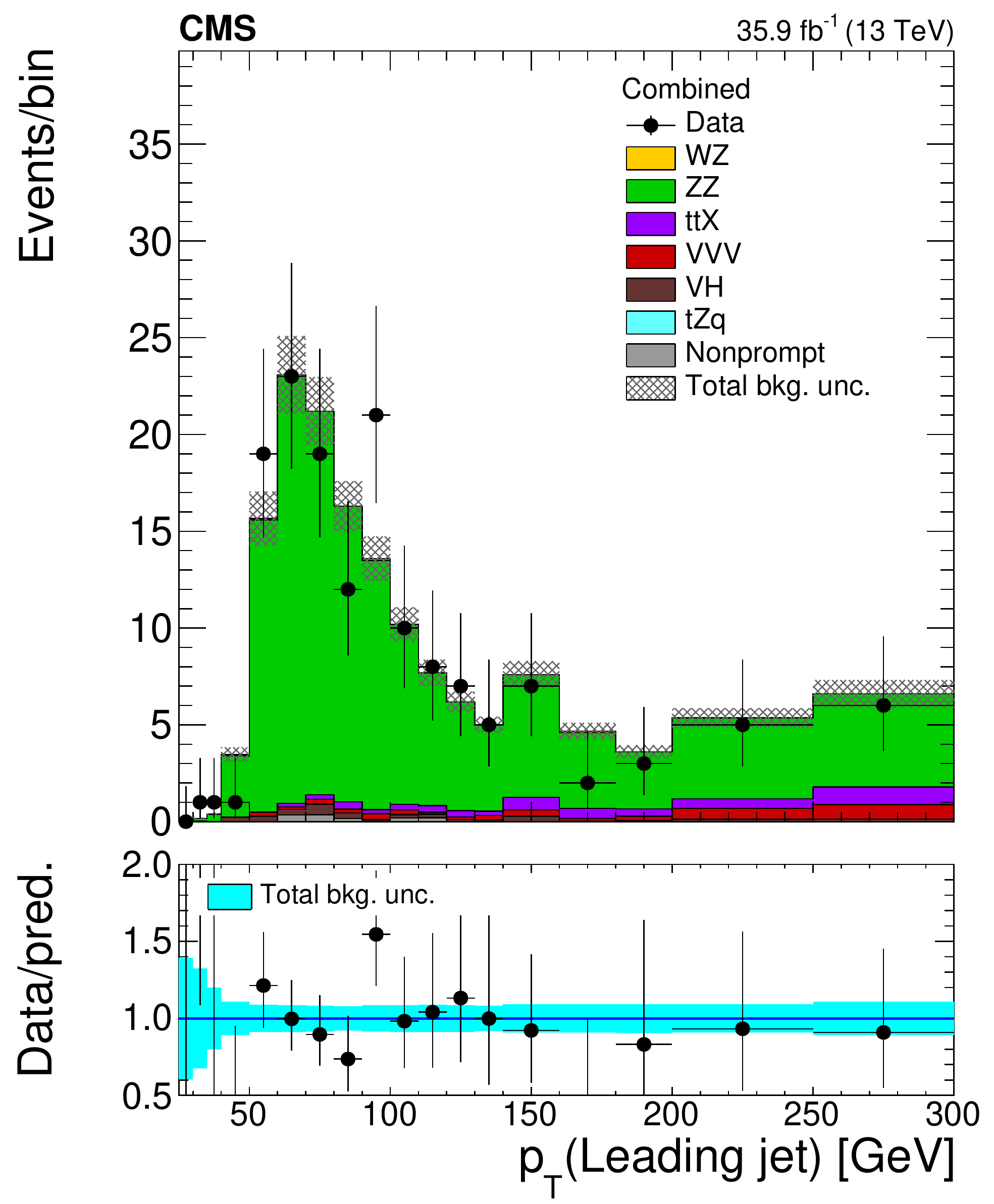}
\caption{Distribution of key observables in the \ZZ control region defined in Table~\ref{tab:SRCR}: flavour composition of the three leading leptons (top left), invariant mass of the three leptons plus missing transverse momentum (top right), transverse momentum of the \PZ\ boson reconstructed from the \pt of the two leptons assigned to it (bottom left), and transverse momentum of the leading jet (bottom right).  Vertical bars on the data points include the statistical uncertainty and shaded bands over the prediction include the contributions of the different sources of uncertainty evaluated after the signal extraction fit.}
\label{fig:crZZ}
\end{figure}

The leading SM prompt background comes from the tetraleptonic decay of \ZZ\ pairs when one of the produced leptons is too soft or does not pass the quality requirements of the identification selection. Our estimation of the contribution of this process is based on MC simulation. To validate the behaviour of this simulation we use a dedicated sideband region (CR-ZZ) that requires exactly four leptons in the final state; the resulting selection is dominated by \ZZ\ production, therefore no additional \ZZ-specific constraint is applied. To illustrate the behaviour of this associated control region, the key observables used in the different measurements of the analysis are shown in Fig.~\ref{fig:crZZ}. As a numerical cross-check, we estimate the possible variations over the simulated prediction in the four flavour-dependent categories. For each category, we subtract the predicted non-\ZZ\ yields from the observed data and divide the result by the expected \ZZ\ contribution. Statistical and normalization uncertainties are propagated to this measurement. The numerical values are consistent with unity for all categories and for the whole region, the value of the data minus the background divided by the predicted \ZZ\ yield is $q_{\ZZ} = 0.99 \pm 0.09$.

\begin{figure}[]
\centering
\includegraphics[width=0.48\textwidth]{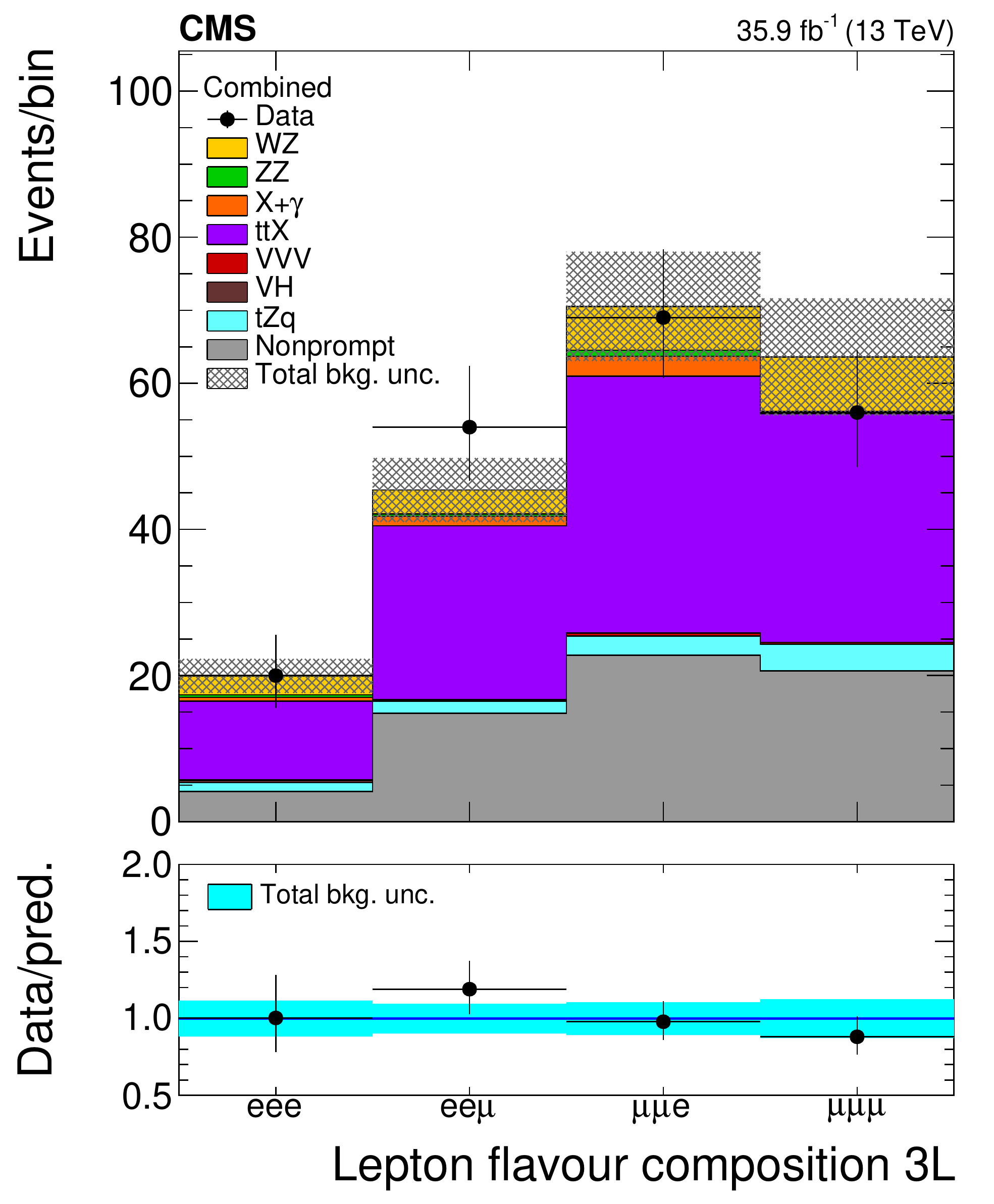}
\includegraphics[width=0.48\textwidth]{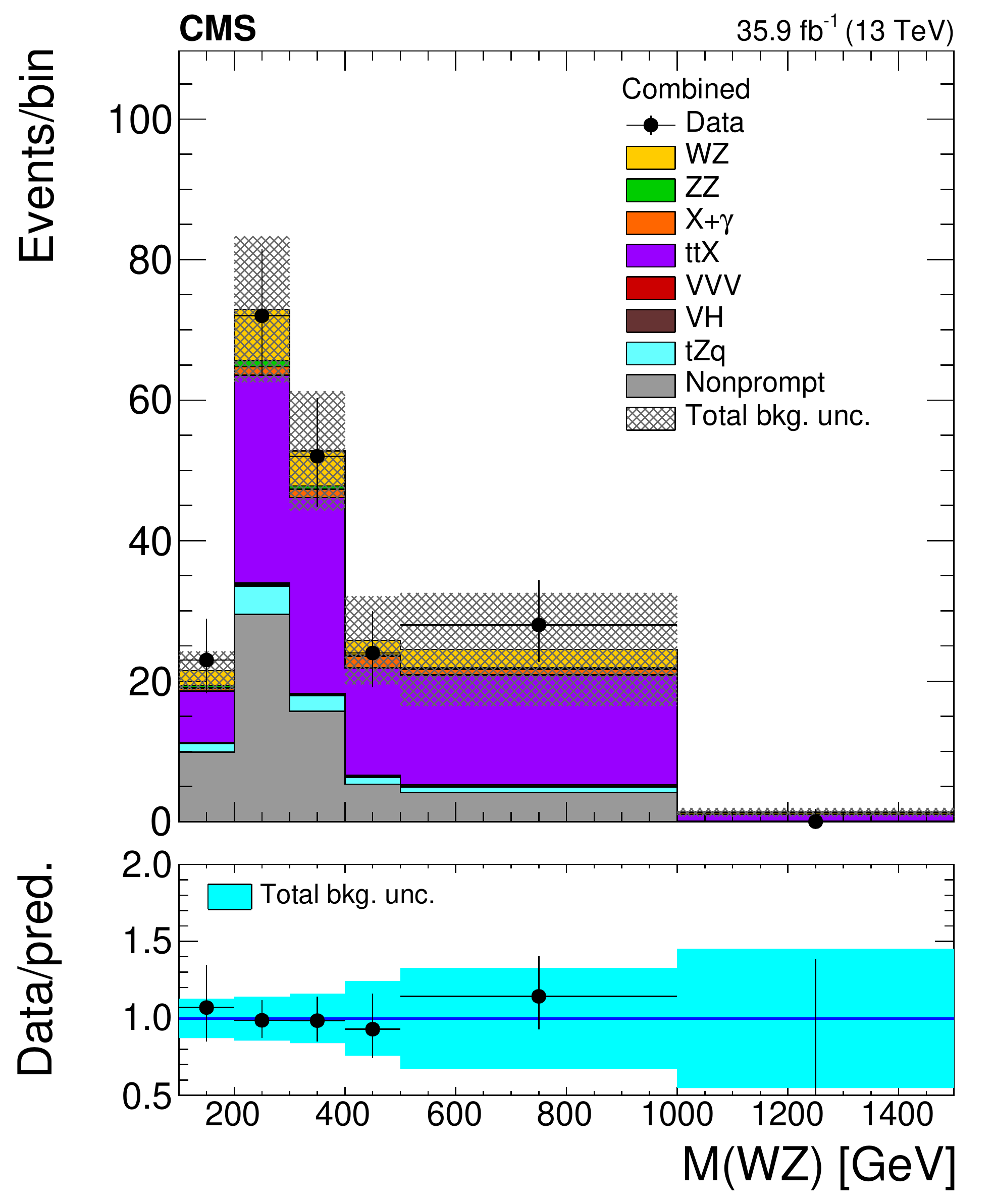}\\
\includegraphics[width=0.48\textwidth]{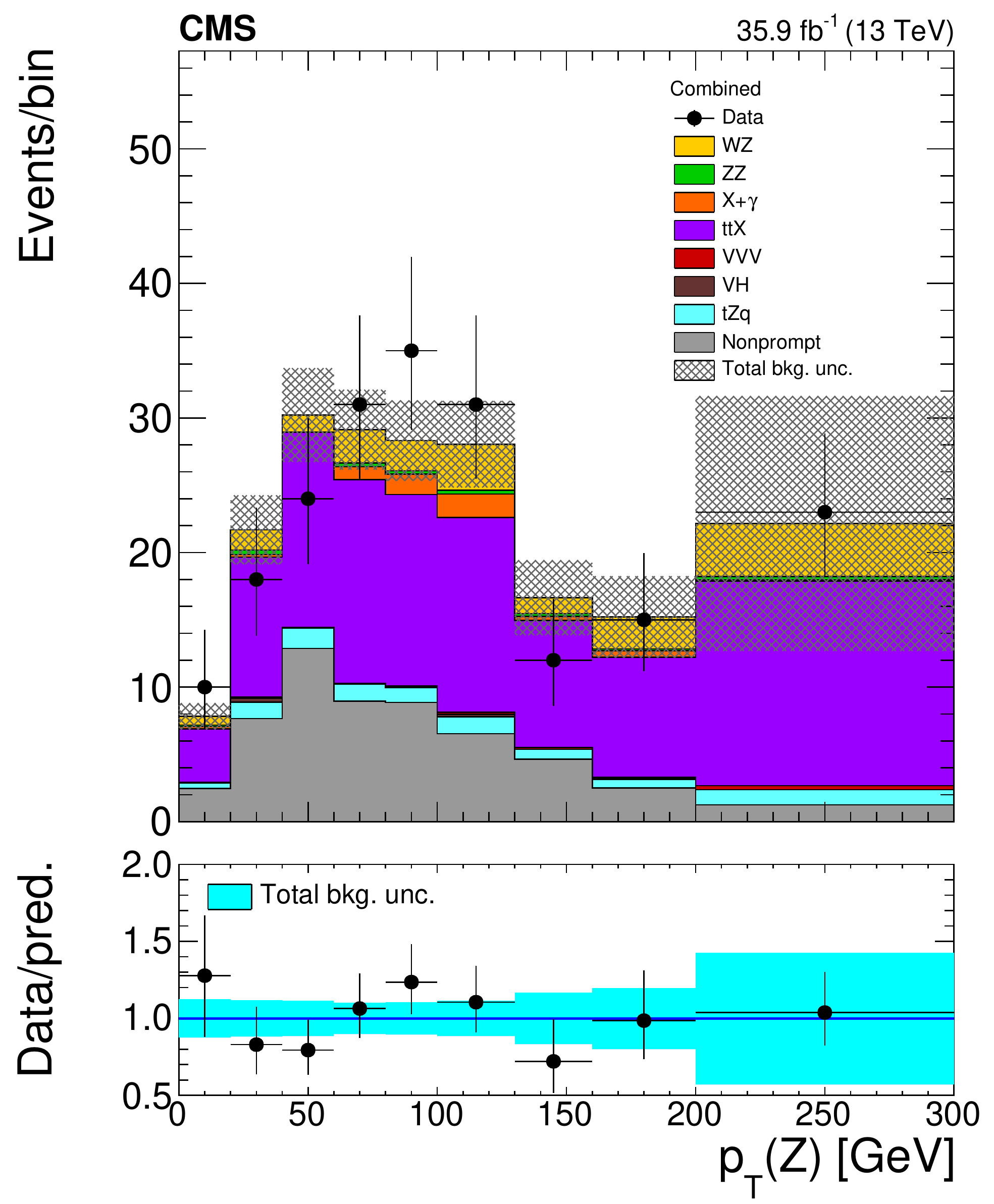}
\includegraphics[width=0.48\textwidth]{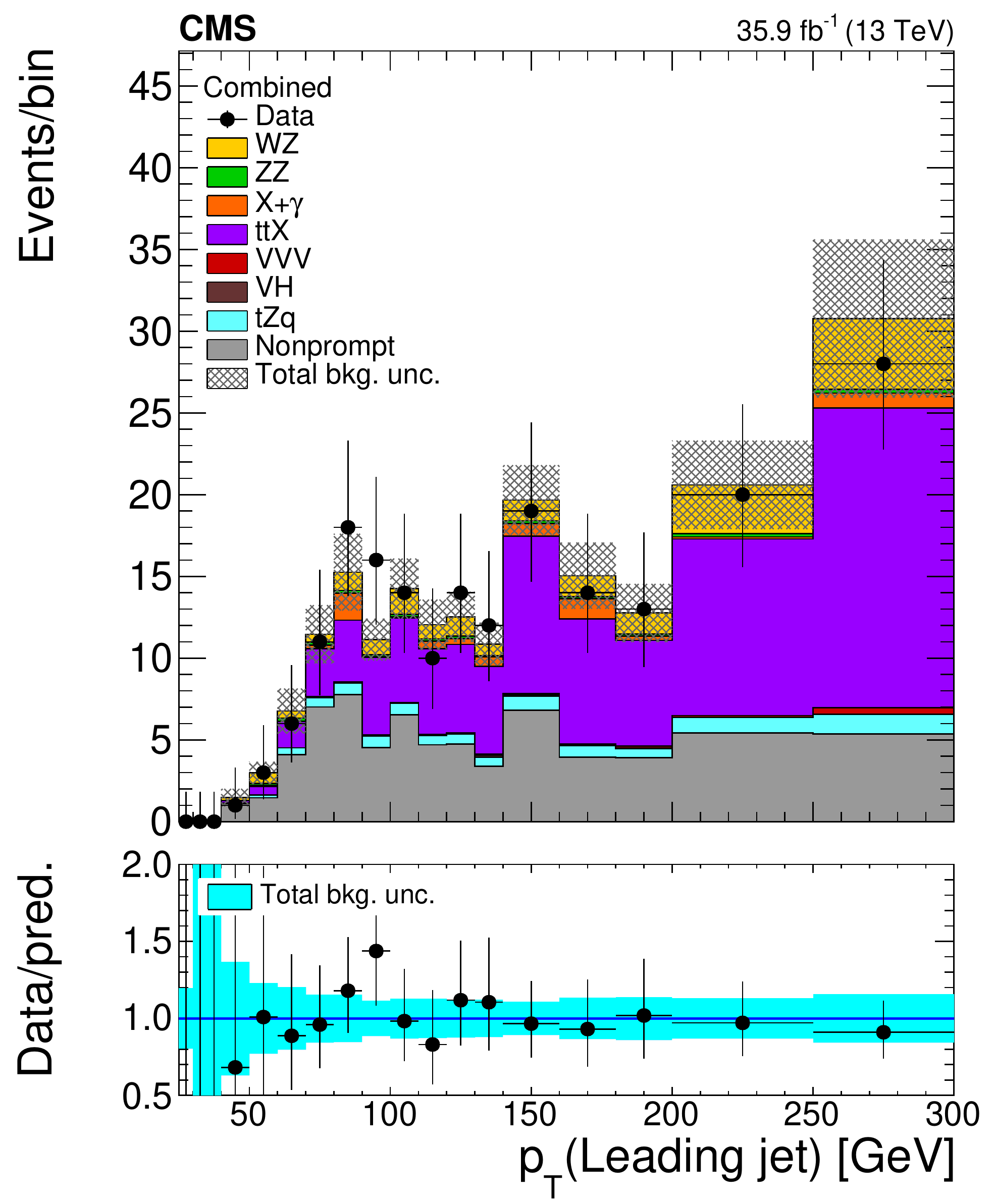}
\caption{Distribution of key observables in the top enriched control region defined in Table~\ref{tab:SRCR}: flavour composition of the three leading leptons (top left), invariant mass of the three lepton plus missing transverse momentum (top right), transverse momentum of the \PZ\ boson reconstructed from the \pt of the two leptons assigned to it (bottom left), and transverse momentum of the leading jet (bottom right). Vertical bars on the data points include the statistical uncertainty and shaded bands over the prediction include the contributions of the different sources of uncertainty evaluated after the signal extraction fit.}
\label{fig:crTT}
\end{figure}

Top quark enriched prompt processes are dominated by \ttZ and \tZq production, where the \cPZ\ boson and one of the top quarks decay leptonically.
A procedure similar to the one used for the estimation of the \ZZ\ background is performed in CR-top region.
The key observables of the analysis in this sideband region are shown in Fig.~\ref{fig:crTT}.
The estimation procedure results in good agreement across the different flavour categories; the global quotient of data minus background over the predicted \ttZ\ plus \tZq\ yields is consistent with unity, $q_{\ttZ + \tZq} = 1.09 \pm 0.20$.

The last major background that contributes to our search is the production of asymmetrical final-state photon conversions.
The production of $\PZ\gamma$ events makes up 99\% of this contribution.
The lepton assignment algorithm tends to match the electron originating from the photon to the \PW\ boson so the contribution in the \eee\ and \emm\ categories is highly enhanced.
The procedure used for the prompt contributions is used to validate the behaviour of the simulated conversion processes in a region denoted as CR-conv and defined in Table~\ref{tab:SRCR}.

\begin{figure}[]
\centering
\includegraphics[width=0.48\textwidth]{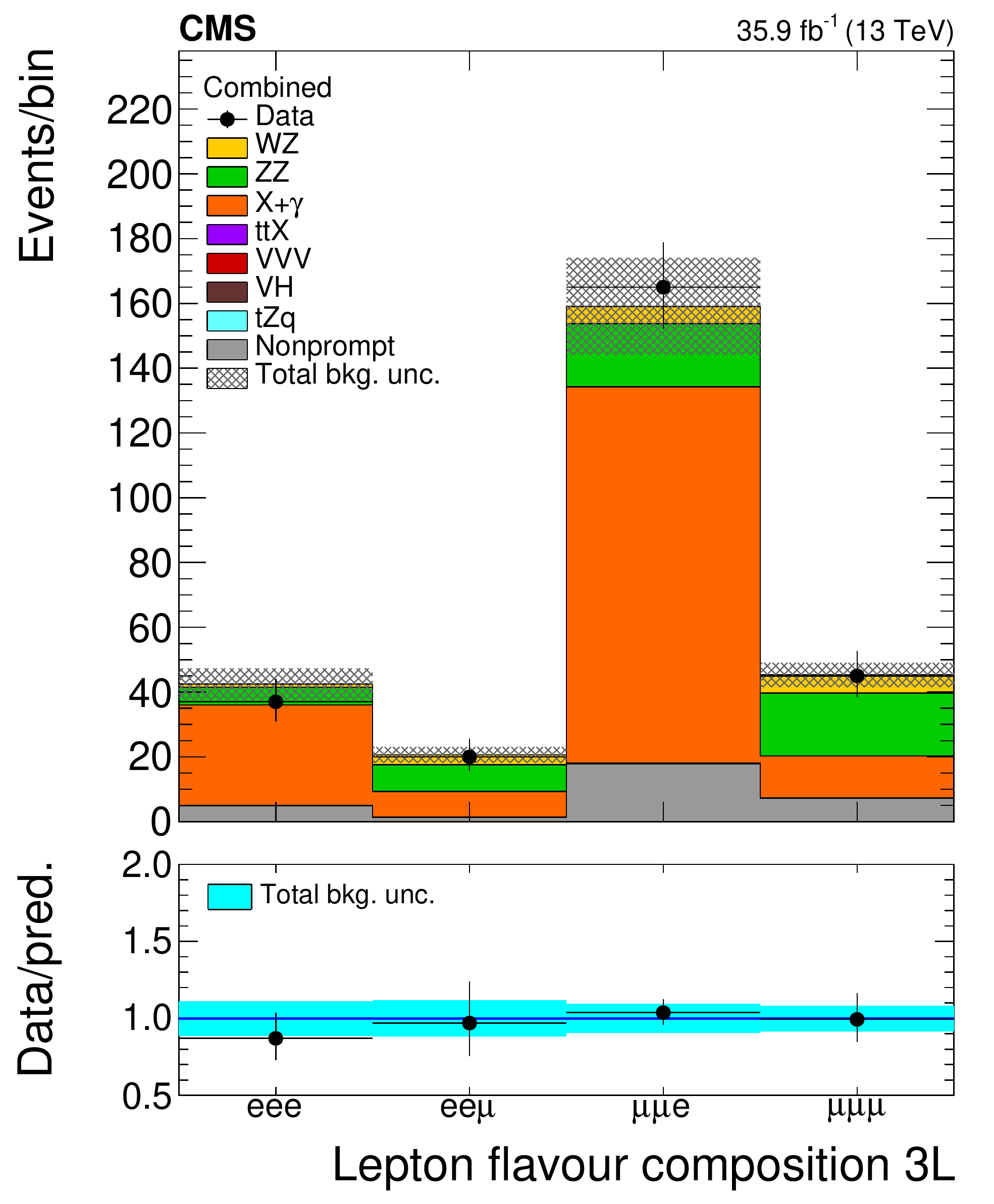}
\includegraphics[width=0.48\textwidth]{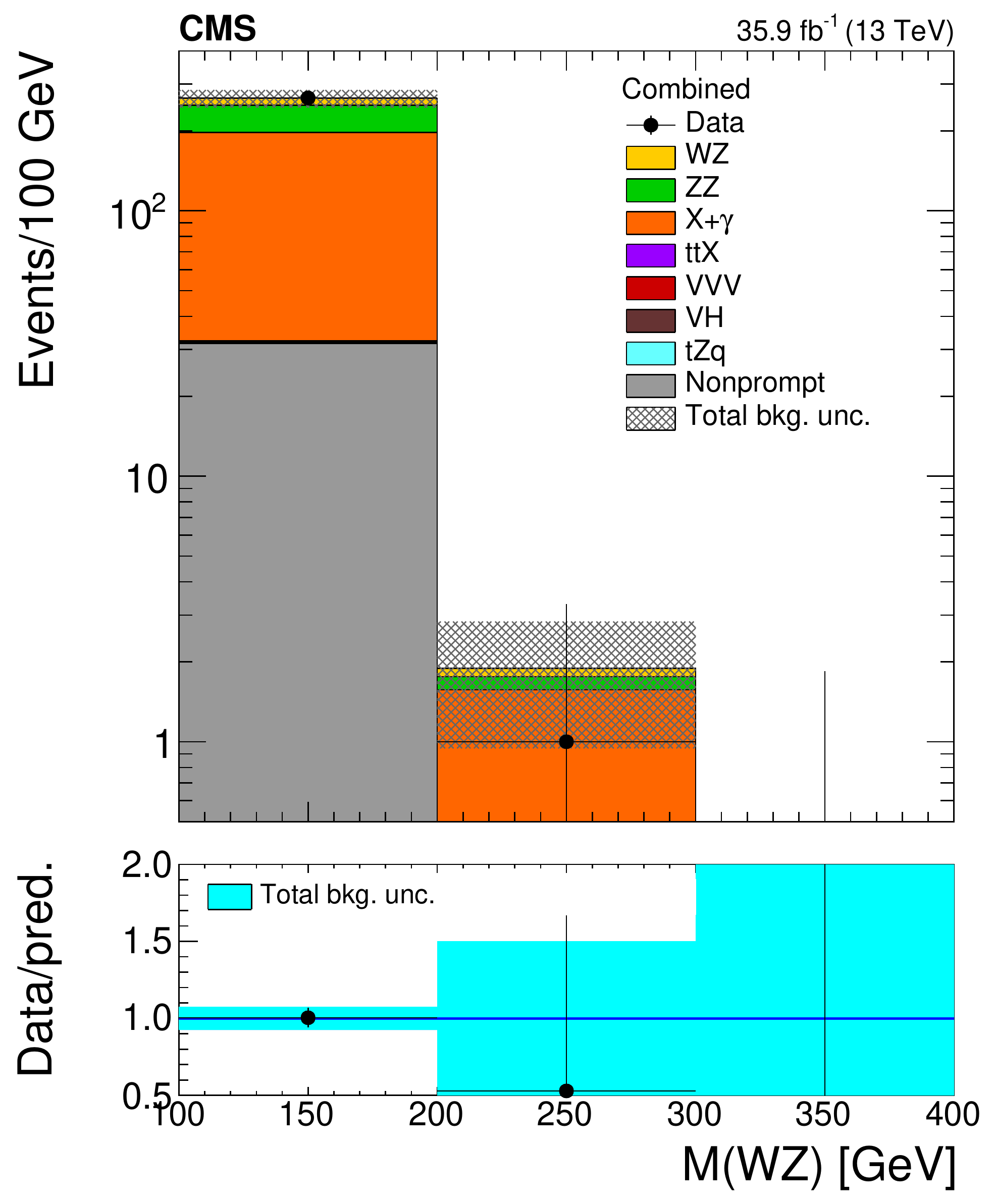}\\
\includegraphics[width=0.48\textwidth]{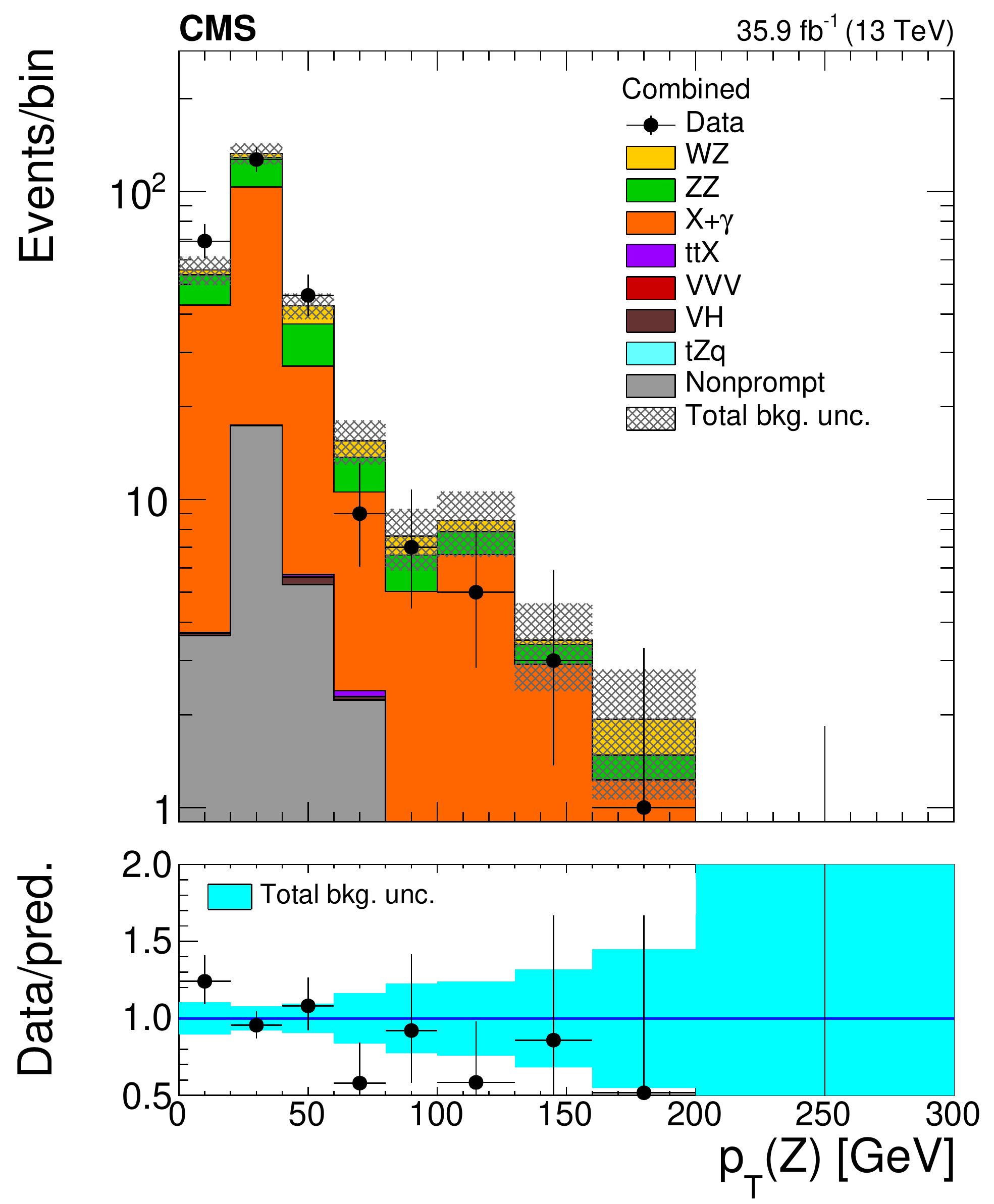}
\includegraphics[width=0.48\textwidth]{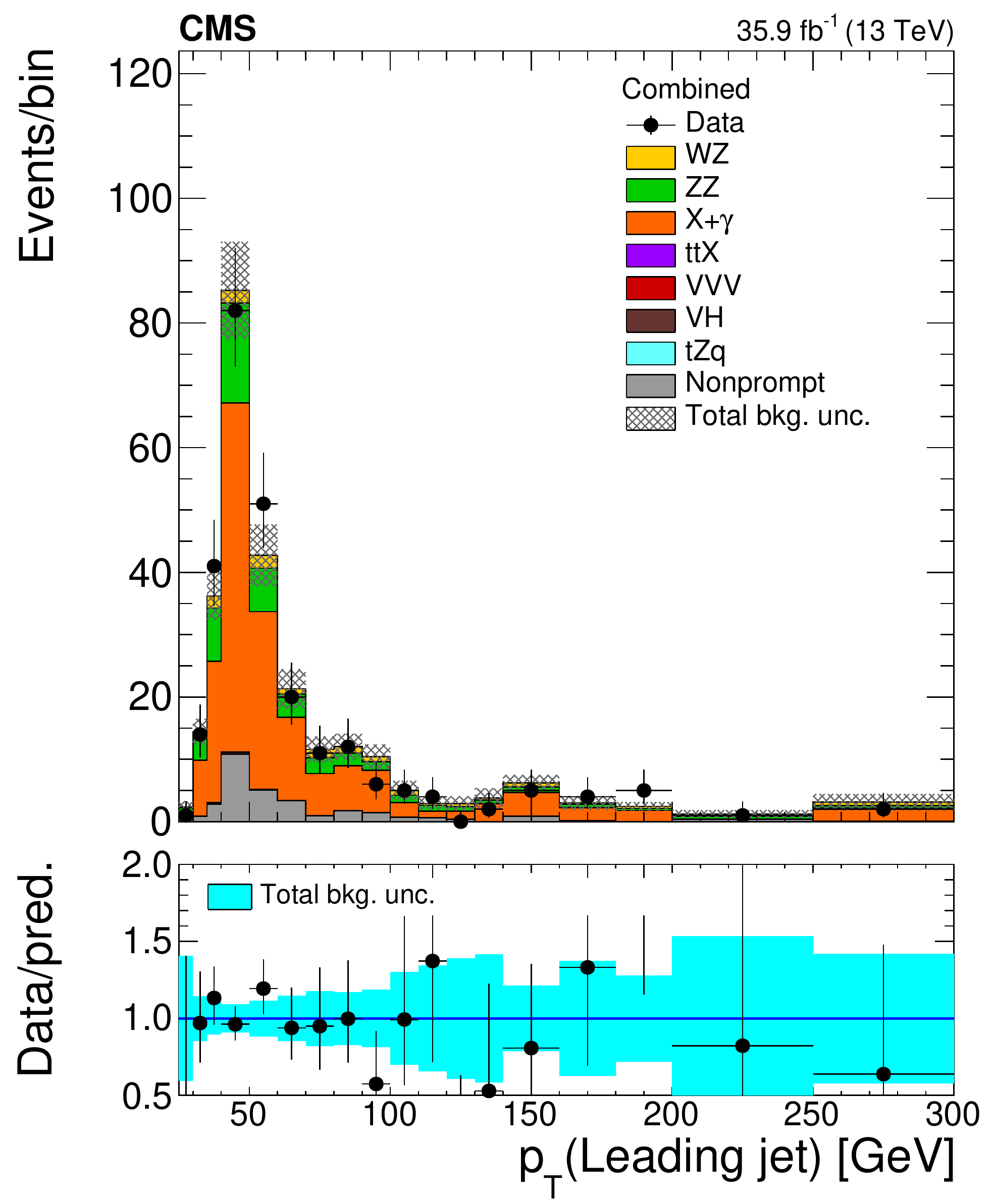}
\caption{Distribution of key observables in the conversion control region defined in Table~\ref{tab:SRCR}: flavour composition of the three leading leptons (top left), invariant mass of the three lepton plus missing transverse momentum (top right), transverse momentum of the \PZ\ boson reconstructed from the \pt of the two leptons assigned to it (bottom left), and transverse momentum of the leading jet (bottom right).  Vertical bars on the data points include the statistical uncertainty and shaded bands over the prediction include the contributions of the different sources of uncertainty evaluated after the signal extraction fit.}
\label{fig:crConv}
\end{figure}

Good agreement is found in the \emm\ and \eee\ categories, where sufficient statistical power is available. The normalization of the \xgamma background is estimated from the difference between the data and the other backgrounds, divided by the \xgamma (X=\ttbar, \PV, \cPqt) SM prediction; the result is $q_{\xgamma} = 1.11 \pm 0.14$, consistent with unity.
Validation plots for key observables used in the analysis are shown in Fig.~\ref{fig:crConv}.

Additional minor background contributions include the leptonic decays of multiboson production processes, dominated by V\PH and \vvv production where \PV\ is either the \cPZ\ or the \PW\ boson and\ \PH\ is the SM Higgs boson. Their contribution is estimated from simulation.

\section{Systematic uncertainties}\label{sec:systematics}

The major sources of systematic uncertainty can be grouped into three different categories: normalization uncertainties that are assigned to each of the background processes individually; global uncertainties related to the definition and energy measurement of the different physical objects, affecting both the background and signal acceptances; and a global uncertainty, correlated across all processes, that accounts for a possible mismeasurement of the total integrated luminosity.

As stated in Section~\ref{sec:background}, the contribution from prompt SM processes is estimated using MC samples and validated in appropriate control regions.
The uncertainties in the normalization of such processes are taken from experimental measurements performed at a centre-of-mass energy of $\sqrts = 13\TeV$, and correspond to assigning flat uncertainties of 7, 15, and 35\% to the contributions of the \ensuremath{\PZ\cPZ}, \ensuremath{\ttbar\PV}, and \ensuremath{\cPqt\cPZ\cPq}\ background processes respectively~\cite{Sirunyan:2017zjc,Sirunyan:2017nbr,TOP-17-005}. The uncertainty in the normalization of the photon conversion background contribution is obtained from the observations in the dedicated control region, and estimated to be about 20\%. The normalization uncertainties applied to the minor contributions of the multiboson production are estimated to be about 25\% for the V\PH process and 50\% for the \vvv ones.

The nonprompt background estimation includes two different sources of systematic uncertainties. First, a 30\% normalization uncertainty is applied to account for the observed variations in the performance of the method when applied to MC simulations. Second, a \pt- and $\eta$-dependent uncertainty that ranges between 5 and 30\% is applied to account for the differences observed amongst different \PW/\PZ background subtraction procedures considered for the tight-to-loose method.

Lepton identification and isolation introduce a sizeable uncertainty in the final measurement. Lepton efficiencies are computed using the tag-and-probe technique~\cite{Chatrchyan:2012jra,Chatrchyan:2012xi,Khachatryan:2015hwa}. Since electron and muon identification efficiencies are computed separately, the uncertainty in their estimate is split by flavour and evaluated separately.
The largest effects are in the \eee\ category for the electron efficiency (about 5\%) and in the \mmm\ category for the muon efficiency (about 3\%). The uncertainty in the energy scale of the leptons is estimated to produce a variation of 1\% in their \pt; the reconstructed muon \pt\ is computed with a different method for high-\pt muons (above 200\GeV), thus a conservative 5\% uncertainty is assigned to each high-\pt\ muon. The uncertainty in the lepton energy scale is  assigned to each lepton---separately for electrons and muons---and propagated to the yields, with effects smaller than 1\% in most cases.

A total trigger efficiency uncertainty is applied across all channels and processes to account for the differences observed between data and MC samples. Two different sources are considered for the estimation of this uncertainty. First, trigger efficiencies are measured in data and simulation samples, with the difference between the two being assigned as a systematic component of the trigger efficiency uncertainty. Second, the effect of limited statistical power in the data measurement is computed using Clopper--Pearson intervals~\cite{doi:10.1093/biomet/26.4.404}, which is an estimation method that yields intervals in the physical region using an estimation that is statistically robust even when the efficiency is close to its extreme values---in this case the value 1---, and added quadratically to the first source. A final asymmetric flat uncertainty of $-1.8$ and $+1.4$\% is applied.

The efficiency of the \cPqb\ tag veto is also corrected by comparing the measurements in data and simulation and propagated to each of the events. Separate uncertainty sources are considered for the \cPqb\ jet identification efficiency and the misidentification of light-flavour jets as \cPqb-tagged jets, with effects of up to 1.6 and 0.7\% in the final signal acceptance, respectively.

Each of the reconstructed jets has an associated energy scale uncertainty of 2--10\% depending on its \pt\ and\ $\eta$. The final measurement is sensitive to this kind of variation through the changes in acceptance that arise in the \ptmiss\ estimations and the \cPqb\ tag veto. The effect on the final signal acceptance amounts to about 1\%.

The pileup modelling uncertainty is evaluated by varying the inelastic cross section up and down by 5\% and propagating the effect to the final signal region, resulting in an uncertainty of up to 1.2\%.

A fiducial region is defined by imposing requirements that mimic the lepton kinematic characteristics in the signal region. The acceptance $\mathcal{A}$ is defined as the fraction of events in the total phase space that pass the requirements of the fiducial region. The efficiency $\epsilon$ is estimated as a transfer factor from the fiducial region to the signal region. Both acceptance and efficiency are estimated using generator-level information; details on the fiducial region, the acceptance, and the efficiency are provided in Section~\ref{sec:inclusive}. Two sources of theoretical uncertainty in $\mathcal{A}$ and $\epsilon$ are considered. Effects due to factorization ($\mu_{\mathrm{F}}$) and renormalization ($\mu_{\mathrm{R}}$) scale choices are evaluated with \POWHEG by varying the scales up and down independently by a factor of two around the nominal value $\mu_0 = (\mZ + \mW)/2$, under the constraint $0.5 < \mu_{\mathrm{F}}/\mu_{\mathrm{R}} < 2.0$. The envelope of the set of variations is assigned as a systematic uncertainty on the yields. Parametric (PDF +\alpS) uncertainties are estimated using the {\textsc{PDF4LHC}}\xspace prescription~\cite{Butterworth:2015oua} with the NNPDF3.0 set~\cite{Ball:2014uwa}.

Finally, a 2.5\% correlated normalization uncertainty is applied to all signal and background processes to account for the variations in the measurement of the total integrated luminosity~\cite{CMS-PAS-LUM-17-001}.

\section{Inclusive measurement}
\label{sec:inclusive}

\begin{figure}[]
\centering
\includegraphics[width=0.5\textwidth]{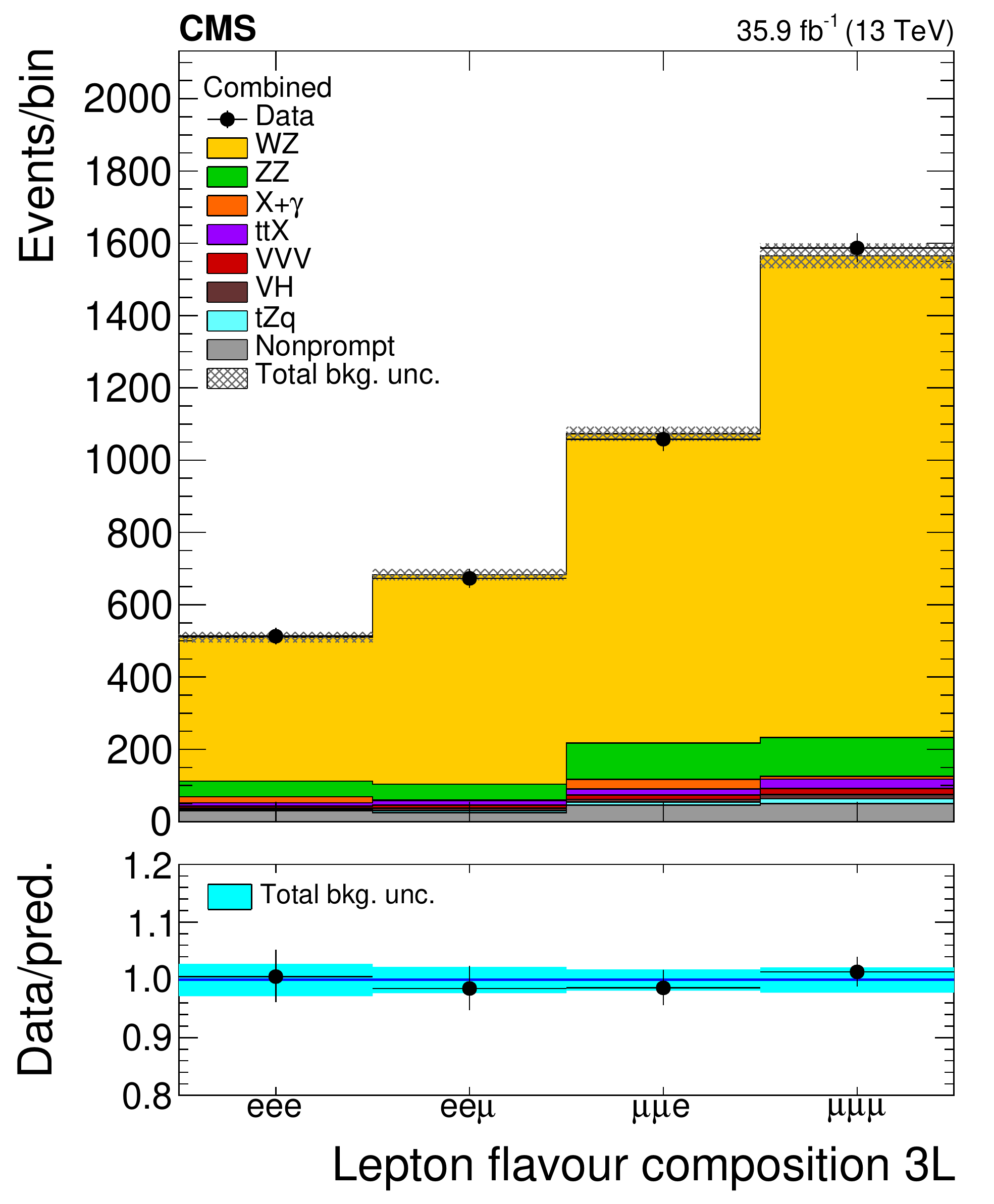}
\caption{Distribution of expected and observed event yields in the four flavour categories used for the cross section measurement.
  Vertical bars on the data points include the statistical uncertainty and shaded bands over the prediction include the contributions of the different sources of uncertainty evaluated after the signal extraction fit.}
\label{fig:srBlind}
\end{figure}

The inclusive \WZ\ production cross section is measured by performing a simultaneous maximum likelihood fit to the total yields in the four flavour categories of the signal region, as presented in Fig.~\ref{fig:srBlind}. The normalization of the \WZ\ signal process is modelled via a parameter representing a multiplicative factor for the total NLO production cross section; the parameter is referred to as signal strength $r_{\PW\PZ}$ and is a free parameter in the fit.

The contributions from the background processes are allowed to vary around the predicted yields, according to the systematic contributions described in Section~\ref{sec:systematics}. The systematic contributions are modelled in the likelihood as nuisance parameters with log-normal priors. The expected and observed yields for the processes involved in each of the flavour categories can be seen in Table~\ref{tab:srPos}. The final contribution of the different sources of uncertainty to the measurement is described in Table~\ref{tab:outSysts}.

A fiducial region is defined by imposing requirements that mimic the lepton kinematic characteristics in the signal region. We require three light leptons located inside the detector acceptance, $\abs{\eta^\ell} < 2.5 (2.4)$ for electrons (muons), with at least one OSSF pair. Electrons and muons from  \PW/\cPZ$\to$\Pgt+X$\to\ell$+X decays are included in this selection. These leptons are assigned to the \PW\ and \PZ\ bosons using the algorithm described in Section~\ref{sec:selection}. Minimum transverse momenta requirements of $\ptzone > 25\GeV$, $\ptztwo > 10\GeV$, and $\ptw > 25\GeV$ are applied. We also apply the two additional criteria $M(\ell_{\PZ 1}\ell_{\PZ 2}\ell_{\PW} ) > 100\GeV$ and $\abs{M(\ell_{\PZ 1}\ell_{\PZ 2}) - \mZ} < 15\GeV$. The total yields in the signal region for the expected background, \nbkg, and observed data, \nobs, are used to obtain the fiducial cross section of the \WZ\ process through the expression:
\begin{equation}
\sigma_{\text{fid}}(\Pp\Pp \to \WZ) = \frac{\nobs - \nbkg}{\epsilon\mathcal{L}},
\end{equation}
where the efficiency $\epsilon$ is estimated as a transfer factor from the fiducial region to the signal region using MC truth and the integrated luminosity $\mathcal{L}$ amounts to \fulllumi. Scale and PDF uncertainties are considered in the computed efficiency and are propagated to the final cross section measurement. Table~\ref{tab:transferfactors} summarizes the efficiencies and their uncertainties. Final state generator-level leptons are \emph{dressed} by adding to their momenta those of generator-level photons within a cone of $\Delta R(\ell,\gamma) < 0.1$. The efficiency is estimated from simulation for each of the flavour channels separately, and for the inclusive case, as the ratio of expected reconstructed events in the signal region to the number of generated trilepton events in the fiducial region. The statistical uncertainty in the measurement of the efficiency, which originates from the limited number of simulated events, is below 1\% and is added quadratically to the total sum of statistical uncertainties in the measurement. Theoretical uncertainties in the cross section measurements arise from renormalization and factorization scale and PDF uncertainties and are added quadratically to the experimental uncertainties. The results are presented in Table~\ref{tab:fiducial} and can be compared to the NLO prediction from \POWHEG + \PYTHIA of $\sigma_{\mathrm{fid}}^{\POWHEG} = 227.6 \mathrm{ }^{+8.8}_{-7.3}\,(\text{scale}) \pm   3.2\,(\mathrm{PDF})$\unit{fb}; the NLO prediction is disfavoured by this measurement.

\begin{table}[]
\centering
\topcaption{\label{tab:srPos} Expected and observed yields for each of the relevant processes and flavour categories. Combined statistical and systematic uncertainties are shown for each case except for the observed data yields for which only statistical uncertainties are presented. All expected yields correspond to quantities estimated after the maximum likelihood fit. Uncertainties are computed taking into account the full correlation matrix between sources of uncertainty, processes, and flavour categories.}
\begin{tabular}{lccccc}
\hline
Process        & \eee       & \eem      & \emm     & \mmm       & Total \\ \hline
Nonprompt      & $30.0 \pm 12.4$ & $25.0 \pm 10.4$ & $45.7 \pm 20.7$ & $50.3 \pm 19.3$    & $151 \pm 63$\\
\ZZ       & $43.4 \pm 4.1$  & $44.4 \pm 3.4$  & $100.1 \pm 9.2$  & $107.1 \pm 8.3$    & $295 \pm 24$\\
X$\gamma$      & $16.8 \pm 5.2$  & $2.0  \pm 0.7$  & $26.9 \pm 8.8$  & $7.6 \pm 2.0$      & $53 \pm 16$\\
\ttbar\PV & $8.5  \pm 2.8$  & $11.6 \pm 4.1$  & $16.8 \pm 5.5$  & $25.8 \pm 9.0$     & $63 \pm 21$\\
\vvv     & $6.2  \pm 2.5$  & $8.6 \pm  3.4$  & $11.4 \pm 4.6$  & $16.9 \pm 6.8$     & $43 \pm 17$\\
V\PH      & $3.3  \pm 0.8$  & $6.4 \pm  1.6$  & $7.7 \pm 1.9$   & $12.1 \pm 3.0$     & $29.6 \pm 7.2$\\
t\cPZ\cPq & $3.9  \pm 1.30$ & $5.7 \pm  1.9$  & $8.4 \pm 2.8$   & $12.6 \pm 4.3$      & $31 \pm 10$\\ [\cmsTabSkip]
Total background & $112 \pm 15$   & $104 \pm 15$    & $217 \pm 28$    & $233 \pm 29$       & $666 \pm 45$\\
\WZ          & $398 \pm 18$    & $579 \pm 21$    & $856 \pm 29$    & $1333 \pm 47$      & $3166 \pm 62$\\ [\cmsTabSkip]
Data             & $513 \pm 23$    & $673 \pm 26$    & $1058 \pm 32$   & $1587 \pm 40$      & $3831 \pm 62$\\ \hline
\end{tabular}
\end{table}

\begin{table}[]
\centering
\topcaption{\label{tab:outSysts} Summary of the total postfit impact of each uncertainty source on the uncertainty in the signal strength measurement, for the four flavour categories and their combination. Theoretical uncertainties are only included in the signal acceptance during the extrapolation to the total phase space, so they are not included in the likelihood fit. The values are percentages and correspond to half the difference between the up and down variation of each systematic uncertainty component. }
\begin{tabular}{lccccc}
\hline
Source                    & Combined & \eee & \eem & \emm & \mmm \\ \hline
Electron efficiency       & $1.9$      & $5.9$ & $3.9$ & $1.9$ & \NA   \\
Electron energy scale            & $0.3$      & $0.9$ & $0.2$ & $0.6$ & \NA   \\
Muon efficiency           & $1.9$      & \NA   & $0.8$ & $1.8$ & $2.6$ \\
Muon momentum scale                & $0.5$      & \NA   & $0.7$ & $0.3$ & $0.9$ \\
Trigger efficiency        & $1.9$      & $2.0$ & $1.9$ & $1.9$ & $1.8$ \\
Jet energy scale          & $0.9$      & $1.6$ & $1.0$ & $1.7$ & $0.8$ \\
\cPqb-tagging (id.)           & $2.6$      & $2.7$ & $2.6$ & $2.6$ & $2.4$ \\
\cPqb-tagging (mis-id.)       & $0.9$      & $1.0$ & $0.9$ & $1.0$ & $0.7$ \\
Pileup                    & $0.8$      & $0.9$ & $0.3$ & $1.3$ & $1.4$ \\ [\cmsTabSkip]
\ZZ                  & $0.6$      & $0.7$ & $0.4$ & $0.8$ & $0.5$ \\
Nonprompt norm.           & $1.2$      & $2.0$ & $1.2$ & $1.5$ & $1.0$ \\
Nonprompt (EWK subtr.)     & $1.0$      & $1.5$ & $1.0$ & $1.3$ & $0.8$ \\
\vvv norm.          & $0.5$      & $0.6$ & $0.6$ & $0.6$ & $0.5$ \\
V\PH norm.           & $0.2$      & $0.2$ & $0.3$ & $0.2$ & $0.2$ \\
\ttbar\PV norm.      & $0.5$      & $0.5$ & $0.5$ & $0.5$ & $0.5$ \\
t\cPZ\cPq\ norm.       & $0.1$      & $0.1$ & $0.1$ & $0.1$ & $0.1$ \\
X+$\gamma$ norm.          & $0.3$      & $0.8$ & $<0.1$   & $0.7$ & $<0.1$   \\ [\cmsTabSkip]
Total systematic          & $4.7$      & $7.8$ & $5.8$ & $5.4$ & $4.6$ \\ [\cmsTabSkip]
Integrated luminosity                & $2.8$      & $2.9$ & $2.8$ & $2.9$ & $2.8$ \\ [\cmsTabSkip]
Statistical               & $2.1$      & $6.0$ & $4.8$ & $4.1$ & $3.1$ \\ [\cmsTabSkip]
Total experimental        & $6.0$      & $10.8$& $8.0$ & $7.5$ & $6.3$ \\ [\cmsTabSkip]
Theoretical               & $0.9$      & $0.9$ & $0.9$ & $0.9$ & $0.9$ \\ \hline
\end{tabular}
\end{table}

\begin{table}[]
\centering
\topcaption{\label{tab:transferfactors} Efficiencies estimated as transfer factors from the fiducial region to the signal region using generator-level information, for an integrated luminosity $\mathcal{L}$ of \fulllumi. Statistical, scale, and PDF uncertainties are later propagated to the final cross section measurement.}
\begin{tabular}{cc}
\hline
Category       & $\epsilon$ \\ \hline
\rule{0pt}{3ex}\eee      & $0.1754 \pm 0.0003\stat \mathrm{ }^{+0.0017}_{-0.0015}\,(\text{scale, PDF})$ \\
\rule{0pt}{3ex}\eem      & $0.2618 \pm 0.0004\stat \mathrm{ }^{+0.0025}_{-0.0021}\,(\text{scale, PDF})$ \\
\rule{0pt}{3ex}\emm      & $0.3764 \pm 0.0006\stat \mathrm{ }^{+0.0035}_{-0.0030}\,(\text{scale, PDF})$ \\
\rule{0pt}{3ex}\mmm      & $0.5625 \pm 0.0009\stat \mathrm{ }^{+0.0047}_{-0.0040}\,(\text{scale, PDF})$ \\
\rule[-2ex]{0pt}{0pt}\rule{0pt}{3ex} Combined: & $0.3453 \pm 0.0005\stat \mathrm{ }^{+0.0031}_{-0.0027}\,(\text{scale, PDF})$ \\ \hline
\end{tabular}
\end{table}

\begin{table}[h!]
\centering
\topcaption{\label{tab:fiducial} Measured fiducial cross sections and their corresponding uncertainties for each of the individual flavour categories, as well as for the combination of the four. The combined value is the result of a simultaneous fit to the four categories, therefore both the central value and its total uncertainty differ from the sum of the central values and the quadratic sum of the uncertainties respectively, because of correlations among sources of uncertainty in the categorized values.}
\begin{tabular}{cc}
\hline
Category & $\sigma_{\text{fid}}(\Pp\Pp \to \WZ)$ [fb] \\ \hline
\rule{0pt}{3ex}\eee          & $63.7 \mathrm{ }^{+3.8}_{-3.7}\stat \mathrm{ }^{+0.6}_{-0.6}\thy\mathrm{ }^{+5.3}_{-4.7}\syst  \pm 1.9\lum$ \\
\rule{0pt}{3ex}\eem    & $61.6 \mathrm{ }^{+3.0}_{-2.9}\stat \mathrm{ }^{+0.6}_{-0.5}\thy \mathrm{ }^{+3.7}_{-3.3}\syst  \pm 1.9\lum$ \\
\rule{0pt}{3ex}\emm   & $63.4 \mathrm{ }^{+2.6}_{-2.6}\stat \mathrm{ }^{+0.6}_{-0.5}\thy \mathrm{ }^{+3.5}_{-3.2}\syst  \pm 1.9\lum$ \\
\rule{0pt}{3ex}\mmm & $67.1 \mathrm{ }^{+2.1}_{-2.0}\stat \mathrm{ }^{+0.6}_{-0.5}\thy \mathrm{ }^{+3.3}_{-3.0}\syst  \pm 1.9\lum$ \\
\rule[-2ex]{0pt}{0pt}\rule{0pt}{3ex} Combined    & $257.5 \mathrm{ }^{+5.3}_{-5.0}\stat \mathrm{ }^{+2.3}_{-2.0}\thy \mathrm{ }^{+12.8}_{-11.6}\syst  \pm 7.4\lum$ \\
\hline
\end{tabular}
\end{table}

The phase space used for the computation of the total cross section is defined by having three generated light leptons that pass the requirement $60\GeV < \mZ^{\mathrm{OSSF}} < 120\GeV$, where $\mZ^{\mathrm{OSSF}}$ is the mass closest to the \PZ\ boson mass among those computed with all possible OSSF lepton pairs. Light leptons originating from tau decay are included in the definition of the total region selection. The extrapolation to the total associated production cross section of \WZ\ bosons is computed as:
\begin{equation}
\sigma_{\text{tot}}(\Pp\Pp \to \WZ) = \frac{\nobs - \nbkg}{\mathcal{B}(\PW \to \ell + X)\mathcal{B}(\PZ \to \ell' \ell' + X)\mathcal{A}\epsilon\mathcal{L}},
\end{equation}
where the leptonic branching ratios of the \PW\ and \cPZ\ bosons, $\mathcal{B}(\PW \to \ell +X) = \mathcal{B}(\PW \to \ell +\nu) + \mathcal{B}(\PW \to \tau +\nu)\mathcal{B}(\tau \to \ell +2\nu)$ and $\mathcal{B}(\PZ \to \ell' \ell' +X) = \mathcal{B}(\PZ \to \ell +\ell) + \mathcal{B}(\PZ \to \tau +\tau)\mathcal{B}(\tau \to \ell +2\nu)^2$, are taken from the current world averages~\cite{PDG2016} and include both the direct leptonic decays of the \PW\ and \cPZ\ bosons and their decays to leptonically decaying $\tau$ leptons. The acceptance $\mathcal{A}$ accounts for the fraction of events in the total phase space that pass the requirements of the fiducial region and is estimated using generator-level information. The same procedure used in the fiducial measurement is applied to estimate the effect of theoretical uncertainties and the limited number of simulated events used in the measurement.

\begin{table}[]
\centering
\topcaption{\label{tab:IncXSec} Measured \WZ\ production cross sections computed separately in each of the flavour categories.}
\begin{tabular}{lc}
\hline
Category       & $\sigmatot(\Pp\Pp \to \WZ)$ [pb] \\ \hline
\rule{0pt}{3ex} \eee      & $47.11^{+5.01}_{-4.63}\,(\mathrm{total}) = 47.11 \mathrm{ }^{+2.88}_{-2.79}\stat \mathrm{ }^{+0.46}_{-0.41}\thy \mathrm{ }^{+3.89}_{-3.47}\syst  \pm 1.41\lum$ \\
\rule{0pt}{3ex} \eem     & $47.16^{+3.87}_{-3.61}\,(\mathrm{total}) = 47.16 \mathrm{ }^{+2.31}_{-2.29}\stat \mathrm{ }^{+0.45}_{-0.38}\thy \mathrm{ }^{+2.83}_{-2.52}\syst  \pm 1.33\lum$ \\
\rule{0pt}{3ex} \emm    & $47.70^{+3.58}_{-3.55}\,(\mathrm{total}) = 47.70 \mathrm{ }^{+2.00}_{-1.96}\stat \mathrm{ }^{+0.45}_{-0.39}\thy \mathrm{ }^{+2.66}_{-2.61}\syst  \pm 1.42\lum$ \\
\rule[-2ex]{0pt}{0pt}\rule{0pt}{3ex} \mmm   & $49.00^{+3.18}_{-3.03}\,(\mathrm{total}) = 49.00 \mathrm{ }^{+1.57}_{-1.53}\stat \mathrm{ }^{+0.41}_{-0.35}\thy \mathrm{ }^{+2.42}_{-2.22}\syst  \pm 1.39\lum$ \\ \hline
\end{tabular}
\end{table}

The results obtained for each flavour category are listed in Table~\ref{tab:IncXSec}. The combined measurement is defined as the measurement obtained from a simultaneous fit to the four categories; the resulting value is:
\begin{equation*}
\sigmatot(\Pp\Pp \to \WZ) = 48.09^{+ 2.98}_{-2.78}\,\unit{pb} = 48.09 \mathrm{ }^{+1.00}_{-0.96}\stat \mathrm{ }^{+0.44}_{-0.37}\thy \mathrm{ }^{+2.39}_{-2.17}\syst  \pm 1.39\lum\,\unit{pb},
\end{equation*}
which can be compared to theoretical predictions at parton level~\cite{theoWZ} using\ \textsc{MATRIX}~\cite{theoWZ} at NLO, $\sigma_{\mathrm{NLO}}(\Pp\Pp \to \WZ) = 45.09^{+4.9\%}_{-3.9\%}$\unit{pb}, {\tolerance=1000 and next-to-next-to-leading order (NNLO)~\cite{Grazzini:2017mhc}, $\sigma_{\mathrm{NNLO}}(\Pp\Pp \to \WZ) = 49.98^{+2.2\%}_{-2.0\%}$, in perturbative QCD, as well as the prediction obtained with \POWHEG + \PYTHIA at NLO QCD, of $\sigma_{\text{Pow}}^{\mathrm{NLO}} = 42.5 \mathrm{ }^{+1.6}_{-1.4}\,(\text{scale}) \pm  0.6\,(\mathrm{PDF})$\unit{pb}. Uncertainties in the theoretical values are derived from scale variations.\par}

\subsection{Charge-dependent measurements}
\begin{figure}[!hbtp]
\centering
\includegraphics[width=0.5\textwidth]{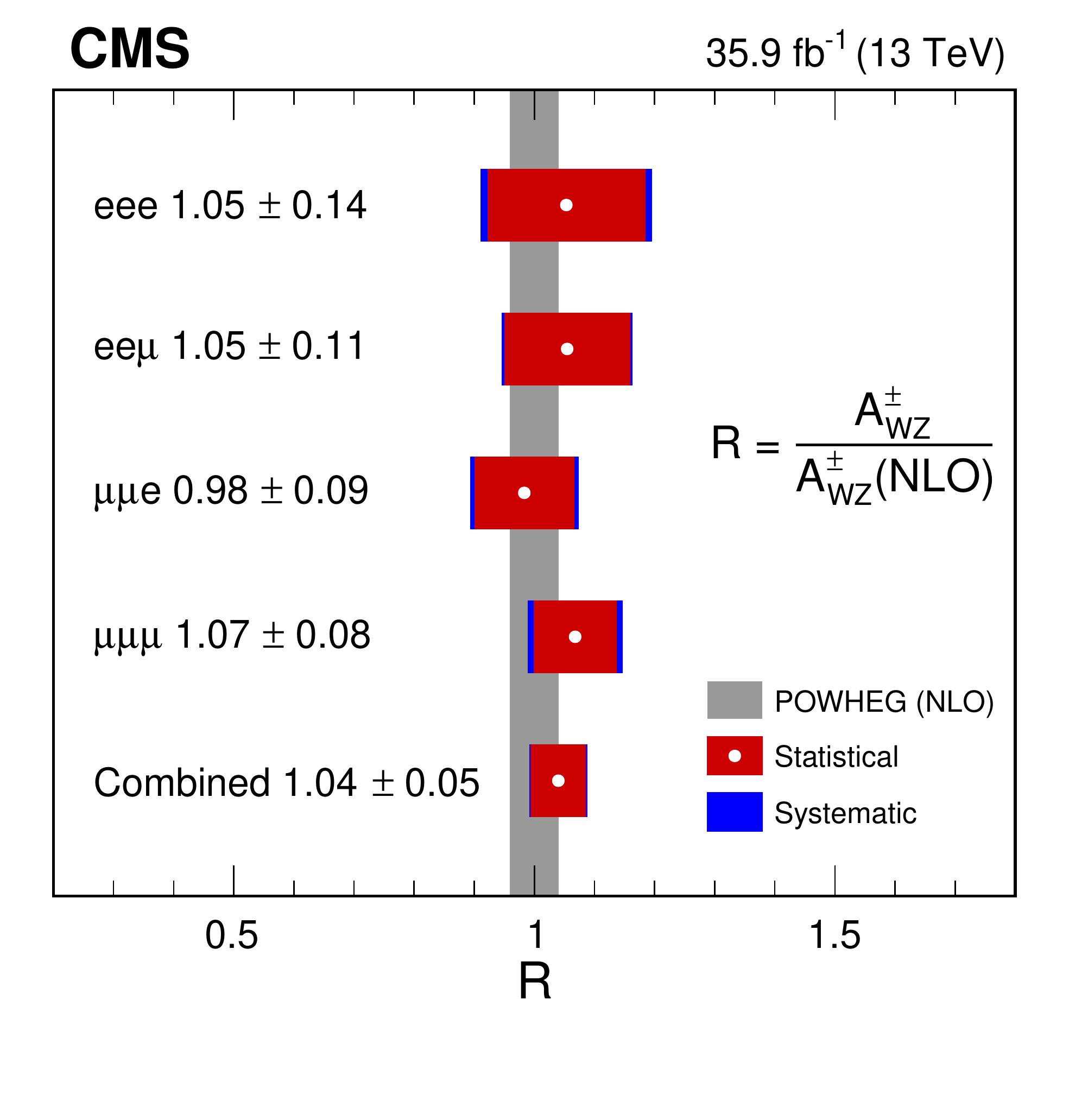}
\caption{Measured ratio of cross sections for the two charge channels for each of the flavour categories and their combination. Values are normalized to the NLO prediction obtained with \POWHEG. Coloured bands for each of the points include both systematic and statistical uncertainties. Shaded bands correspond to the MC prediction from the nominal \POWHEG sample and its associated uncertainty.}
\label{fig:chargeasym}
\end{figure}

The signal process is further divided depending on the charge of the \PW\ boson in order to compute the $\PW^+\cPZ$ and $\PW^-\cPZ$ production cross sections and their ratio; the value obtained for the ratio is then compared with theoretical predictions. The procedure described in the previous section is applied separately for the two categories classified according to the charge of the lepton associated with the \PW\ boson. The results are:
\begin{equation*}
\begin{aligned}
\sigmatot(\Pp\Pp \to \PW^+\cPZ) && = 28.91 \mathrm{ }^{+0.63}_{-0.61}\stat \mathrm{ }^{+0.28}_{-0.25}\thy \mathrm{ }^{+1.43}_{-1.31}\syst  \pm 0.80\lum\,\unit{pb},\\
\sigmatot(\Pp\Pp \to \PW^-\cPZ) && = 19.55 \mathrm{ }^{+0.45}_{-0.44}\stat \mathrm{ }^{+0.17}_{-0.15}\thy \mathrm{ }^{+0.97}_{-0.88}\syst  \pm 0.55\lum\,\unit{pb}.\\
\end{aligned}
\end{equation*}
The ratio between the charge-dependent production cross sections is calculated. Statistical uncertainties are treated as completely uncorrelated between the two values, while the other sources of uncertainty are considered completely correlated in their propagation to the ratio. The final effect is that most of the systematic uncertainties show a similar behaviour in both cases so they are greatly reduced when computing the ratio. The value obtained for the ratio is:
\begin{equation*}
\awz = \frac{\sigmatot(\Pp\Pp \to \PW^+\cPZ)}{\sigmatot(\Pp\Pp \to \PW^-\cPZ)} = 1.48 \pm 0.06\stat \pm 0.02\syst \pm 0.01\thy,
\end{equation*}
{\tolerance=1000 which is compatible within the uncertainties with the \POWHEG + \PYTHIA prediction of $\awz(NLO) = 1.43^{+0.06}_{-0.05}$. The same results, split by flavour category, are shown in Fig.~\ref{fig:chargeasym}.\par}

\section{Differential measurement}
\label{sec:differential}

The differential \WZ\ cross sections are measured in the full volume of the detector as a function of three observables.
To better model the data, in the definition of such observables leptons are \emph{dressed} in simulation by adding to their momenta those of all generator-level photons within a cone of $\Delta R(\ell,\gamma) < 0.1$.

The first observable is the \pt of the \PZ\ boson,
defined as the transverse sum of the momenta of the two final-state leptons assigned to the \PZ\ boson decay.
The second observable is the \pt of the leading jet, which represents a probe for the boost of the \WZ\ system recoiling against initial-state radiation.
The generated leading jet is defined using the anti-\kt algorithm with a cone radius of 0.4,
and by requiring a spatial separation of $\Delta R > 0.5$ from the leptons coming from the \WZ\ decay.
The third observable is the \mwz variable, defined as the invariant mass of the system composed of the three leptons and the \ptmiss.
A general formula for the definition of the variable is:
\begin{equation}
  M(WZ)^2 =  \left[p(\ell_1) + p(\ell_2) + p(\ell_3) + p(\nu)\right]^2,
\end{equation}
where $p(\ell_i)$ is the measured four-momentum of each lepton.
The four-momentum of the neutrino is defined in the (mass, \pt, $\eta$, $\phi$) base as
$p(\nu) = (0, \pt(\ptmiss) , 0, \phi(\ptmiss))$. Slightly different choices (solving the $\PW\to\ell\nu$ system for $\eta(\nu)$ or setting the neutrino four-momentum to zero) were tested as well, giving similar results.

The reconstructed quantities are defined using the objects described in Section~\ref{sec:objects}, and the pair of tight
leptons most likely to come from the \PZ\ decay, as well as the tight lepton most likely to come from the \PW\ boson decay, are selected using the algorithm described in Section~\ref{sec:selection}.

\begin{figure}[!hbtp]
  \centering
  \includegraphics[width=0.45\linewidth]{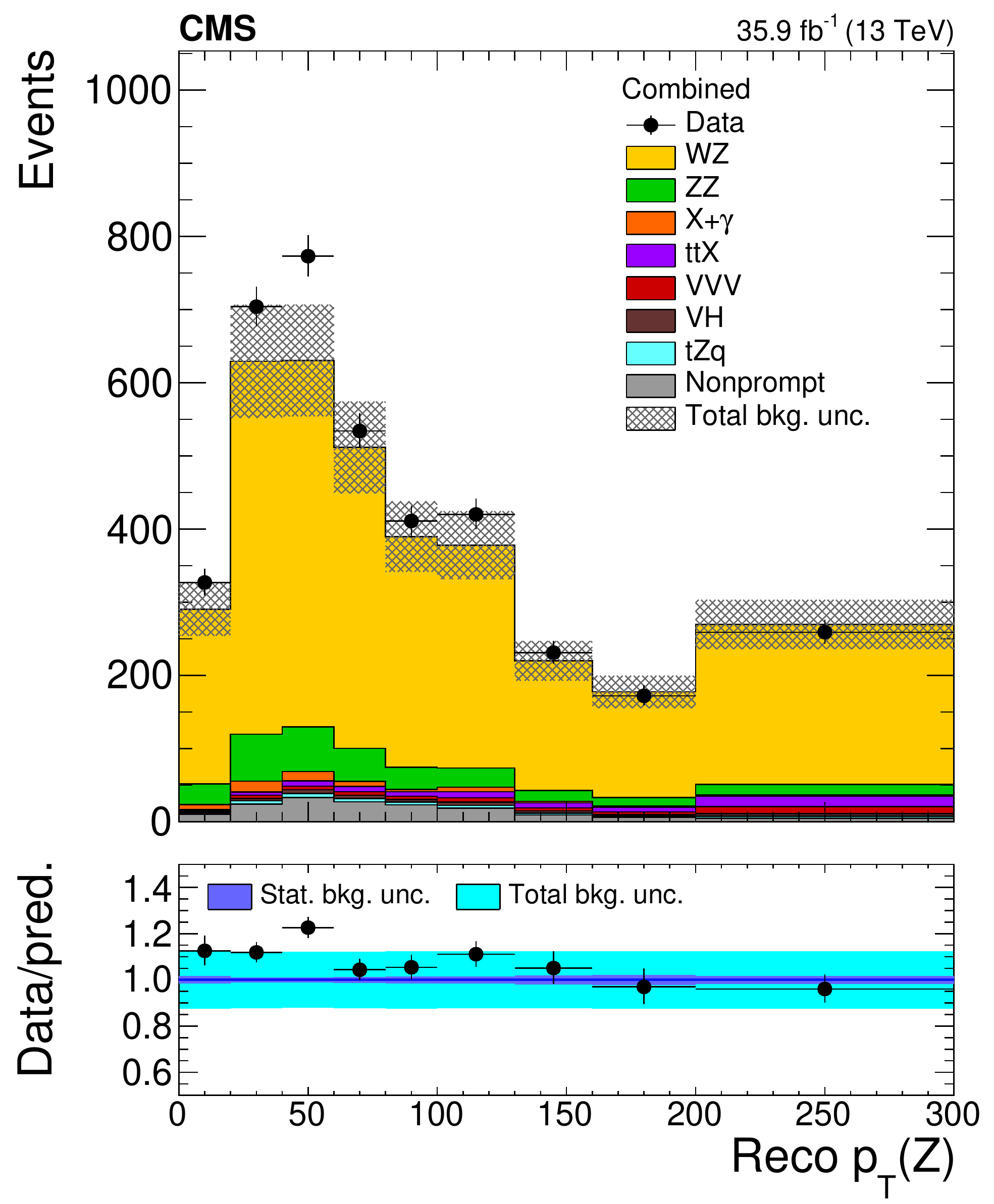}
  \includegraphics[width=0.45\linewidth]{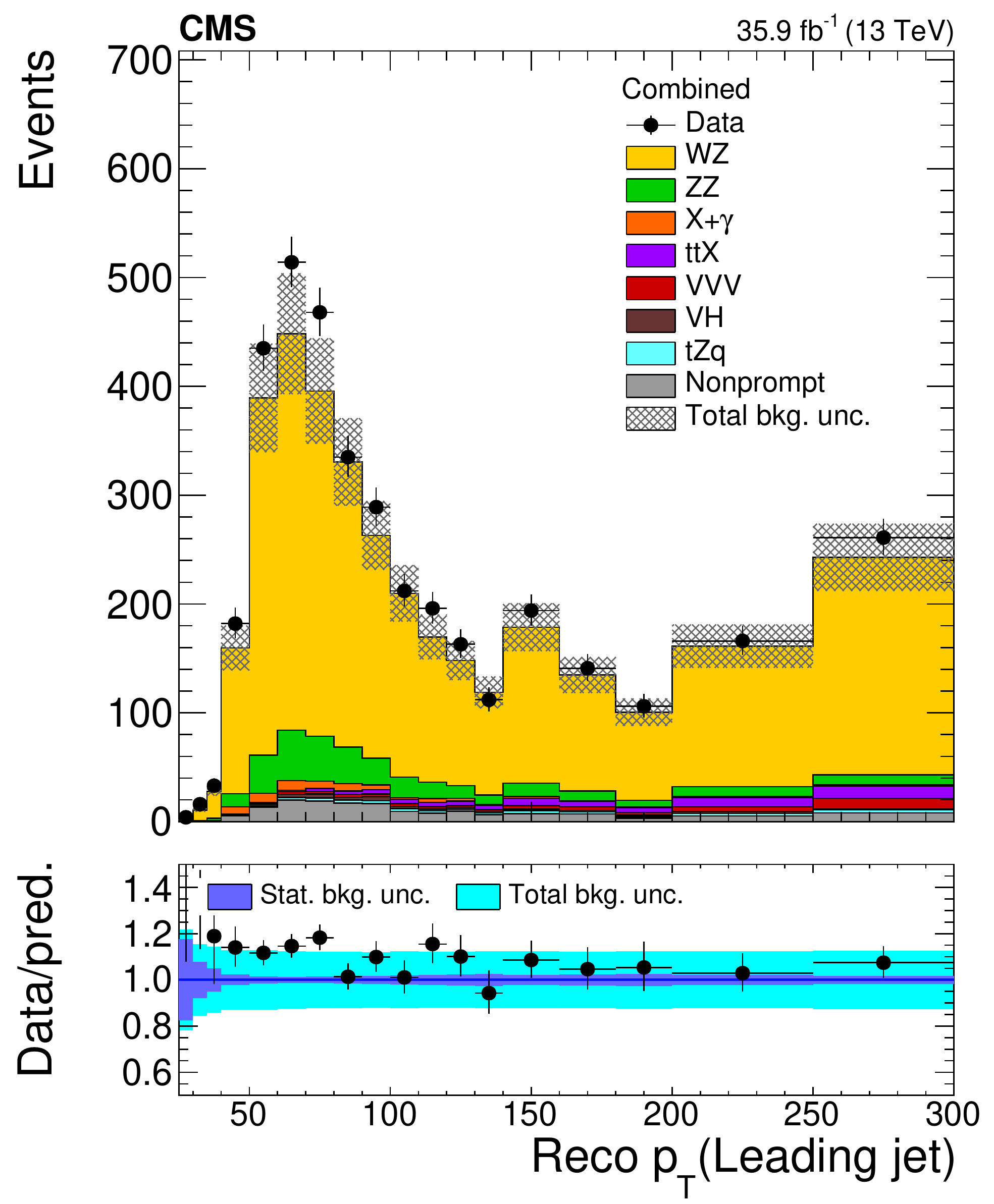}
  \includegraphics[width=0.45\linewidth]{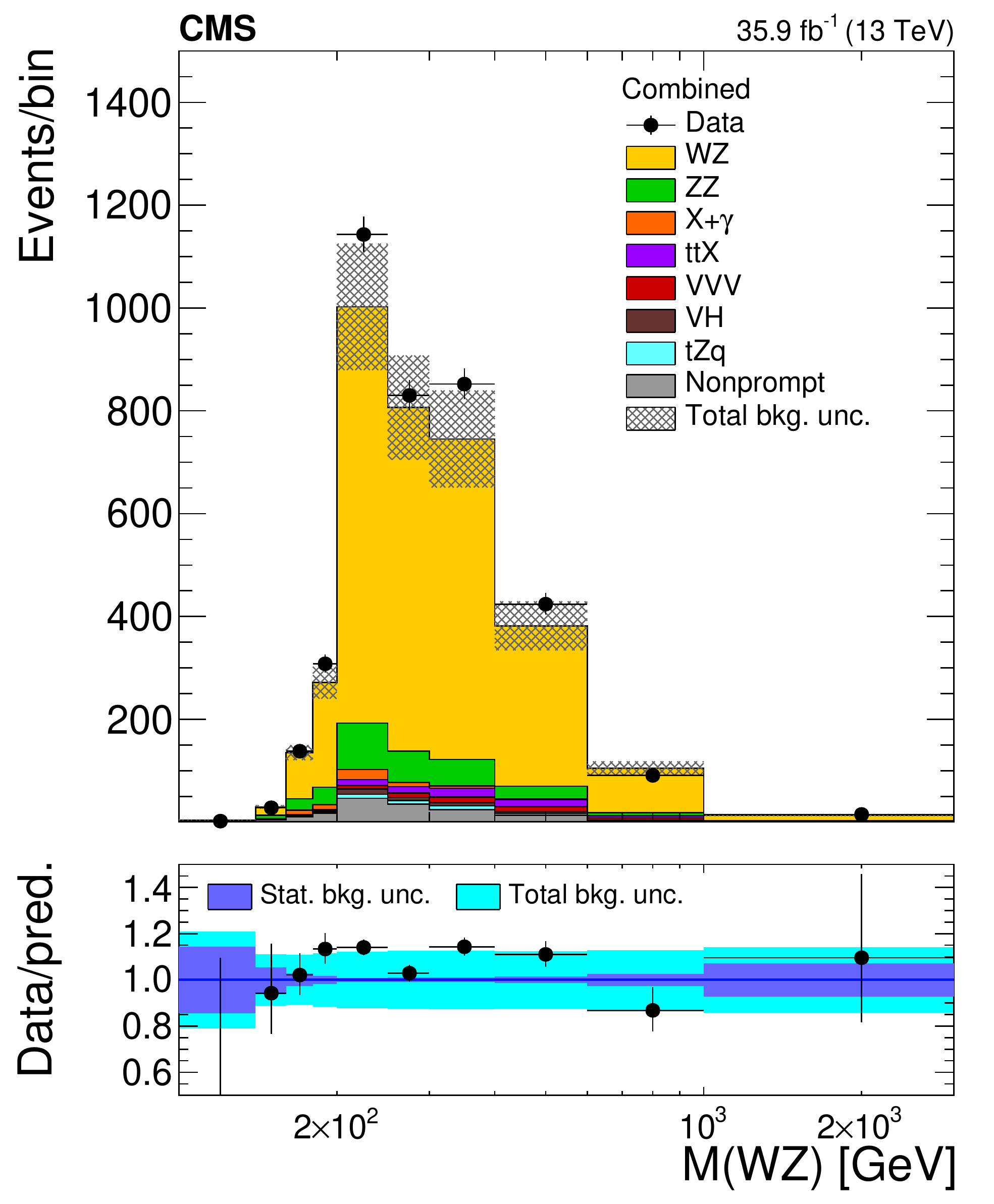}\\
  \caption{Prefit distributions of key observables in the signal region. The transverse momentum of the \cPZ\ boson (top left), the transverse momentum of the leading jet (top right), and the mass of the \WZ\ system (bottom). The last bin contains the overflow. Vertical bars on the data points include the statistical uncertainty and the shaded band over the MC prediction include both the statistical and the systematic uncertainties in the normalization of each of the background processes. An additional 15\% uncertainty is assigned to the signal \WZ\ process in the figures to account for the NLO/NNLO normalization differences.}
\label{fig:inputFancy}
\end{figure}

For these three measurements, events must pass the selection used for the inclusive cross section measurement,
which is described in Section~\ref{sec:selection}. The resulting reconstructed (\emph{reco}) level distributions are shown in Fig.~\ref{fig:inputFancy}.

The differential \WZ\ cross section is measured in the signal region of the inclusive measurement, here referred to as the inclusive final state,
and in four exclusive categories corresponding to a classification by lepton flavour (\eee, \eem, \emm, and \mmm), referred to as \emph{exclusive} final states.

\begin{figure}[!hbtp]
  \centering
  \includegraphics[width=0.45\linewidth]{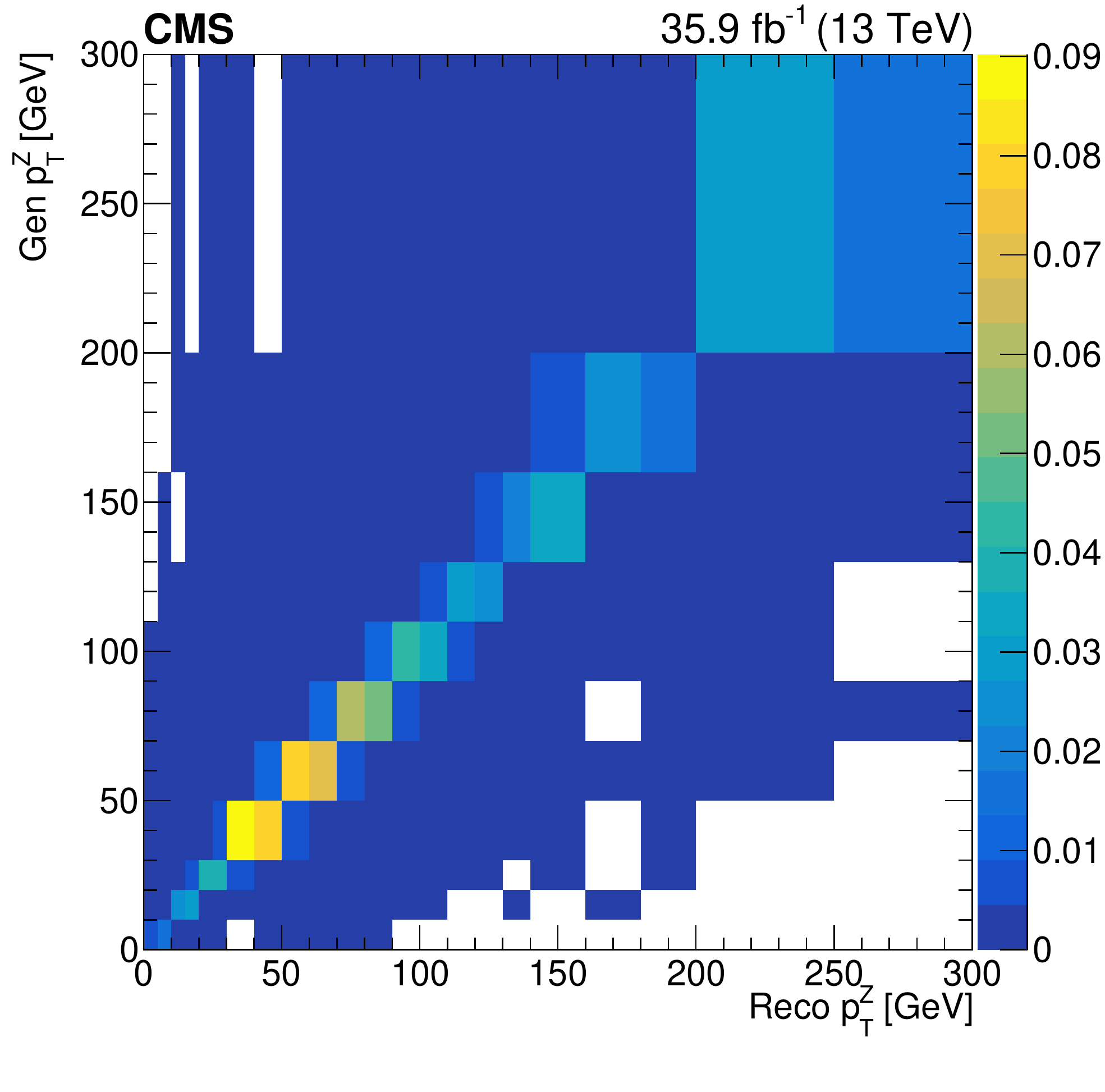}
  \includegraphics[width=0.45\linewidth]{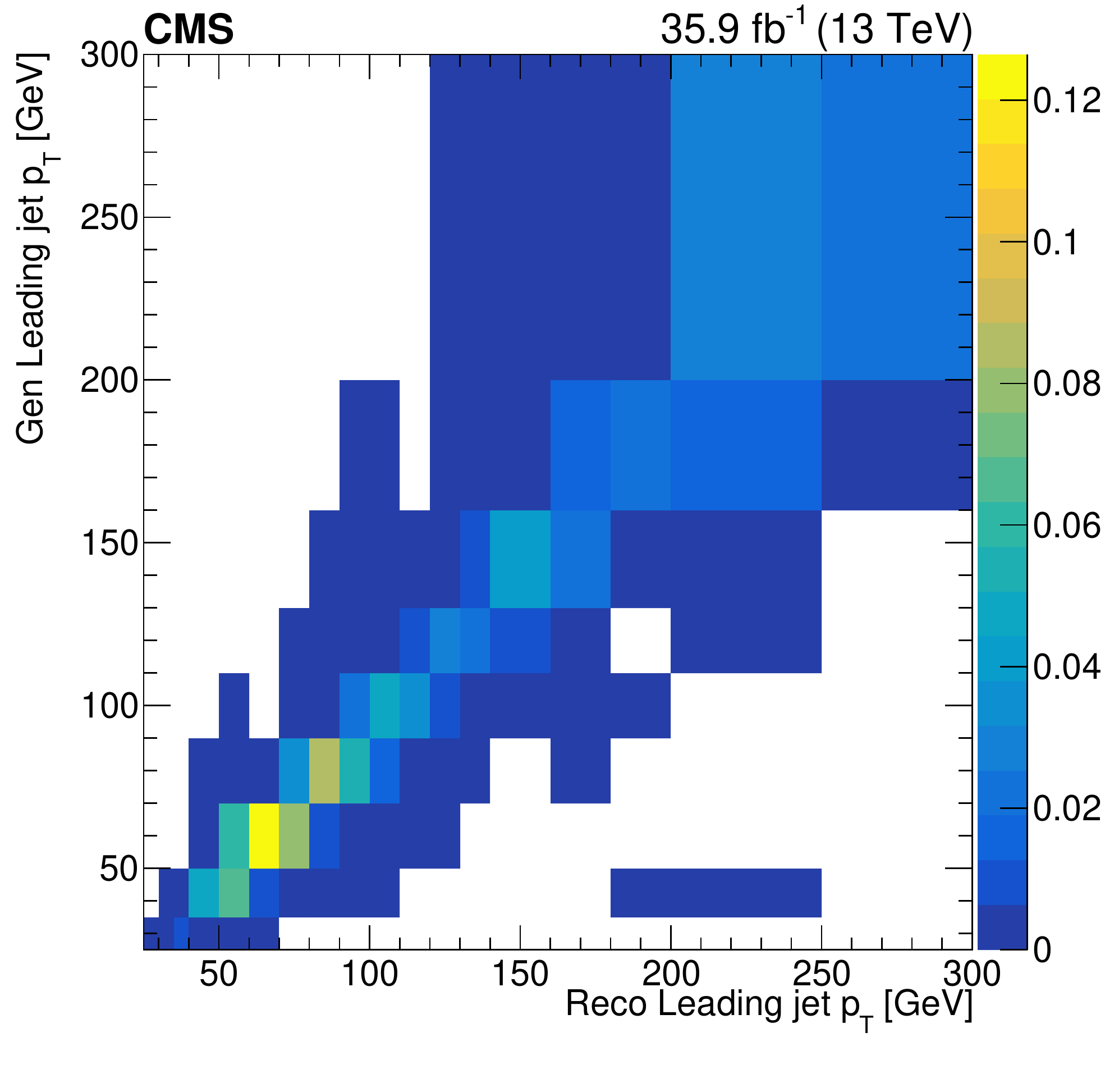}
  \includegraphics[width=0.45\linewidth]{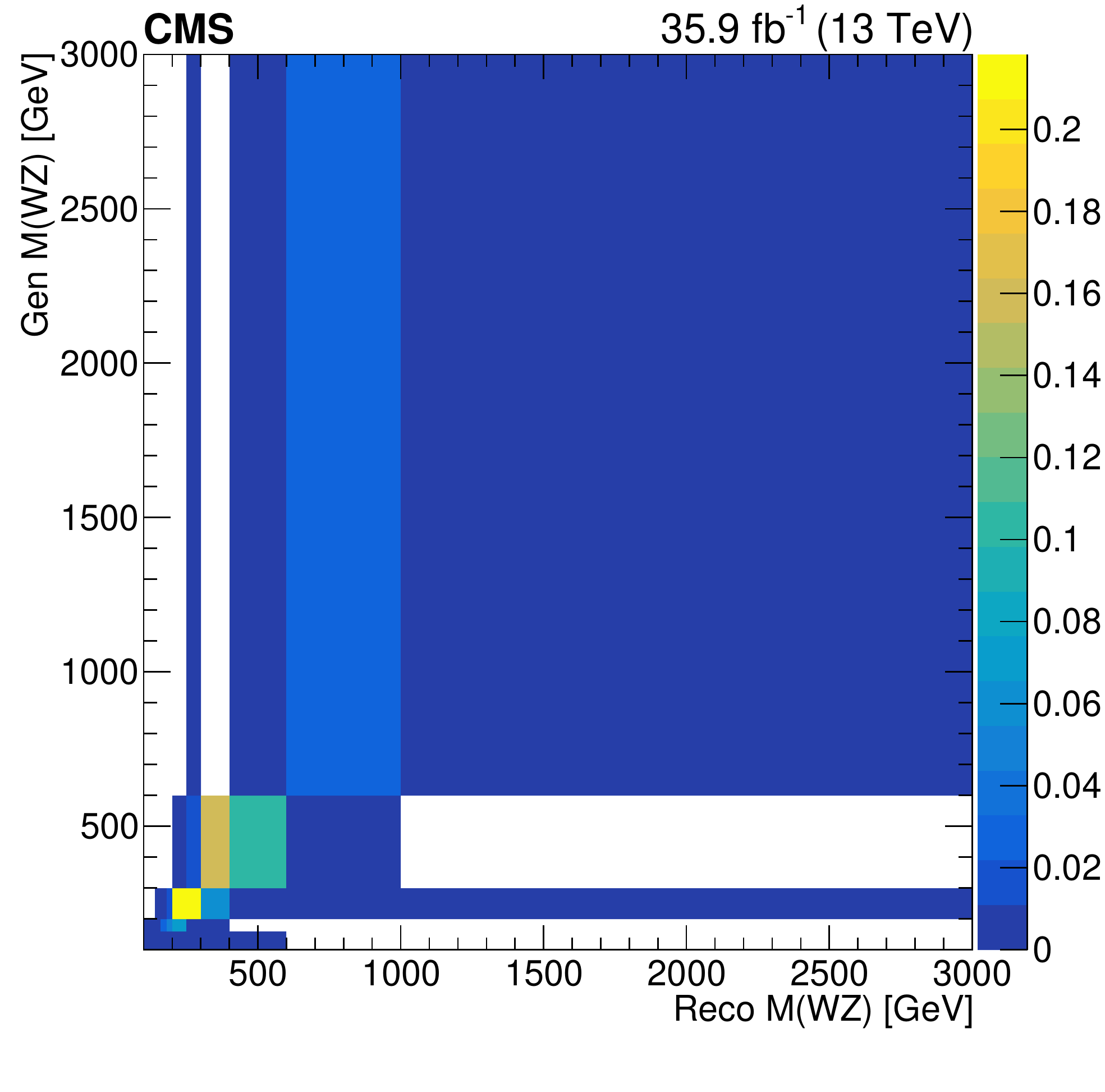}\\
  \caption{Response matrices obtained using NLO samples, simulated with the \POWHEG generator and normalized to unity. The transverse momentum of the \cPZ\ boson (top left), the leading jet transverse momentum (top right) and the mass of the \WZ\ system (bottom) are shown.}
  \label{fig:nloResponses}
\end{figure}

The reconstructed and generated distributions are assumed to differ by the effects of the detector response.
This response can be modelled using a two-dimensional matrix that summarizes the bin migration effects induced by the detector on the target observables.
Response matrices are obtained in the signal region for the inclusive and exclusive categories, using the \POWHEG and \MGvATNLO NLO generators. These matrices are shown for the inclusive selection in Fig.~\ref{fig:nloResponses},
where the bin contents are normalized to the expected NLO yield for the integrated luminosity analyzed in this paper.
The binning scheme is chosen such that for all the matrices the width of the diagonal bins in each dimension are larger than the standard deviation of the average of the bin content across the orthogonal axis.

The process of inverting the detector response matrix is known as \emph{unfolding}~\cite{Cowan}, and several techniques
are available in the literature for solving the problem~\cite{Schmitt:2016orm}, although in many cases it may be argued that the best strategy would be to perform any comparison in the reconstructed space.
In the following, the space populated by reconstructed events (the reco-level distributions) is denoted as \emph{folded space} (\emph{folded distributions}), while the generator-level space (distributions) is denoted as \emph{unfolded space} (\emph{unfolded distributions}).

The unfolding procedure consists of performing a least-squares fit with optional Tikhonov regularization~\cite{Tikhonov,ridge}, as implemented in the \textsc{TUnfold} software package~\cite{Schmitt:2012kp}. The unfolding problem, and the least-squares fit used to solve it, are modelled according to:
\begin{equation}
\begin{aligned}
  \mathcal{L}(\mathbf{x},\lambda) &= \mathcal{L}_1 + \mathcal{L}_2 + \mathcal{L}_3, \qquad\qquad\qquad\\
  \mathcal{L}_1 &= (\mathbf{y} - \mathbf{Ax})^T\mathbf{V_{yy}}(\mathbf{y} - \mathbf{Ax}),\qquad  \\
  \mathcal{L}_2 &= \tau^2(\mathbf{x} - f_b\mathbf{x_0})^T(\mathbf{L}^T\mathbf{L})(\mathbf{x} - f_b\mathbf{x_0}),\qquad \\
  \mathcal{L}_3 &= \lambda(Y - \mathbf{e}^T\mathbf{x}),\qquad \\
  Y &= \sum_i y_i,\\
  e_j &= \sum_i A_{ij}.
  \label{eq:unfold}
\end{aligned}
\end{equation}
Here $\mathbf{y}$ is the vector of observed yields, $\mathbf{A}$ is the response matrix, and $\mathbf{x}$ is the unfolded result. $\mathcal{L}_1$ models the least-squares minimization, where $\mathbf{V_{yy}}$ is the covariance matrix of $\mathbf{y}$, with elements $V_{ij}$ defined by the correlation coefficients obtained by rescaling each covariance $e_{ij}$ by the variances $e_{ii}$ and $e_{jj}$, $V_{ij} = e_{ij}/e_{ii}e_{jj}$. The regularization is described by $\mathcal{L}_2$, which reduces the fluctuations in $\mathbf{x}$---induced by the statistical fluctuations of $\mathbf{y}$---via the regularization conditions defined in the matrix $\mathbf{L}$. The strength of the regularization is described by the parameter $\tau$, and a bias vector $f_b\mathbf{x_0}$ defines the reference with respect to which large deviations are suppressed. An optional area constraint governs whether the normalization of the unfolded result is bound to the total yield in the folded space, as modelled by $\mathcal{L}_3$.

To evaluate the performance of the algorithm, the data counts $\mathbf{y}$ are substituted with a pseudodata distribution obtained by sampling from the signal plus background MC distributions. This distribution is then unfolded and folded back; the resulting distribution agrees with the MC truth in the unfolded space and the original sampled distribution, respectively, within uncertainties.

The default configuration for the unfolding performed in this paper is as follows. The \POWHEG generator is used to model the response matrix and the area constraint is applied; such constraint accounts for the difference between the expected yields, from the NLO predictions, and the observed yields, which were shown by the inclusive measurement to be more compatible with the NNLO predictions. The bias vector is the generator-level distribution rescaled to the NNLO prediction by a bias scale of 1.13. By default, no regularization is performed. These settings are chosen following a series of checks using the pseudodata distributions, to evaluate the effect of the area constraint, the bias scale and vector, and the regularization scheme.

In particular, the effect of regularization has been checked by applying Tikhonov regularization to the curvature of the unfolded distribution $\mathbf{x}$. The best value for the regularization parameter $\tau$ is chosen using the well-established L-curve method~\cite{Hansen00thel-curve}. The regularization process is applied for each of the variables and in no case is there an appreciable gain. No regularization is thus applied to obtain the final result.

\begin{figure}[!hbtp]
  \centering
  \includegraphics[width=0.45\linewidth]{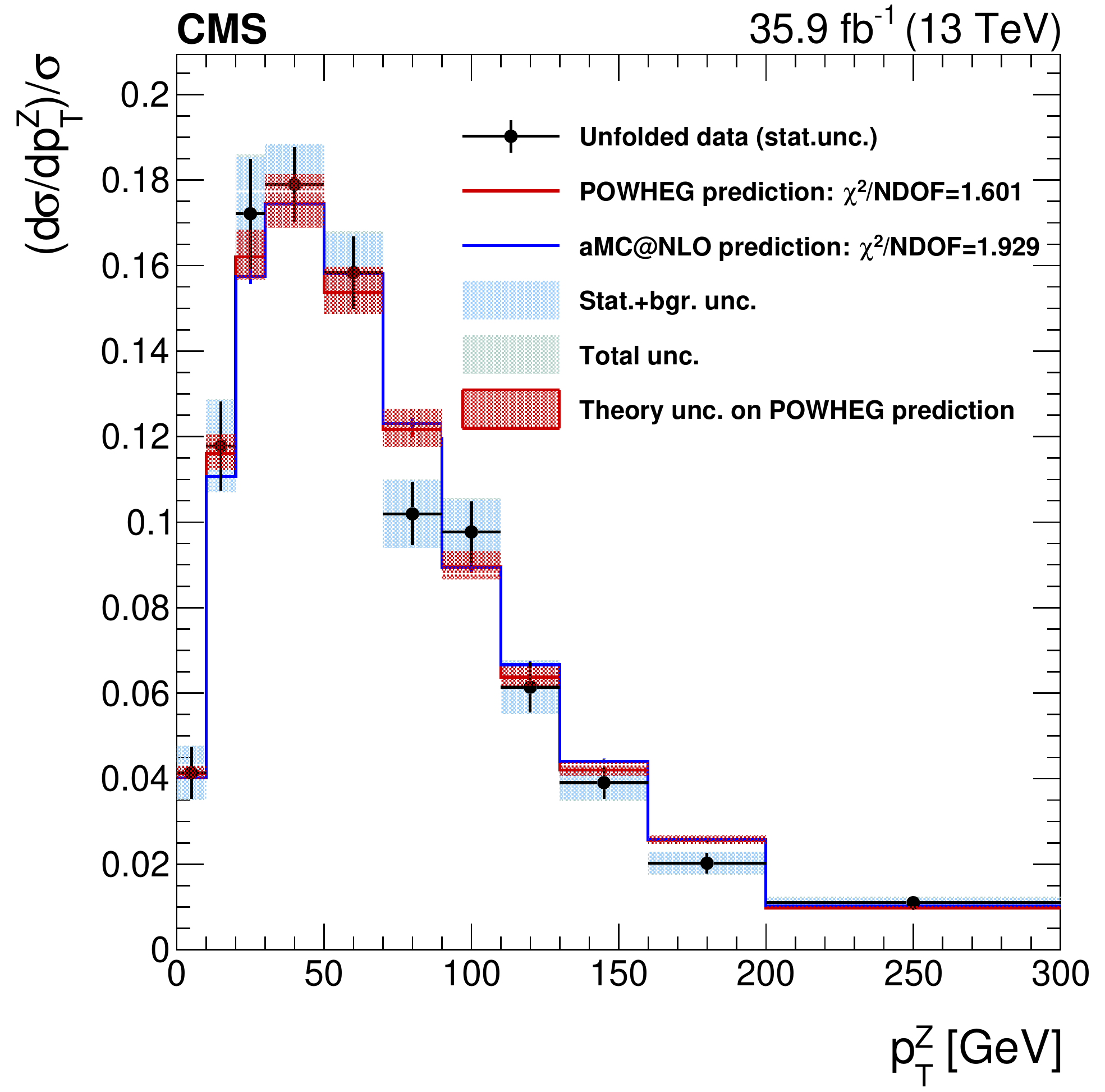}
  \includegraphics[width=0.45\linewidth]{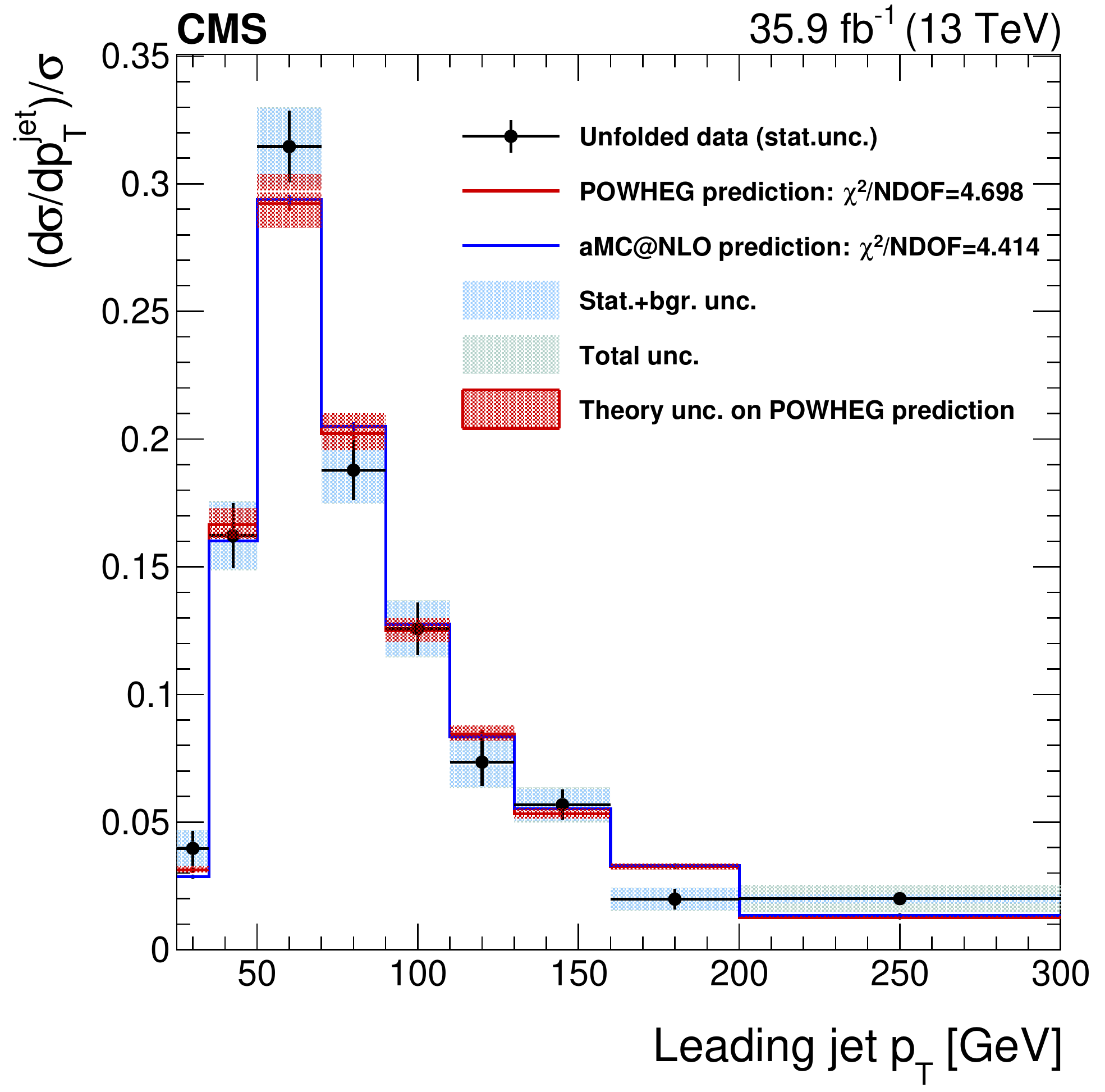}\\
  \includegraphics[width=0.45\linewidth]{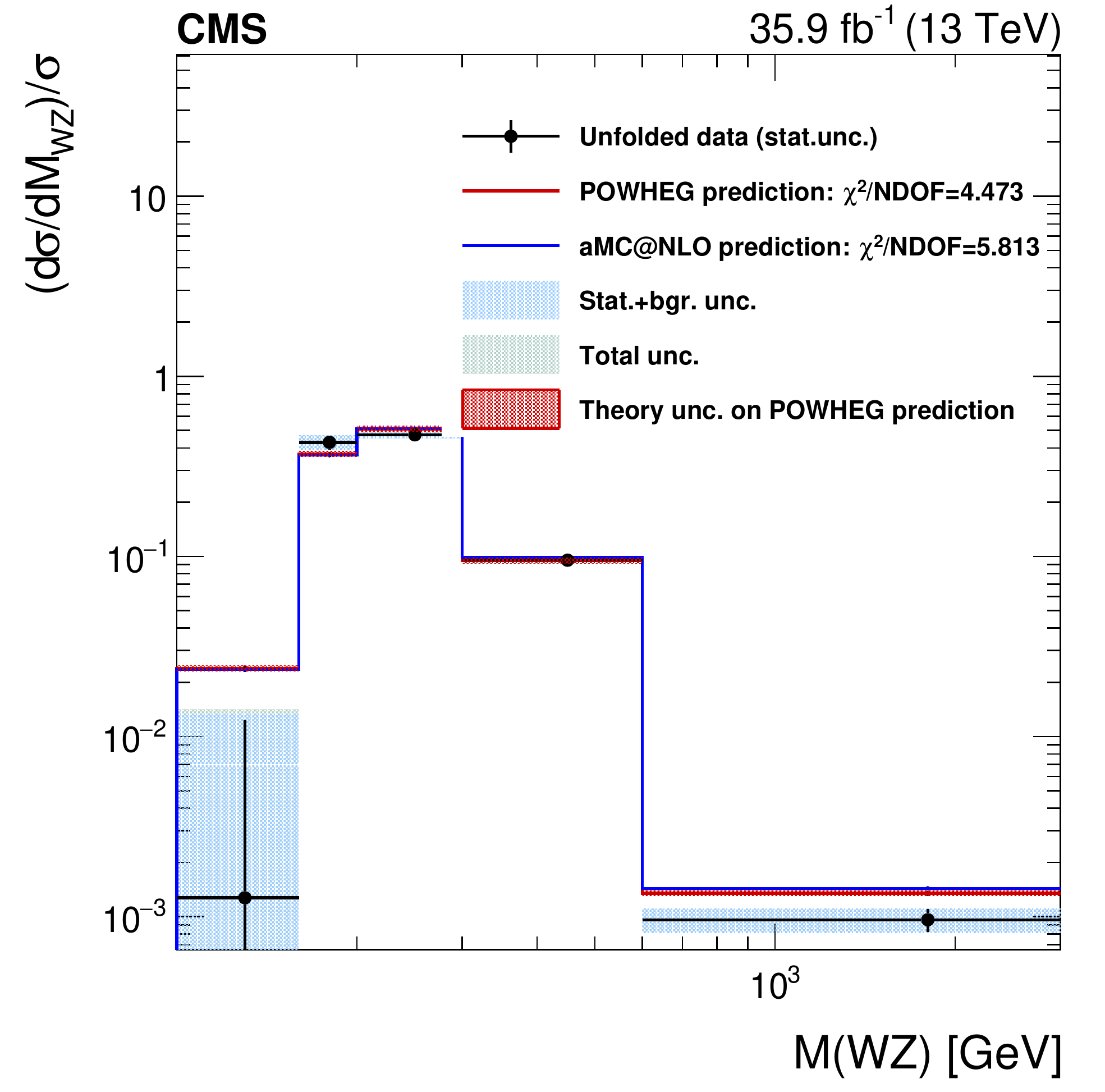}\\
  \caption{Differential distributions for the \cPZ\ boson \pt (top left), leading jet \pt (top right), and mass of the \WZ\ system (bottom). Data distributions are unfolded at the dressed leptons level and compared with the \POWHEG, \MGvATNLO NLO generators, and \PYTHIA predictions, as described in the text. The red band around the \POWHEG prediction represents the theory uncertainty in it; the effect on the unfolded data of this uncertainty, through the unfolding matrix, is included in the shaded bands described in the legend.}
  \label{fig:result_inclusive}
\end{figure}

Figure~\ref{fig:result_inclusive} shows the results in the inclusive final state for the \cPZ\ boson \pt\ distribution (top left), leading jet \pt\ distribution (top right), and mass of the \WZ\ system (bottom). Good agreement is found between the unfolded data distribution and the MC predictions at particle level, and is quantified by $\chi^{2}/NDOF$ values given in the plot legends.  Results in the four different flavour channels are compatible. The results for the differential cross section in the inclusive and exclusive final states are expressed as a fraction of the total cross section and tabulated in Tables~\ref{tab:unfolddiffZpt_1},~\ref{tab:unfolddiffZpt_2}, and~\ref{tab:unfolddiffZpt_3} for the \cPZ\ boson \pt, Tables~\ref{tab:unfolddiffLeadJetPt_1} and~\ref{tab:unfolddiffLeadJetPt_2} for the leading jet \pt, and Table~\ref{tab:unfolddiffMWZ} for the mass of the \WZ\ system. The total cross section is constrained by the aforementioned area constraint. The \emph{bottom line test}~\cite{Cousins:2016ksu} is performed, in which goodness of fit tests are performed in the folded and in the unfolded space to ensure that the agreement between the data and the model does not become worse after unfolding. The purpose is to check that the unfolding procedure is not enhancing the ability to reject incorrect models. The test shows a substantial agreement, giving further confidence in the unfolding procedure.

The results are derived using all the systematic uncertainties described in Section~\ref{sec:systematics}, including their effect on the response matrix. In addition, a systematic uncertainty due to the unfolding procedure is defined as the difference between the nominal result and the result obtained by unfolding the nominal shape using an alternative response matrix. Such alternative matrix is modelled using the \MGvATNLO generator. The effect of such uncertainty on the result is smaller than the effect of statistical fluctuation and of the background subtraction, and is included in the tables together with all the other sources of uncertainty within the \emph{other syst} category.

\begin{table}[!hptb]
\centering
\topcaption{\label{tab:unfolddiffZpt_1}Differential cross section in bins of \pt(\PZ). Values are expressed as a fraction of the total cross section. The \eee and \eem final states are shown.}
\begin{tabular}{cccccc}
  \hline
   & \multicolumn{5}{c}{Cross section [fraction of the total cross section]} \\
  Bin \pt(\PZ) [\GeVns{}] & Central value & (stat) & (bgr) & (other syst) & (total) \\
  \hline
  \multicolumn{6}{c}{\eee} \\
  [\cmsTabSkip]
 $[0, 10]$    & 0.024 & $\pm$ 0.016 & $\pm$ 0.002 & $\pm$ 0.005 & ($\pm$ 0.016)\\
 $[10, 20]$   & 0.102 & $\pm$ 0.030 & $\pm$ 0.006 & $\pm$ 0.005 & ($\pm$ 0.031)\\
 $[20, 30]$   & 0.169 & $\pm$ 0.039 & $\pm$ 0.010 & $\pm$ 0.007 & ($\pm$ 0.041)\\
 $[30, 50]$   & 0.122 & $\pm$ 0.024 & $\pm$ 0.007 & $\pm$ 0.006 & ($\pm$ 0.025)\\
 $[50, 70]$   & 0.180 & $\pm$ 0.027 & $\pm$ 0.006 & $\pm$ 0.005 & ($\pm$ 0.028)\\
 $[70, 90]$   & 0.132 & $\pm$ 0.023 & $\pm$ 0.005 & $\pm$ 0.004 & ($\pm$ 0.024)\\
 $[90, 110]$  & 0.092 & $\pm$ 0.020 & $\pm$ 0.006 & $\pm$ 0.004 & ($\pm$ 0.022)\\
 $[110, 130]$ & 0.078 & $\pm$ 0.018 & $\pm$ 0.004 & $\pm$ 0.005 & ($\pm$ 0.019)\\
 $[130, 160]$ & 0.053 & $\pm$ 0.012 & $\pm$ 0.002 & $\pm$ 0.004 & ($\pm$ 0.013)\\
 $[160, 200]$ & 0.037 & $\pm$ 0.008 & $\pm$ 0.001 & $\pm$ 0.001 & ($\pm$ 0.008)\\
 $[200, 300]$ & 0.010 & $\pm$ 0.003 & $\pm$ 0.000 & $\pm$ 0.001 & ($\pm$ 0.003)\\
  [\cmsTabSkip]
  \multicolumn{6}{c}{\eem} \\
  [\cmsTabSkip]
 $[0, 10]$    & 0.033 & $\pm$ 0.013 & $\pm$ 0.002 & $\pm$ 0.005 & ($\pm$ 0.014)\\
 $[10, 20]$   & 0.101 & $\pm$ 0.023 & $\pm$ 0.003 & $\pm$ 0.005 & ($\pm$ 0.024)\\
 $[20, 30]$   & 0.177 & $\pm$ 0.030 & $\pm$ 0.004 & $\pm$ 0.005 & ($\pm$ 0.030)\\
 $[30, 50]$   & 0.188 & $\pm$ 0.020 & $\pm$ 0.003 & $\pm$ 0.003 & ($\pm$ 0.021)\\
 $[50, 70]$   & 0.148 & $\pm$ 0.019 & $\pm$ 0.003 & $\pm$ 0.003 & ($\pm$ 0.019)\\
 $[70, 90]$   & 0.103 & $\pm$ 0.016 & $\pm$ 0.003 & $\pm$ 0.005 & ($\pm$ 0.017)\\
 $[90, 110]$  & 0.080 & $\pm$ 0.015 & $\pm$ 0.003 & $\pm$ 0.003 & ($\pm$ 0.015)\\
 $[110, 130]$ & 0.090 & $\pm$ 0.015 & $\pm$ 0.002 & $\pm$ 0.003 & ($\pm$ 0.015)\\
 $[130, 160]$ & 0.049 & $\pm$ 0.009 & $\pm$ 0.001 & $\pm$ 0.002 & ($\pm$ 0.009)\\
 $[160, 200]$ & 0.015 & $\pm$ 0.005 & $\pm$ 0.001 & $\pm$ 0.001 & ($\pm$ 0.005)\\
 $[200, 300]$ & 0.015 & $\pm$ 0.003 & $\pm$ 0.000 & $\pm$ 0.000 & ($\pm$ 0.003)\\
  \hline
\end{tabular}
\end{table}

\begin{table}[!hptb]
\centering
\topcaption{\label{tab:unfolddiffZpt_2}Differential cross section in bins of \pt(\PZ). Values are expressed as a fraction of the total cross section. The \emm and \mmm final states are shown.}
\begin{tabular}{cccccc}
  \hline
   & \multicolumn{5}{c}{Cross section [fraction of the total cross section]} \\
  Bin \pt(\PZ) [\GeVns{}] & Central value & (stat) & (bgr) & (other syst) & (total) \\
  \hline
  \multicolumn{6}{c}{\emm} \\
  [\cmsTabSkip]
 $[0, 10]$    & 0.052 & $\pm$ 0.012 & $\pm$ 0.001 & $\pm$ 0.003 & ($\pm$ 0.013)\\
 $[10, 20]$   & 0.132 & $\pm$ 0.021 & $\pm$ 0.004 & $\pm$ 0.005 & ($\pm$ 0.021)\\
 $[20, 30]$   & 0.175 & $\pm$ 0.024 & $\pm$ 0.005 & $\pm$ 0.007 & ($\pm$ 0.026)\\
 $[30, 50]$   & 0.186 & $\pm$ 0.017 & $\pm$ 0.005 & $\pm$ 0.004 & ($\pm$ 0.018)\\
 $[50, 70]$   & 0.149 & $\pm$ 0.015 & $\pm$ 0.005 & $\pm$ 0.002 & ($\pm$ 0.016)\\
 $[70, 90]$   & 0.083 & $\pm$ 0.013 & $\pm$ 0.004 & $\pm$ 0.004 & ($\pm$ 0.014)\\
 $[90, 110]$  & 0.108 & $\pm$ 0.014 & $\pm$ 0.004 & $\pm$ 0.006 & ($\pm$ 0.016)\\
 $[110, 130]$ & 0.043 & $\pm$ 0.010 & $\pm$ 0.002 & $\pm$ 0.005 & ($\pm$ 0.011)\\
 $[130, 160]$ & 0.041 & $\pm$ 0.008 & $\pm$ 0.002 & $\pm$ 0.003 & ($\pm$ 0.008)\\
 $[160, 200]$ & 0.020 & $\pm$ 0.005 & $\pm$ 0.001 & $\pm$ 0.002 & ($\pm$ 0.005)\\
 $[200, 300]$ & 0.010 & $\pm$ 0.002 & $\pm$ 0.000 & $\pm$ 0.001 & ($\pm$ 0.002)\\
  [\cmsTabSkip]
  \multicolumn{6}{c}{\mmm} \\
  [\cmsTabSkip]
 $[0, 10]$    & 0.039 & $\pm$ 0.009 & $\pm$ 0.001 & $\pm$ 0.003 & ($\pm$ 0.010)\\
 $[10, 20]$   & 0.122 & $\pm$ 0.016 & $\pm$ 0.002 & $\pm$ 0.005 & ($\pm$ 0.017)\\
 $[20, 30]$   & 0.171 & $\pm$ 0.020 & $\pm$ 0.002 & $\pm$ 0.004 & ($\pm$ 0.020)\\
 $[30, 50]$   & 0.182 & $\pm$ 0.013 & $\pm$ 0.003 & $\pm$ 0.003 & ($\pm$ 0.014)\\
 $[50, 70]$   & 0.165 & $\pm$ 0.013 & $\pm$ 0.003 & $\pm$ 0.007 & ($\pm$ 0.015)\\
 $[70, 90]$   & 0.108 & $\pm$ 0.012 & $\pm$ 0.002 & $\pm$ 0.003 & ($\pm$ 0.012)\\
 $[90, 110]$  & 0.102 & $\pm$ 0.011 & $\pm$ 0.001 & $\pm$ 0.002 & ($\pm$ 0.011)\\
 $[110, 130]$ & 0.051 & $\pm$ 0.009 & $\pm$ 0.002 & $\pm$ 0.002 & ($\pm$ 0.009)\\
 $[130, 160]$ & 0.031 & $\pm$ 0.006 & $\pm$ 0.001 & $\pm$ 0.001 & ($\pm$ 0.006)\\
 $[160, 200]$ & 0.019 & $\pm$ 0.004 & $\pm$ 0.001 & $\pm$ 0.001 & ($\pm$ 0.004)\\
 $[200, 300]$ & 0.011 & $\pm$ 0.001 & $\pm$ 0.000 & $\pm$ 0.001 & ($\pm$ 0.002)\\
  \hline
\end{tabular}
\end{table}

\begin{table}[!hptb]
\centering
\topcaption{\label{tab:unfolddiffZpt_3}Differential cross section in bins of \pt(\PZ). Values are expressed as a fraction of the total cross section. The inclusive final state is shown}
\begin{tabular}{cccccc}
  \hline
   & \multicolumn{5}{c}{Cross section [fraction of the total cross section]} \\
  Bin \pt(\PZ) [\GeVns{}] & Central value & (stat) & (bgr) & (other syst) & (total) \\
  \hline
  \multicolumn{6}{c}{Inclusive} \\
  [\cmsTabSkip]
 $[0, 10]$    & 0.041 & $\pm$ 0.006 & $\pm$ 0.001 & $\pm$ 0.001 & ($\pm$ 0.006)\\
 $[10, 20]$   & 0.118 & $\pm$ 0.010 & $\pm$ 0.002 & $\pm$ 0.002 & ($\pm$ 0.011)\\
 $[20, 30]$   & 0.172 & $\pm$ 0.013 & $\pm$ 0.003 & $\pm$ 0.004 & ($\pm$ 0.014)\\
 $[30, 50]$   & 0.179 & $\pm$ 0.009 & $\pm$ 0.003 & $\pm$ 0.001 & ($\pm$ 0.009)\\
 $[50, 70]$   & 0.158 & $\pm$ 0.008 & $\pm$ 0.003 & $\pm$ 0.003 & ($\pm$ 0.010)\\
 $[70, 90]$   & 0.102 & $\pm$ 0.007 & $\pm$ 0.003 & $\pm$ 0.001 & ($\pm$ 0.008)\\
 $[90, 110]$  & 0.098 & $\pm$ 0.007 & $\pm$ 0.002 & $\pm$ 0.002 & ($\pm$ 0.008)\\
 $[110, 130]$ & 0.061 & $\pm$ 0.006 & $\pm$ 0.001 & $\pm$ 0.001 & ($\pm$ 0.006)\\
 $[130, 160]$ & 0.039 & $\pm$ 0.004 & $\pm$ 0.001 & $\pm$ 0.002 & ($\pm$ 0.004)\\
 $[160, 200]$ & 0.020 & $\pm$ 0.002 & $\pm$ 0.001 & $\pm$ 0.001 & ($\pm$ 0.003)\\
 $[200, 300]$ & 0.011 & $\pm$ 0.001 & $\pm$ 0.000 & $\pm$ 0.001 & ($\pm$ 0.001)\\
  \hline
\end{tabular}
\end{table}

\begin{table}[!hptb]
\centering
\topcaption{\label{tab:unfolddiffLeadJetPt_1}Differential cross section in bins of  \pt(Leading jet). Values are expressed as a fraction of the total cross section. The \eee, \eem, \emm, and \mmm final states are shown.}
\begin{tabular}{cccccr}
  \hline
   & \multicolumn{5}{c}{Cross section [fraction of the total cross section]} \\
  Bin  \pt(Leading jet) \pt [\GeVns{}]  & Central value & (stat) & (bgr) & (other syst) & (total) \\
\hline
  \multicolumn{6}{c}{\eee}  \\
 [\cmsTabSkip]
 $[25, 35]$   & 0.022 & $\pm$ 0.015 & $\pm$ 0.002 & $\pm$ 0.008 & ($\pm$ 0.017)\\
 $[35, 50]$   & 0.189 & $\pm$ 0.038 & $\pm$ 0.008 & $\pm$ 0.006 & ($\pm$ 0.039)\\
 $[50, 70]$   & 0.257 & $\pm$ 0.039 & $\pm$ 0.012 & $\pm$ 0.007 & ($\pm$ 0.041)\\
 $[70, 90]$   & 0.194 & $\pm$ 0.035 & $\pm$ 0.010 & $\pm$ 0.011 & ($\pm$ 0.038)\\
 $[90, 110]$  & 0.140 & $\pm$ 0.033 & $\pm$ 0.007 & $\pm$ 0.008 & ($\pm$ 0.034)\\
 $[110, 130]$ & 0.109 & $\pm$ 0.030 & $\pm$ 0.005 & $\pm$ 0.006 & ($\pm$ 0.031)\\
 $[130, 160]$ & 0.031 & $\pm$ 0.016 & $\pm$ 0.004 & $\pm$ 0.004 & ($\pm$ 0.017)\\
 $[160, 200]$ & 0.035 & $\pm$ 0.013 & $\pm$ 0.002 & $\pm$ 0.003 & ($\pm$ 0.013)\\
 $[200, 300]$ & 0.023 & $\pm$ 0.005 & $\pm$ 0.001 & $\pm$ 0.006 & ($\pm$ 0.007)\\
 [\cmsTabSkip]
 \multicolumn{6}{c}{\eem} \\
 [\cmsTabSkip]
 $[25, 35]$   & 0.059 & $\pm$ 0.025 &  $\pm$ 0.001 & $\pm$ 0.008 & ($\pm$ 0.026)\\
 $[35, 50]$   & 0.146 & $\pm$ 0.031 &  $\pm$ 0.003 & $\pm$ 0.011 & ($\pm$ 0.033)\\
 $[50, 70]$   & 0.286 & $\pm$ 0.032 &  $\pm$ 0.005 & $\pm$ 0.007 & ($\pm$ 0.033)\\
 $[70, 90]$   & 0.224 & $\pm$ 0.028 &  $\pm$ 0.005 & $\pm$ 0.006 & ($\pm$ 0.029)\\
 $[90, 110]$  & 0.111 & $\pm$ 0.023 &  $\pm$ 0.002 & $\pm$ 0.005 & ($\pm$ 0.024)\\
 $[110, 130]$ & 0.083 & $\pm$ 0.022 &  $\pm$ 0.004 & $\pm$ 0.007 & ($\pm$ 0.024)\\
 $[130, 160]$ & 0.055 & $\pm$ 0.013 &  $\pm$ 0.002 & $\pm$ 0.003 & ($\pm$ 0.014)\\
 $[160, 200]$ & 0.017 & $\pm$ 0.009 &  $\pm$ 0.001 & $\pm$ 0.002 & ($\pm$ 0.010)\\
 $[200, 300]$ & 0.019 & $\pm$ 0.004 &  $\pm$ 0.001 & $\pm$ 0.004 & ($\pm$ 0.006)\\
 [\cmsTabSkip]
 \multicolumn{6}{c}{\emm} \\
 [\cmsTabSkip]
 $[25, 35]$   & 0.037 & $\pm$ 0.013 &  $\pm$ 0.002 & $\pm$ 0.007 & ($\pm$ 0.015)\\
 $[35, 50]$   & 0.166 & $\pm$ 0.026 &  $\pm$ 0.005 & $\pm$ 0.009 & ($\pm$ 0.028)\\
 $[50, 70]$   & 0.329 & $\pm$ 0.029 &  $\pm$ 0.007 & $\pm$ 0.005 & ($\pm$ 0.030)\\
 $[70, 90]$   & 0.181 & $\pm$ 0.024 &  $\pm$ 0.007 & $\pm$ 0.006 & ($\pm$ 0.026)\\
 $[90, 110]$  & 0.121 & $\pm$ 0.021 &  $\pm$ 0.005 & $\pm$ 0.010 & ($\pm$ 0.024)\\
 $[110, 130]$ & 0.067 & $\pm$ 0.019 &  $\pm$ 0.005 & $\pm$ 0.009 & ($\pm$ 0.022)\\
 $[130, 160]$ & 0.060 & $\pm$ 0.012 &  $\pm$ 0.002 & $\pm$ 0.004 & ($\pm$ 0.013)\\
 $[160, 200]$ & 0.015 & $\pm$ 0.008 &  $\pm$ 0.001 & $\pm$ 0.003 & ($\pm$ 0.009)\\
 $[200, 300]$ & 0.023 & $\pm$ 0.003 &  $\pm$ 0.001 & $\pm$ 0.005 & ($\pm$ 0.006)\\
 [\cmsTabSkip]
 \multicolumn{6}{c}{\mmm} \\
 [\cmsTabSkip]
 $[25, 35]$   & 0.042 & $\pm$ 0.011 & $\pm$ 0.000 & $\pm$ 0.003 & ($\pm$ 0.011)\\
 $[35, 50]$   & 0.155 & $\pm$ 0.019 & $\pm$ 0.002 & $\pm$ 0.008 & ($\pm$ 0.021)\\
 $[50, 70]$   & 0.333 & $\pm$ 0.021 & $\pm$ 0.004 & $\pm$ 0.004 & ($\pm$ 0.022)\\
 $[70, 90]$   & 0.176 & $\pm$ 0.017 & $\pm$ 0.004 & $\pm$ 0.006 & ($\pm$ 0.019)\\
 $[90, 110]$  & 0.132 & $\pm$ 0.015 & $\pm$ 0.003 & $\pm$ 0.004 & ($\pm$ 0.016)\\
 $[110, 130]$ & 0.062 & $\pm$ 0.013 & $\pm$ 0.002 & $\pm$ 0.004 & ($\pm$ 0.014)\\
 $[130, 160]$ & 0.062 & $\pm$ 0.009 & $\pm$ 0.002 & $\pm$ 0.004 & ($\pm$ 0.010)\\
 $[160, 200]$ & 0.020 & $\pm$ 0.006 & $\pm$ 0.001 & $\pm$ 0.003 & ($\pm$ 0.007)\\
 $[200, 300]$ & 0.018 & $\pm$ 0.002 & $\pm$ 0.001 & $\pm$ 0.005 & ($\pm$ 0.006)\\
 \hline
\end{tabular}
\end{table}

\begin{table}[!hptb]
\centering
\topcaption{\label{tab:unfolddiffLeadJetPt_2}Differential cross section in bins of  \pt(Leading jet). Values are expressed as a fraction of the total cross section. The inclusive final state is shown.}
\begin{tabular}{cccccr}
  \hline
   & \multicolumn{5}{c}{Cross section [fraction of the total cross section]} \\
  Bin  \pt(Leading jet) \pt [\GeVns{}]  & Central value & (stat) & (bgr) & (other syst) & (total) \\
\hline
 \multicolumn{6}{c}{Inclusive} \\
[\cmsTabSkip]
 $[25, 35]$   & 0.040 & $\pm$ 0.007 & $\pm$ 0.001 & $\pm$ 0.002 &  ($\pm$ 0.007)\\
 $[35, 50]$   & 0.162 & $\pm$ 0.013 & $\pm$ 0.003 & $\pm$ 0.004 &  ($\pm$ 0.014)\\
 $[50, 70]$   & 0.315 & $\pm$ 0.014 & $\pm$ 0.005 & $\pm$ 0.003 &  ($\pm$ 0.015)\\
 $[70, 90]$   & 0.188 & $\pm$ 0.012 & $\pm$ 0.005 & $\pm$ 0.003 &  ($\pm$ 0.013)\\
 $[90, 110]$  & 0.126 & $\pm$ 0.010 & $\pm$ 0.003 & $\pm$ 0.003 &  ($\pm$ 0.011)\\
 $[110, 130]$ & 0.073 & $\pm$ 0.009 & $\pm$ 0.002 & $\pm$ 0.003 &  ($\pm$ 0.010)\\
 $[130, 160]$ & 0.057 & $\pm$ 0.006 & $\pm$ 0.001 & $\pm$ 0.003 &  ($\pm$ 0.007)\\
 $[160, 200]$ & 0.020 & $\pm$ 0.004 & $\pm$ 0.001 & $\pm$ 0.002 &  ($\pm$ 0.004)\\
 $[200, 300]$ & 0.020 & $\pm$ 0.002 & $\pm$ 0.001 & $\pm$ 0.005 &  ($\pm$ 0.005)\\
 \hline
\end{tabular}
\end{table}

\begin{table}[!hptb]
\centering
\topcaption{\label{tab:unfolddiffMWZ}Differential cross section in bins of mass of the \WZ\ system. Values are expressed as a fraction of the total cross section.}
\begin{tabular}{cccccc}
  \hline
   & \multicolumn{5}{c}{Cross section [fraction of the inclusive cross section]} \\
 Bin \mwz [\GeVns{}]   & Central value & (stat) & (bgr) & (other syst) & (total) \\
\hline
\multicolumn{6}{c}{\eee} \\
[\cmsTabSkip]
 $[100, 160]$ & 0.000 & $\pm$ 0.035 & $\pm$ 0.011 & $\pm$ 0.010 & ($\pm$ 0.038)\\
 $[160, 200]$ & 0.515 & $\pm$ 0.120 & $\pm$ 0.034 & $\pm$ 0.024 & ($\pm$ 0.127)\\
 $[200, 300]$ & 0.370 & $\pm$ 0.050 & $\pm$ 0.013 & $\pm$ 0.009 & ($\pm$ 0.053)\\
 $[300, 600]$ & 0.118 & $\pm$ 0.012 & $\pm$ 0.003 & $\pm$ 0.002 & ($\pm$ 0.012)\\
 $[600, 3000]$ & 0.001 &  $\pm$ 0.000 & $\pm$ 0.000 & $\pm$ 0.000 & ($\pm$ 0.000)\\
[\cmsTabSkip]
\multicolumn{6}{c}{\eem} \\
[\cmsTabSkip]
 $[100, 160]$ & 0.000  &  $\pm$ 0.029 & $\pm$ 0.006 & $\pm$ 0.010 & ($\pm$ 0.031)\\
 $[160, 200]$ & 0.458  & $\pm$ 0.097 & $\pm$ 0.014 & $\pm$ 0.037 & ($\pm$ 0.105)\\
 $[200, 300]$ & 0.465  & $\pm$ 0.041 & $\pm$ 0.009 & $\pm$ 0.013 & ($\pm$ 0.044)\\
 $[300, 600]$ & 0.083  & $\pm$ 0.009 & $\pm$ 0.002 & $\pm$ 0.002 & ($\pm$ 0.009)\\
 $[600, 3000]$ & 0.001 &  $\pm$ 0.000 & $\pm$ 0.000 & $\pm$ 0.000 & ($\pm$ 0.000)\\
[\cmsTabSkip]
\multicolumn{6}{c}{\emm} \\
[\cmsTabSkip]
 $[100, 160]$ & 0.006  & $\pm$ 0.024 & $\pm$ 0.006 & $\pm$ 0.014 & ($\pm$ 0.028)\\
 $[160, 200]$ & 0.415  & $\pm$ 0.075 & $\pm$ 0.017 & $\pm$ 0.024 & ($\pm$ 0.081)\\
 $[200, 300]$ & 0.489  & $\pm$ 0.035 & $\pm$ 0.013 & $\pm$ 0.008 & ($\pm$ 0.038)\\
 $[300, 600]$ & 0.090  & $\pm$ 0.008 & $\pm$ 0.002 & $\pm$ 0.001 & ($\pm$ 0.008)\\
 $[600, 3000]$ & 0.001 &  $\pm$ 0.000 & $\pm$ 0.000 & $\pm$ 0.000 & ($\pm$ 0.000)\\
[\cmsTabSkip]
\multicolumn{6}{c}{\mmm} \\
[\cmsTabSkip]
 $[100, 160]$ & 0.009  & $\pm$ 0.016 & $\pm$ 0.004 & $\pm$ 0.010 & ($\pm$ 0.019)\\
 $[160, 200]$ & 0.384  & $\pm$ 0.056 & $\pm$ 0.010 & $\pm$ 0.021 & ($\pm$ 0.061)\\
 $[200, 300]$ & 0.507  & $\pm$ 0.028 & $\pm$ 0.007 & $\pm$ 0.008 & ($\pm$ 0.030)\\
 $[300, 600]$ & 0.099  & $\pm$ 0.006 & $\pm$ 0.002 & $\pm$ 0.002 & ($\pm$ 0.007)\\
 $[600, 3000]$ & 0.001 &  $\pm$ 0.000 & $\pm$ 0.000 & $\pm$ 0.000 & ($\pm$ 0.000)\\
 [\cmsTabSkip]
\multicolumn{6}{c}{Inclusive} \\
[\cmsTabSkip]
 $[100, 160]$  & 0.001 & $\pm$ 0.011 & $\pm$ 0.005 & $\pm$ 0.005 & ($\pm$ 0.013)\\
 $[160, 200]$  & 0.430 & $\pm$ 0.038 & $\pm$ 0.009 & $\pm$ 0.012 & ($\pm$ 0.041)\\
 $[200, 300]$  & 0.473 & $\pm$ 0.018 & $\pm$ 0.009 & $\pm$ 0.005 & ($\pm$ 0.020)\\
 $[300, 600]$  & 0.095 & $\pm$ 0.004 & $\pm$ 0.002 & $\pm$ 0.001 & ($\pm$ 0.004)\\
 $[600, 3000]$ & 0.001 & $\pm$ 0.000 & $\pm$ 0.000 & $\pm$ 0.000 & ($\pm$ 0.000)\\
\hline
\end{tabular}
\end{table}

\subsection{Differential measurement split per \PW\ boson charge}
\label{sec:differentialCharge}
The differential \WZ\ cross section is computed as a function of the same observables as in Section~\ref{sec:differential} and categorized according to the sign of the charge of the lepton associated with the \PW\ boson. Additionally, the differential cross section is computed as a function of the momentum of the lepton that is assigned to the \PW\ boson using the procedure outlined in Section~\ref{sec:selection}.

The charge of the \PW\ boson is estimated using as a proxy the charge of the lepton associated to the \PW\ boson. Results are shown here for the inclusive final state, but similar results have been obtained in the four exclusive categories (\mmm, \emm, \eem, and \eee).

\begin{figure}[!hbtp]
  \centering
  \includegraphics[width=0.45\linewidth]{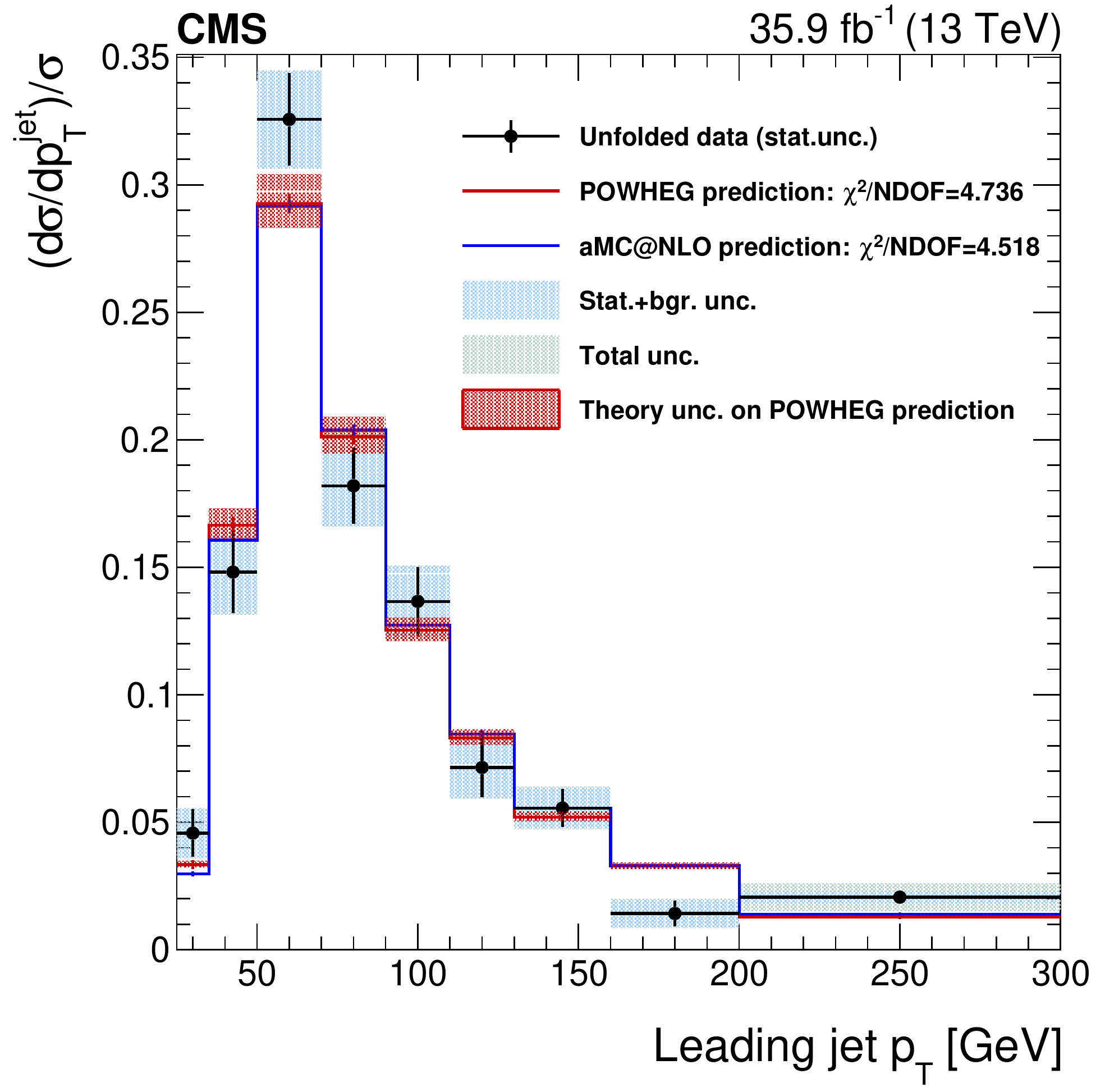}
  \includegraphics[width=0.45\linewidth]{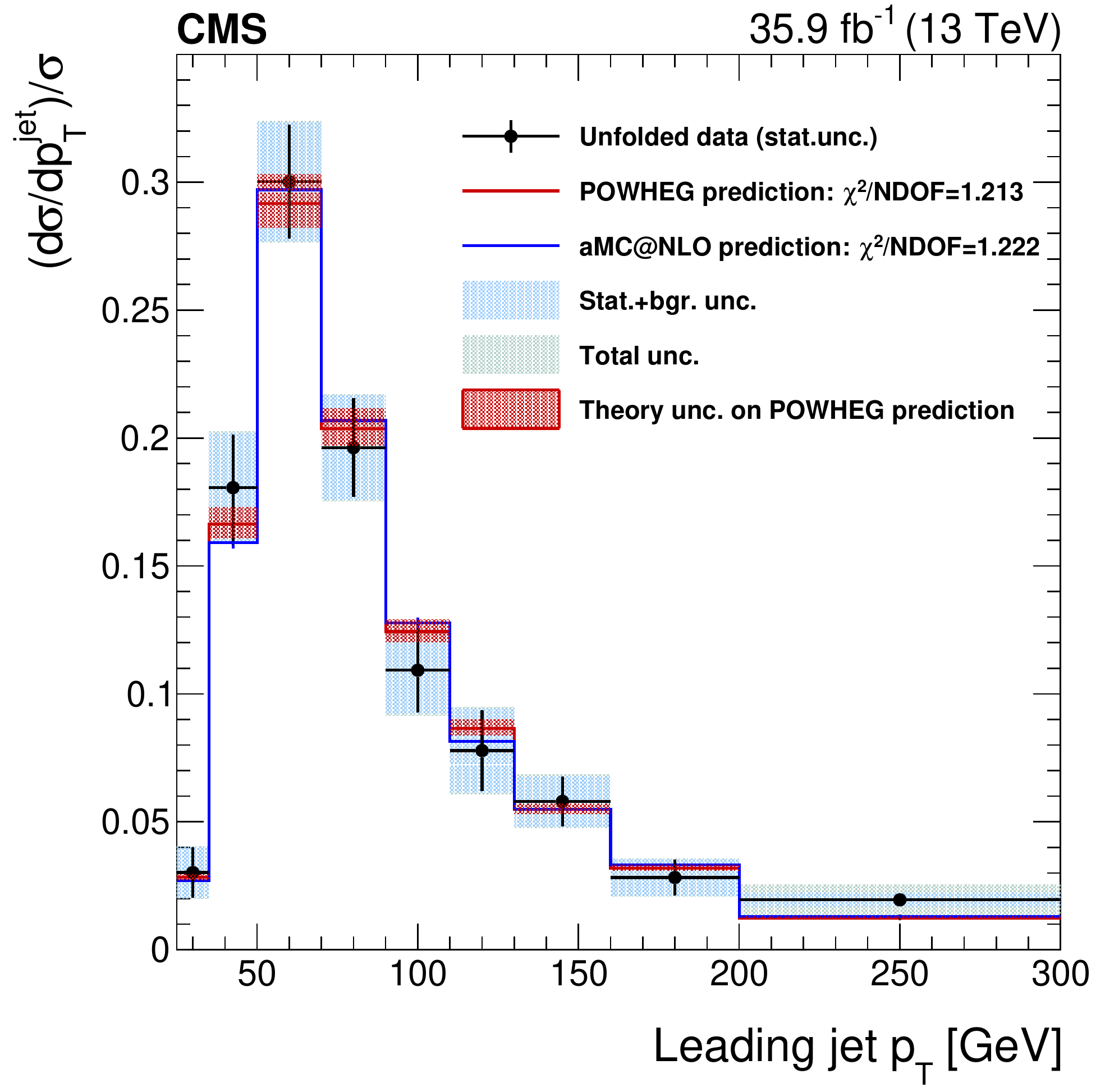}\\
  \caption{Differential distributions for \PW$^{+}$ (left) and \PW$^{-}$ (right), in the full SR. The leading jet transverse momentum is unfolded at the dressed leptons level, as described in the text. The red band around the \POWHEG prediction represents the theory uncertainty in it. The effect on the unfolded data of this uncertainty, through the unfolding matrix, is included in the shaded bands described in the legend.}
  \label{fig:result_inclusive_LeadJetPt_summary}
\end{figure}

\begin{figure}[!hbtp]
  \centering
  \includegraphics[width=0.45\linewidth]{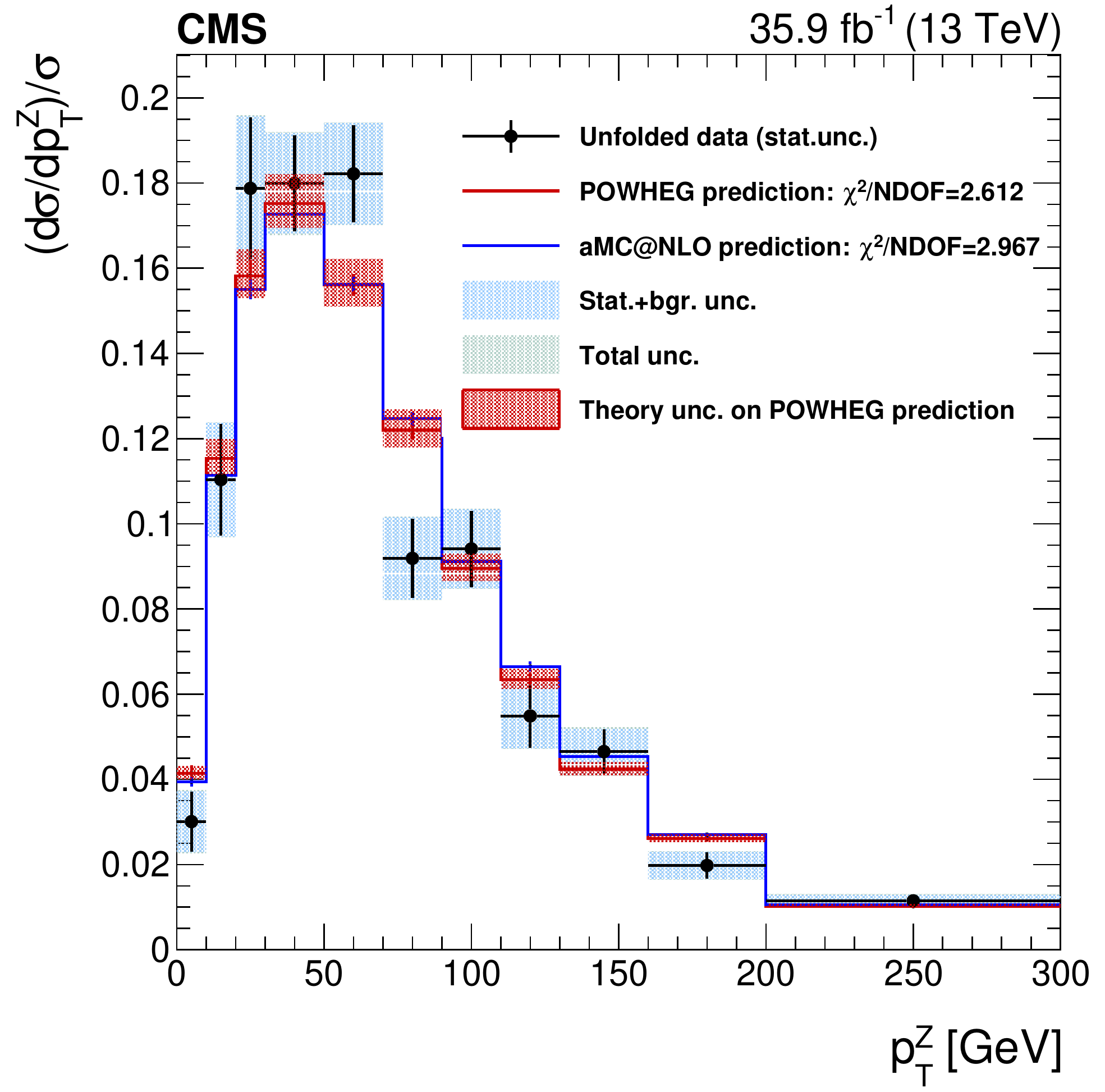}
  \includegraphics[width=0.45\linewidth]{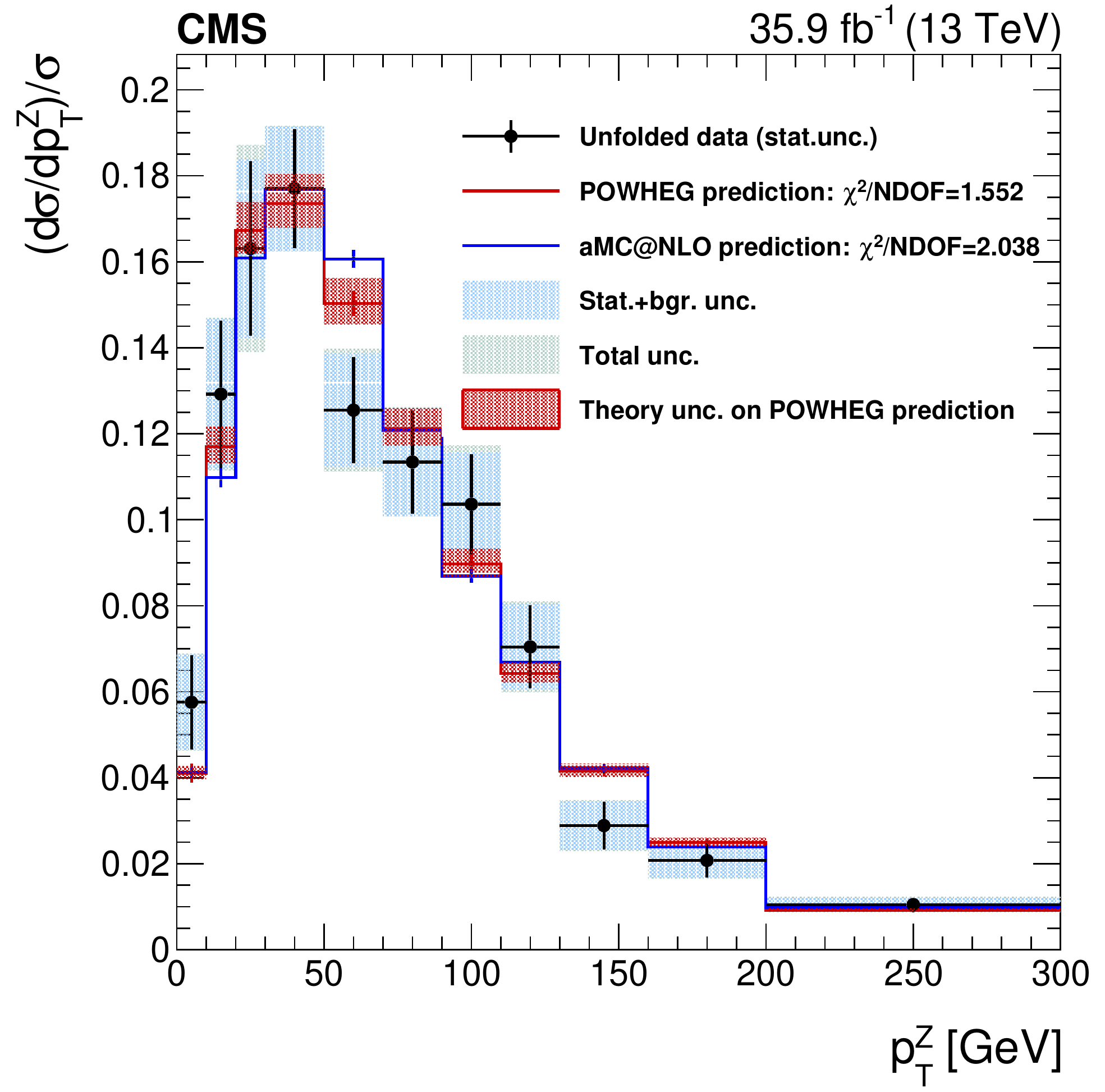}\\
  \caption{Differential distributions for \PW$^{+}$ (left) and \PW$^{-}$ (right), in the full SR. The transverse momentum of the \PZ\ boson is unfolded at the dressed leptons level, as described in the text. The red band around the \POWHEG prediction represents the theory uncertainty in it; the effect on the unfolded data of this uncertainty, through the unfolding matrix, is included in the shaded bands described in the legend.}
  \label{fig:result_inclusive_Zpt_summary}
\end{figure}

\begin{figure}[!hbtp]
  \centering
  \includegraphics[width=0.45\linewidth]{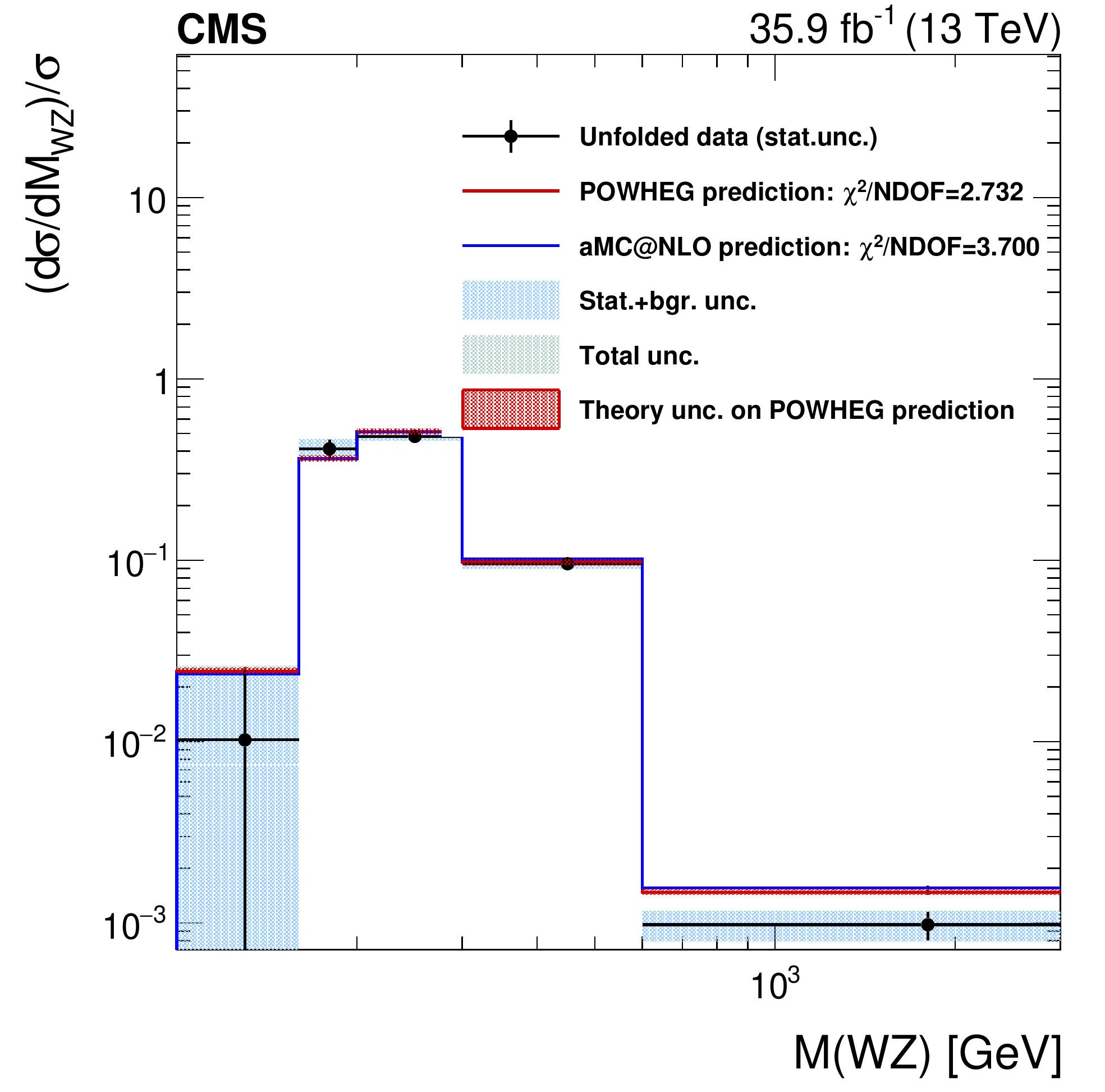}
  \includegraphics[width=0.45\linewidth]{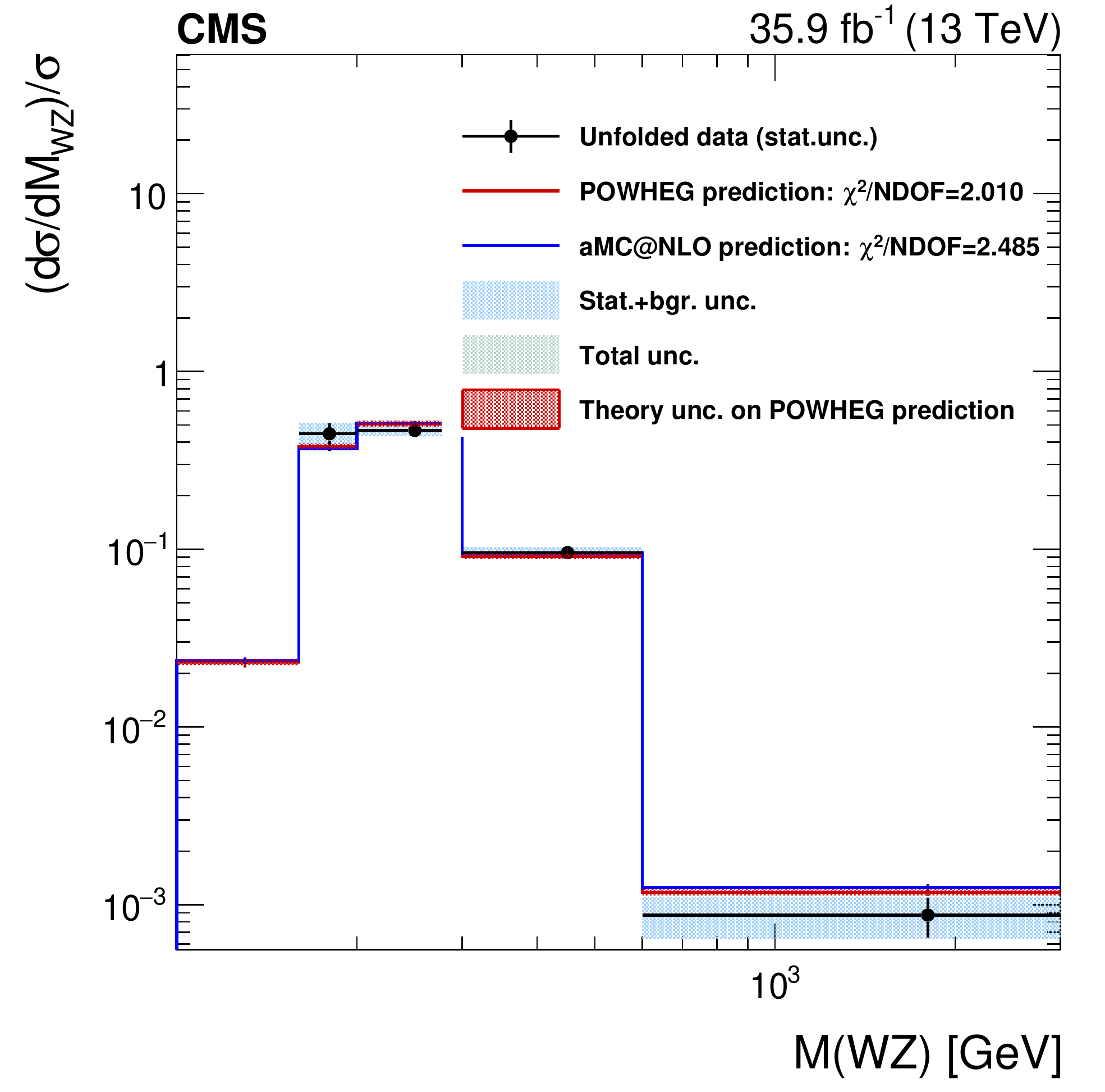}\\
  \caption{Differential distributions for \PW$^{+}$ (left) and \PW$^{-}$ (right), in the full SR. The mass of the \WZ\ system data distribution is unfolded at the dressed leptons level, as described in the text. The red band around the \POWHEG prediction represents the theory uncertainty in it; the effect on the unfolded data of this uncertainty, through the unfolding matrix, is included in the shaded bands described in the legend.}
  \label{fig:result_inclusive_MWZ_summary}
\end{figure}

\begin{figure}[!hbtp]
  \centering
  \includegraphics[width=0.45\linewidth]{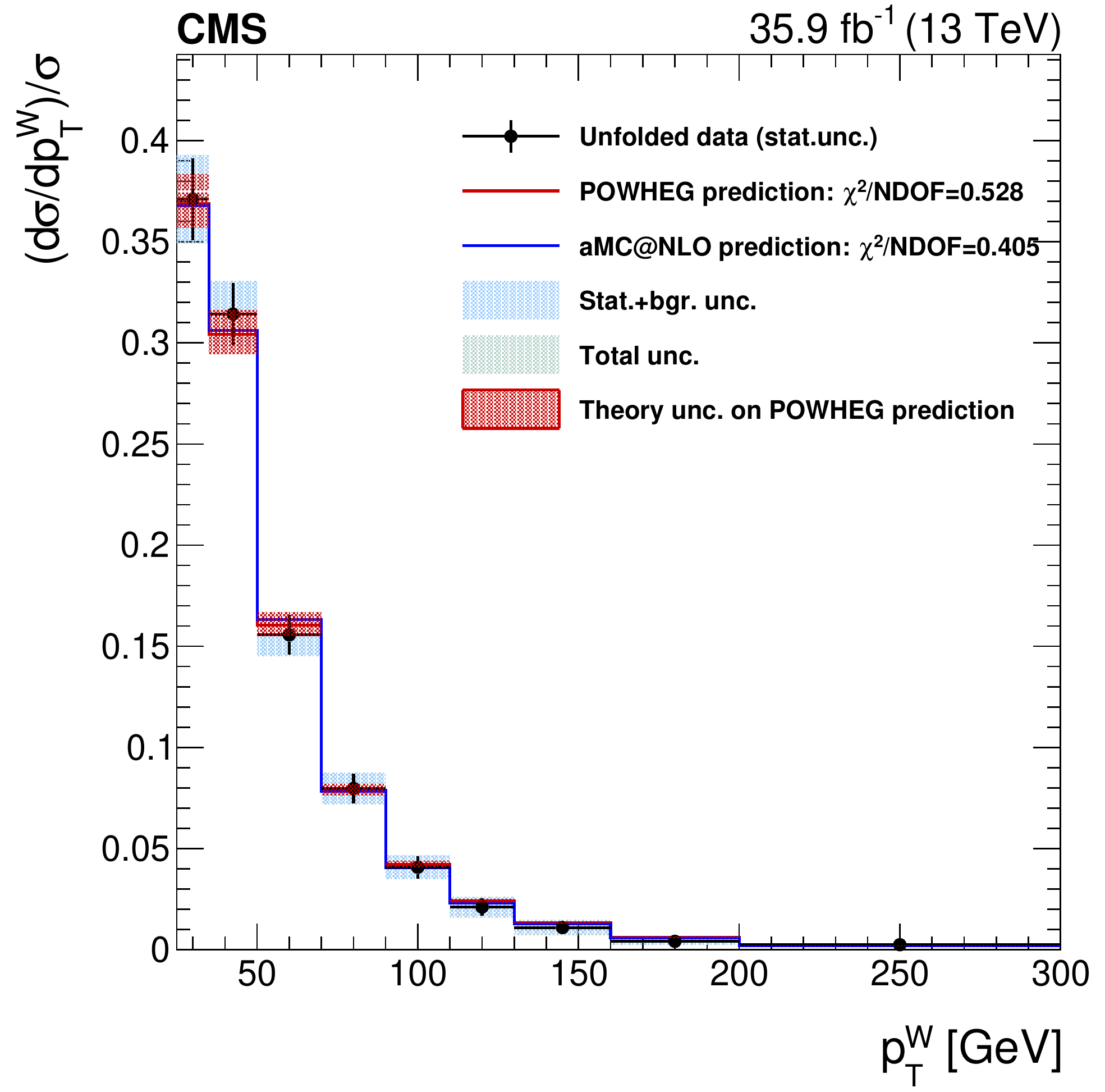}
  \includegraphics[width=0.45\linewidth]{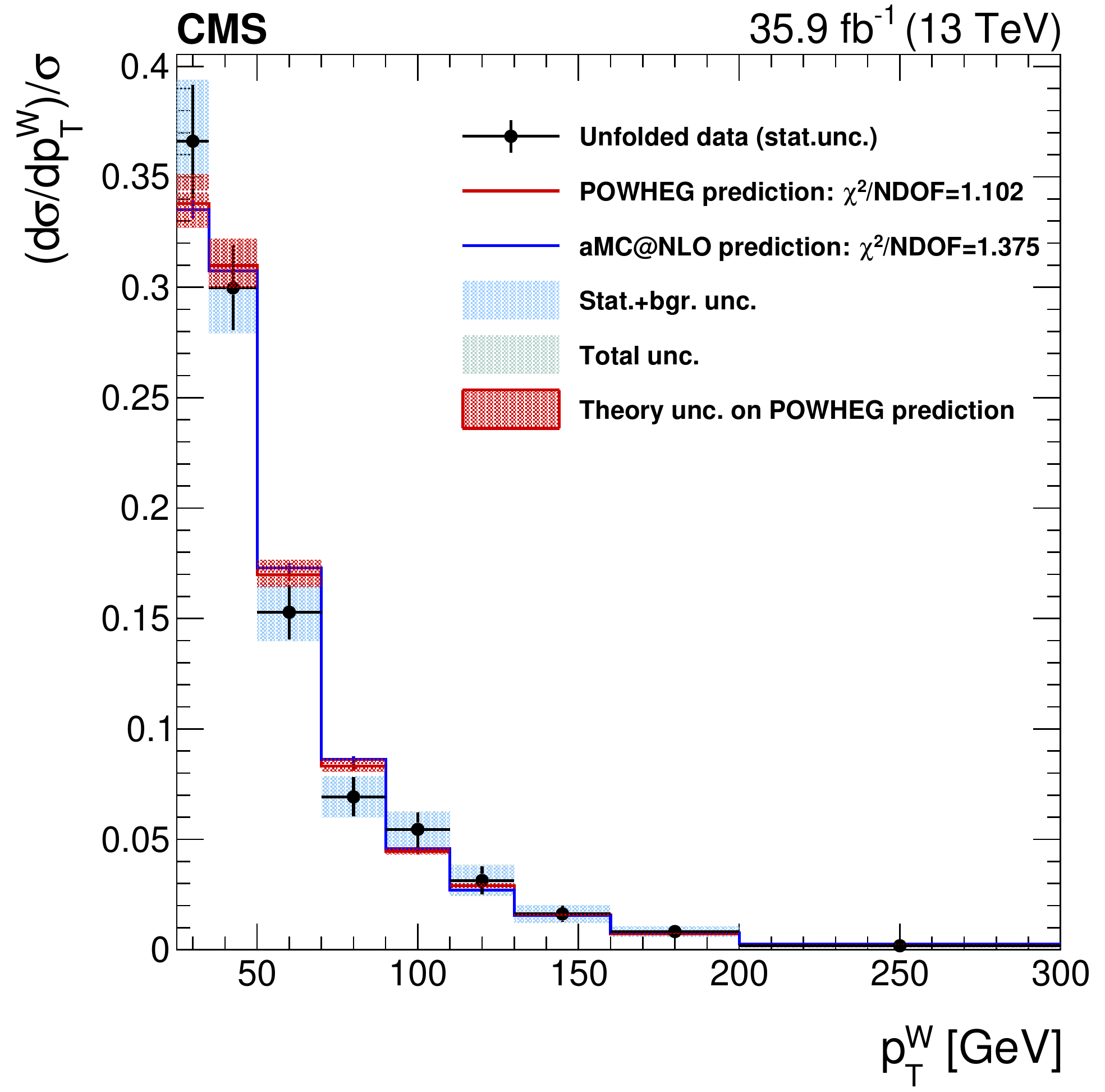}\\
  \caption{Differential distributions for \PW$^{+}$ (left) and \PW$^{-}$ (right), in the full SR. The \PW\ boson transverse momentum is unfolded at the dressed leptons level, as described in the text. The red band around the \POWHEG prediction represents the theory uncertainty in it; the effect on the unfolded data of this uncertainty, through the unfolding matrix, is included in the shaded bands described in the legend.}
  \label{fig:result_inclusive_Wpt_summary}
\end{figure}

Results for the leading jet \pt are shown in Fig.~\ref{fig:result_inclusive_LeadJetPt_summary}, results for the \PZ\ boson \pt are shown in Fig.~\ref{fig:result_inclusive_Zpt_summary}, results for the mass of the \WZ\ system are shown in Fig.~\ref{fig:result_inclusive_MWZ_summary}, and results for the \PW\ boson \pt are shown in Fig.~\ref{fig:result_inclusive_Wpt_summary}. The overall description of the data by the simulation is good. The agreement is quantified by $\chi^{2}/NDOF$ values that are given in the plot legends. As in the case of the measurement not split by charge, the total uncertainty is dominated by the statistical and background subtraction uncertainties. The remaining uncertainties include the one due to the unfolding procedure and are grouped into the \emph{other syst.} category.

\section{Confidence regions for anomalous triple gauge couplings}\label{sec:ano}

The \WZ\ production process is sensitive to the presence of BSM physics through the presence of deviations from the SM predictions of the coupling constants between the SM vector bosons. Because of the dominant SM production modes, the process is expected to be particularly influenced by TGCs of the \PW\ and \cPZ\ bosons. Such couplings are called \emph{anomalous} when they assume values different from the SM predictions. The total set of allowed operators of dimension six can be summarized in three independent parameters~\cite{Degrande:2012wf}. Usually the choice of a basis for these parameters is based on the effective field theory (EFT) approach, where the anomalous coupling Lagrangian can be written as:
\begin{equation}
\delta \mathcal{L}_{\mathrm{AC}} = \frac{\cwww}{\Lambda^{2}} \mathrm{Tr[}\PW_{\mu\nu}^{} \PW^{ \nu\rho} \PW^{\mu}_{\rho}\mathrm{]} + \frac{\cw}{\Lambda^{2}} \left(\PD_{\mu} \PH\right)^{\dagger} \PW^{\mu\nu} \left( \PD_{\nu} \PH \right) + \frac{\cb}{\Lambda^{2}} \left(\PD_{\mu} \PH \right)^{\dagger} \PB^{\mu\nu} \left( \PD_{\nu} \PH \right),
\end{equation}
where $\PW_{\mu\nu}^{\pm}, \PB_{\mu\nu}$ are the field strengths associated to the SM electroweak bosons and $\PH$ is the SM Higgs field. The parameters representing different aTGC effects are noted as $\{\cw, \cwww, \cb\}$. Values predicted by the SM are $\cw = \cwww = \cb = 0$. The typical energy scale at which BSM physics are dominant is represented by $\Lambda^2$ and it is usually absorbed in the definition of the aTGC parameters.

\begin{figure}[]
\centering
\includegraphics[width=0.48\textwidth]{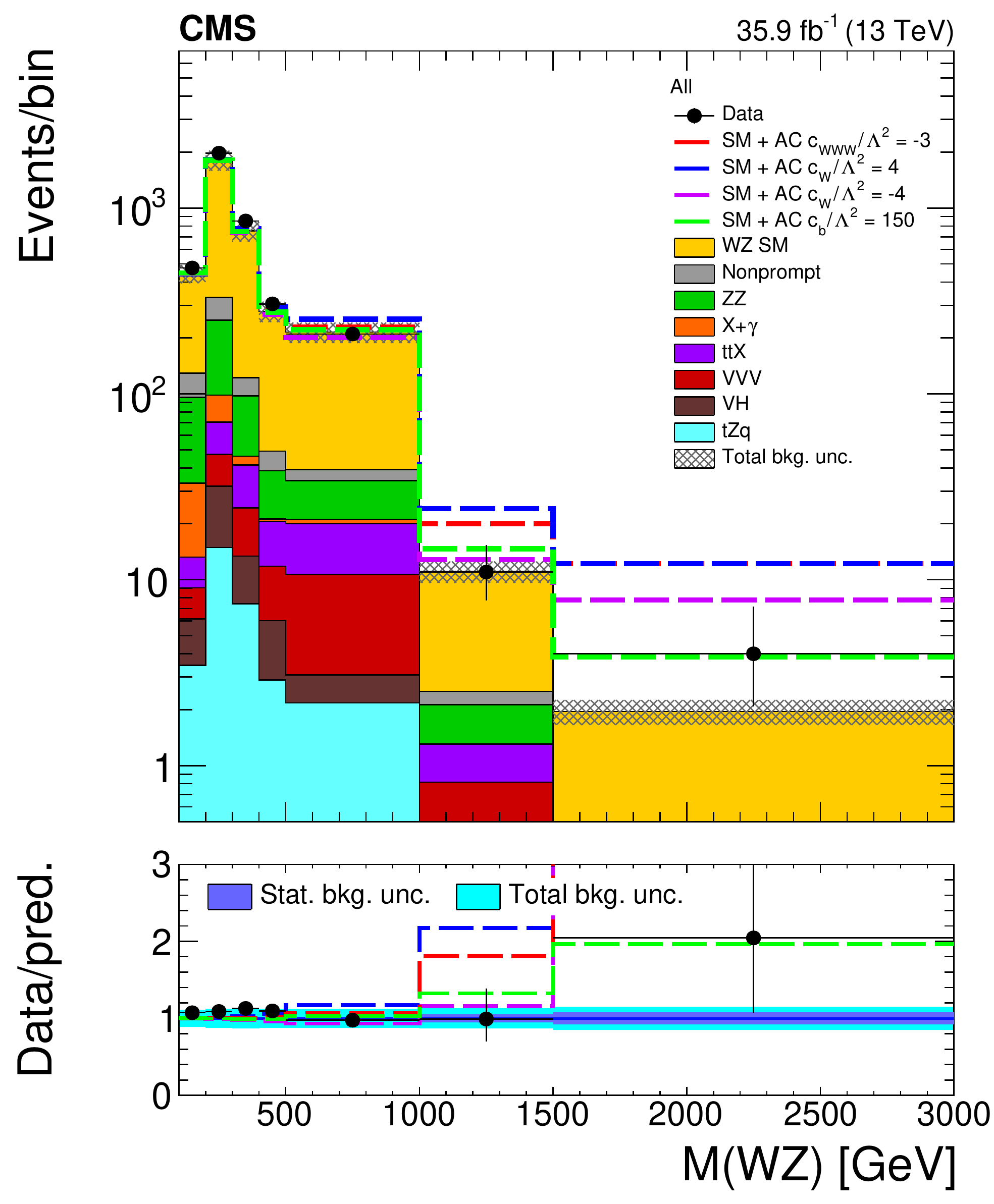}
\includegraphics[width=0.48\textwidth]{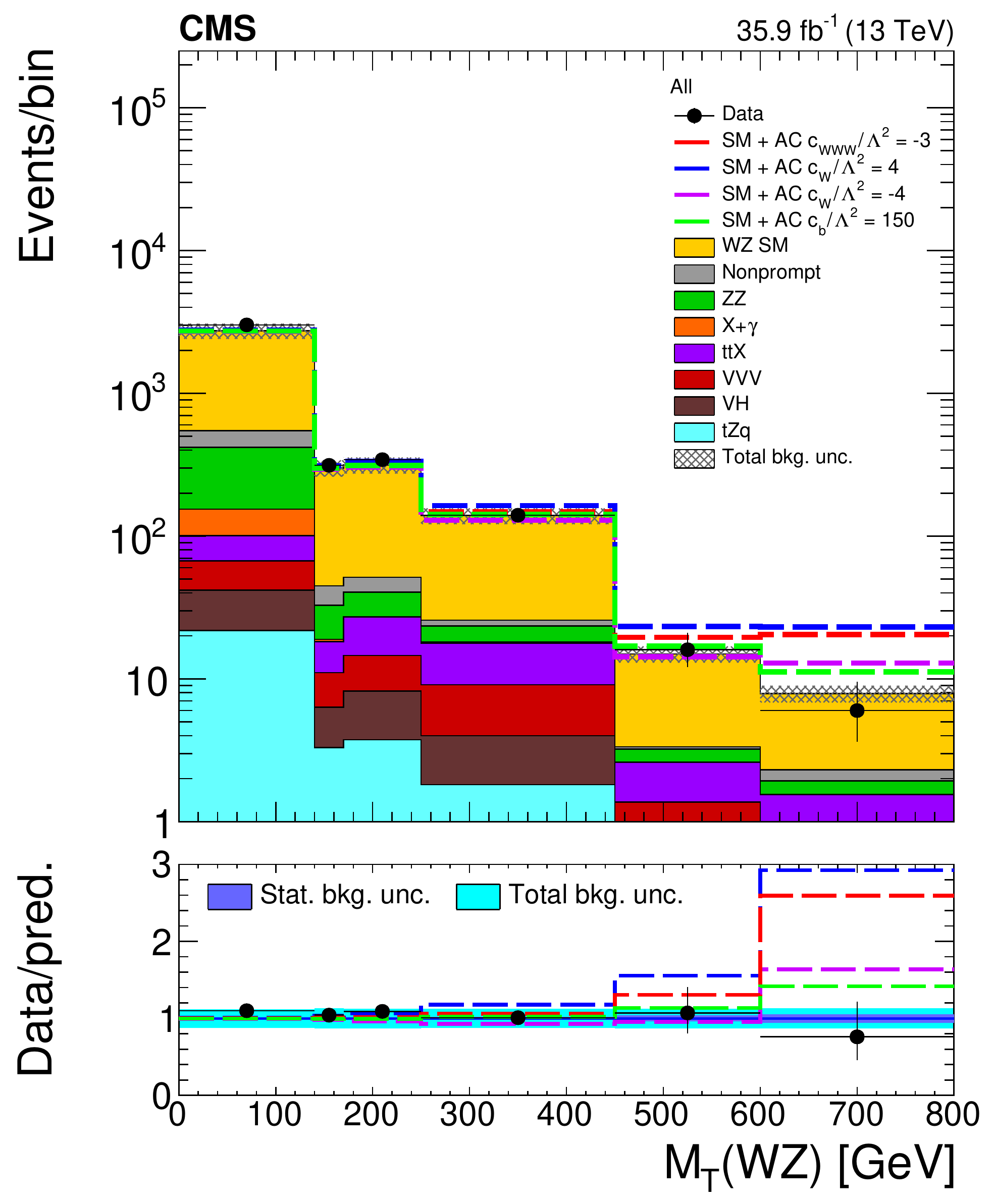}
\caption{Distributions of discriminant observables in the anomalous triple gauge couplings searches, before the fit used to determine confidence regions on the couplings. The invariant mass of the three lepton and missing transverse momentum system (left) and the transverse mass of the same configuration (right). The dashed lines represent the total yields expected from the sum of the SM processes, with the total \WZ\ yields for different values of the associated anomalous coupling (AC) parameters. The SM prediction for the \WZ\ process is obtained from the aTGC simulated sample with the AC parameters set to 0.}
\label{fig:acSMPretty}
\end{figure}

The behaviour of the SM prediction and those of different configurations of anomalous couplings values are compared in Fig.~\ref{fig:acSMPretty} for two different observables that aim to reconstruct the mass of a hypothetical BSM particle decaying to a \WZ\ pair. The predictions corresponding to four different anomalous couplings are drawn for comparison to outline the behaviour of the most asymmetric one (\cwww).

The \mwz variable, defined in Section~\ref{sec:differential}, is chosen to determine confidence regions for each of the anomalous parameters considered. A different behaviour as a function of this variable is expected at high energy values in the presence of anomalous couplings, because of the nature of the proper anomalous terms, which include the momenta of the bosons through the field strength terms.

For each of the bins presented in Fig.~\ref{fig:acSMPretty} (left), a three-dimensional quadratic fit is performed to the predicted yields of the anomalous couplings in a grid of simulated points in order to extrapolate the prediction to the continuous space of parameter values. A binned likelihood function is built with the signal yields for each bin depending on the values of each of the three anomalous coupling parameters obtained from the fit. The uncertainties described in Section~\ref{sec:systematics} are included as additional nuisance parameters correlated across the bins. Confidence regions for each parameter and each combination of two parameters are derived using a multidimensional likelihood fit to the relevant parameters, with the remaining ones set to the SM values. Nuisance parameters are profiled in the likelihood fit. Appropriate confidence levels (\CLs) are derived assuming the distribution of the log-likelihood function is half a $\chi^2$ distribution with degrees of freedom equal to the number of free parameters.

\begin{figure}[!hptb]
\centering
\includegraphics[width=0.65\textwidth]{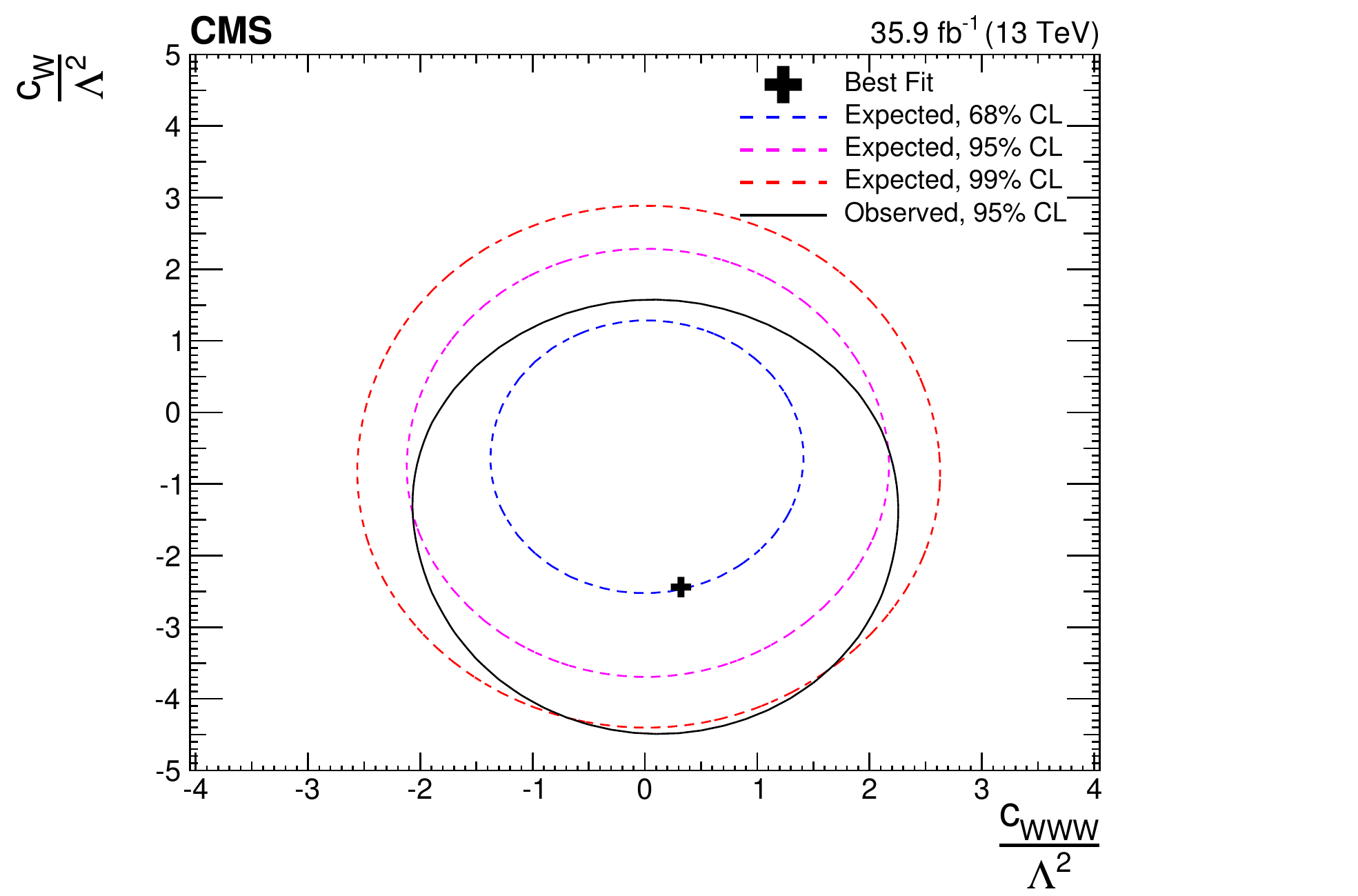}
\includegraphics[width=0.65\textwidth]{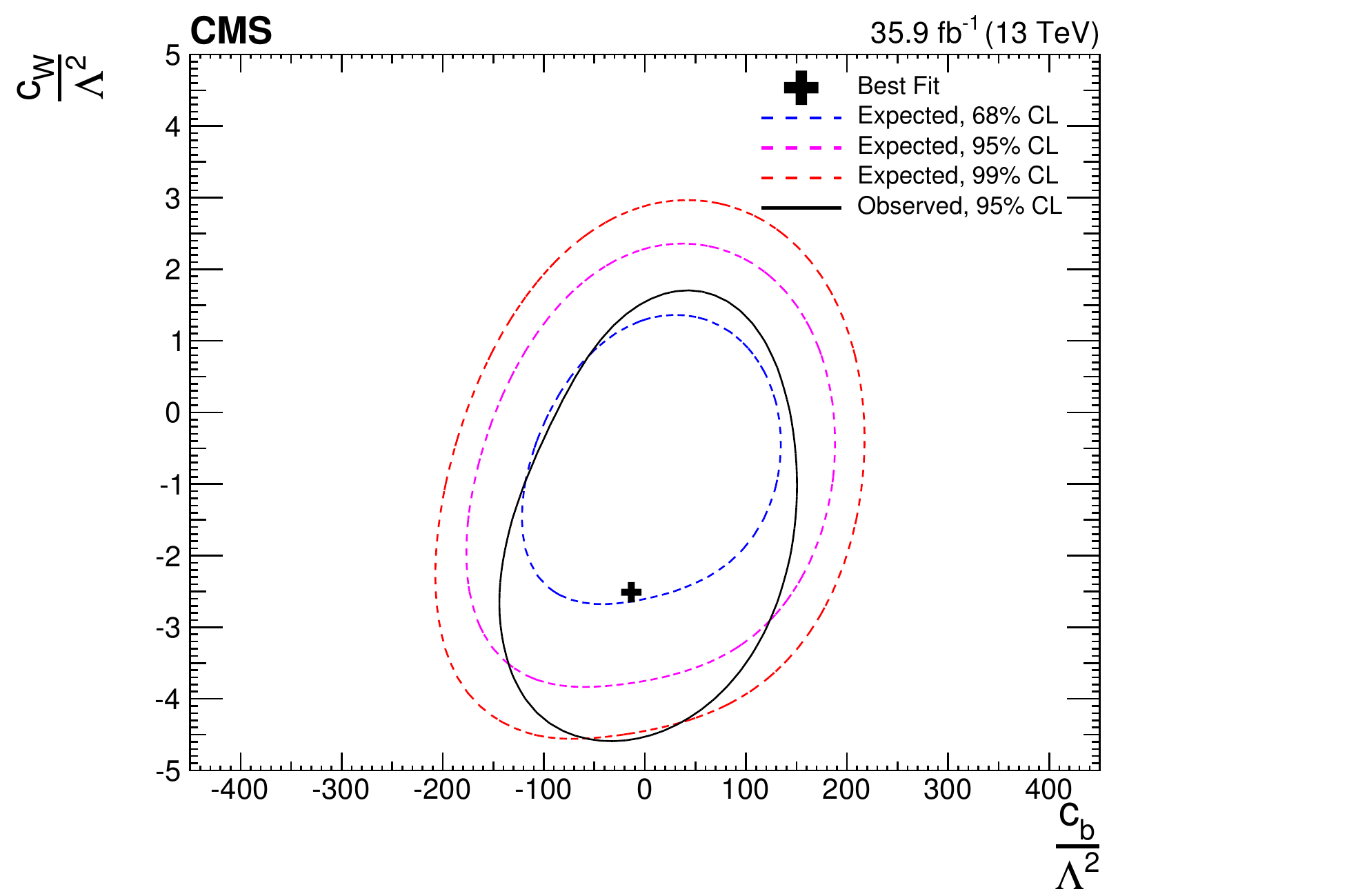}
\includegraphics[width=0.65\textwidth]{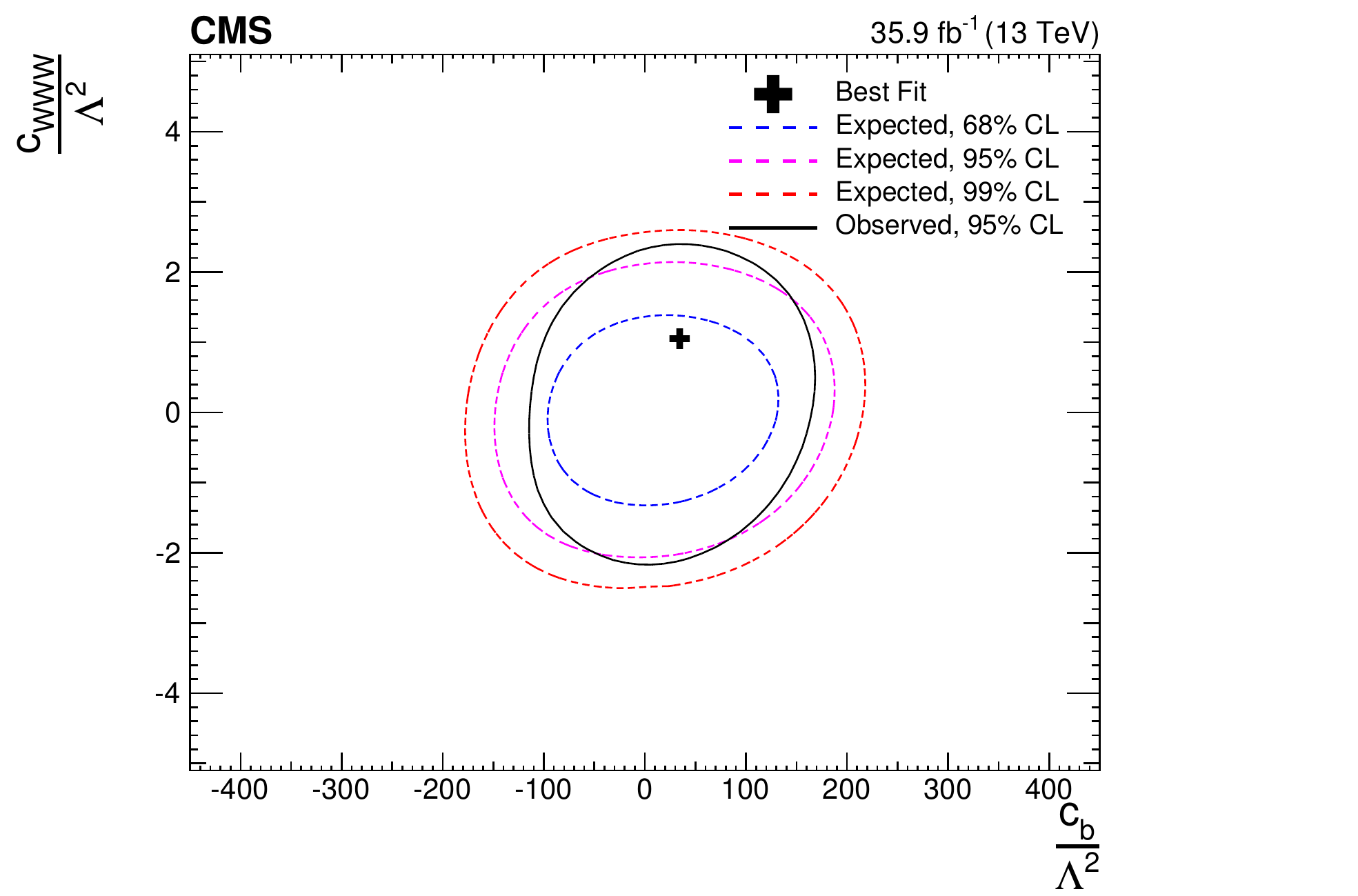}
\caption{Two-dimensional confidence regions for each of the possible combinations of the considered aTGC parameters. The contours of the expected confidence regions for 68\% and 95\% confidence level are presented in each case. The parameters considered in each plot are \cw--\cwww (top), \cw--\cb (middle) and \cwww--\cb (bottom).}
\label{fig:2DAC}
\end{figure}

The full procedure is applied to derive one-dimensional confidence intervals in each of the anomalous couplings parameters, fixing the other two parameters to zero---the SM value. The results are shown in Table~\ref{tab:1DAC}. For each pair of parameters, a two-dimensional confidence region is derived, as shown in Fig.~\ref{fig:2DAC}.

The procedure we described includes both the interference term between the SM amplitude and the BSM one, and for the square of the dimension-6 contribution. If the quadratic term used to build the statistical model is suppressed in the fit, the resulting confidence intervals include the interference term between the SM amplitude and the BSM one only, neglecting the square of the dimension-6 contributions. The results corresponding to this approximation are tabulated in Table~\ref{tab:1DAC_linear}.

\begin{table}[]
\centering
\topcaption{\label{tab:1DAC} Expected and observed one-dimensional confidence intervals (CI) at 95\% confidence level for each of the considered EFT parameters. Both the square matrix of the dimension-6 contribution and the interference term between the SM amplitude and the BSM one are accounted for. The one-dimensional intervals for each parameter are computed fixing the other two parameters to zero, the SM value.}
\begin{tabular}{lcc}
\hline
Parameter                       & 95\% CI (expected) $[\TeVns^{-2}]$ & 95\% CI (observed) $[\TeVns^{-2}]$ \\ \hline
$\cw/\Lambda^{2}$      & $[-3.3,2.0]$    & $[-4.1,1.1]$    \\
$\cwww/\Lambda^{2}$    & $[-1.8,1.9]$    & $[-2.0,2.1]$     \\
$\cb/\Lambda^{2}$      & $[-130,170] $     & $[-100,160]$     \\
\hline
\end{tabular}
\end{table}

\begin{table}[]
\centering
\topcaption{\label{tab:1DAC_linear} Expected and observed one-dimensional confidence intervals (CI) at 95\% confidence level for each of the considered EFT parameters, accounting only for the interference term between the SM amplitude and the BSM one. The one-dimensional intervals for each parameter are computed fixing the other two parameters to zero, the SM value.}
\begin{tabular}{lcc}
\hline
Parameter                       & 95\% CI (expected) $[\TeVns^{-2}]$ & 95\% CI (observed) $[\TeVns^{-2}]$ \\ \hline
$\cw/\Lambda^{2}$      & $[-2.3,3.4]$    & $[-2.2,2.7]$    \\
$\cwww/\Lambda^{2}$    & $[-33.2,28.6]$    & $[-13.8,41.2]$     \\
$\cb/\Lambda^{2}$      & $[-360,300] $     & $[-230,390]$     \\
\hline
\end{tabular}
\end{table}

\begin{figure}[!hptb]
\centering
\includegraphics[width=0.65\textwidth]{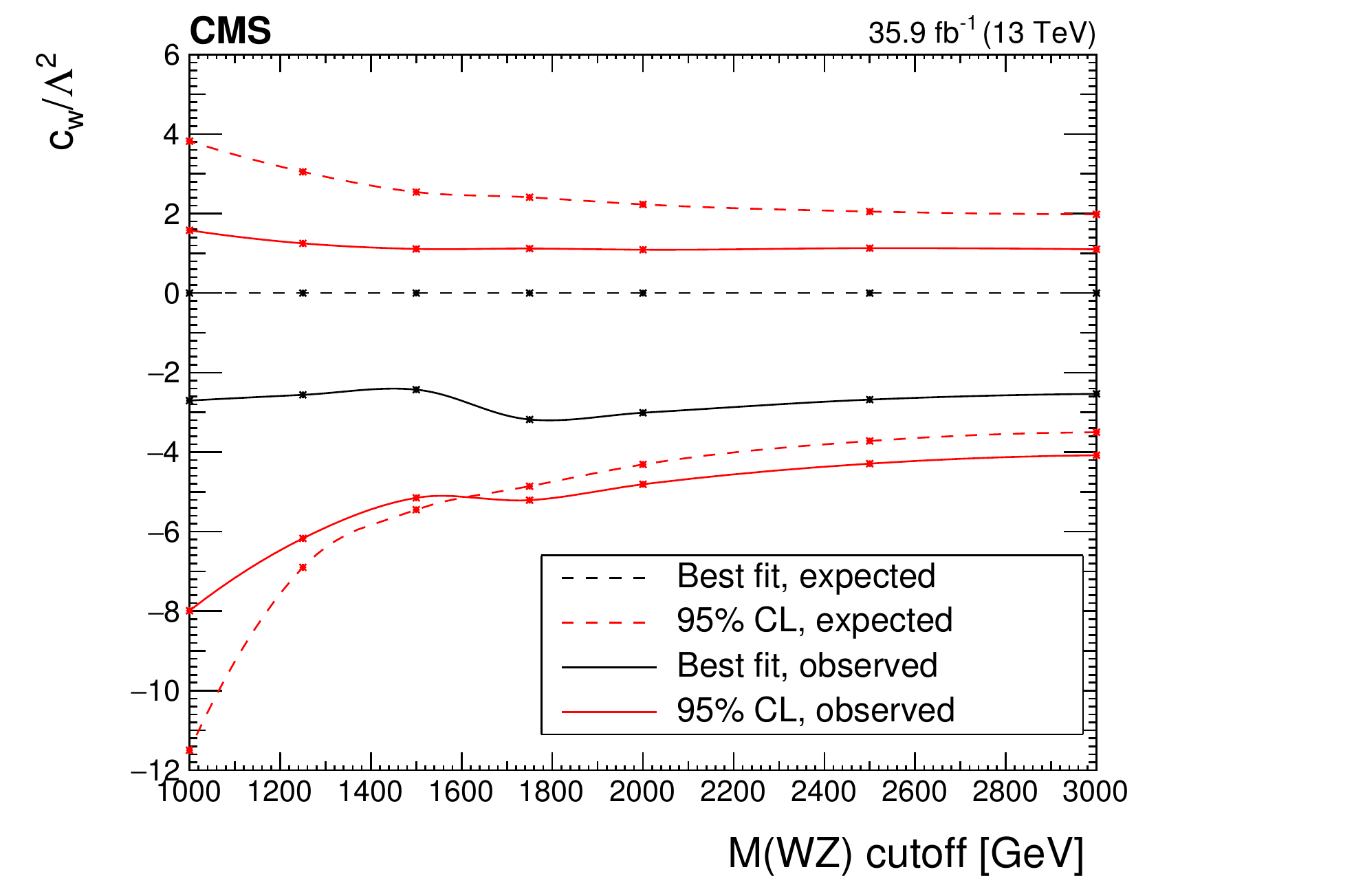}
\includegraphics[width=0.65\textwidth]{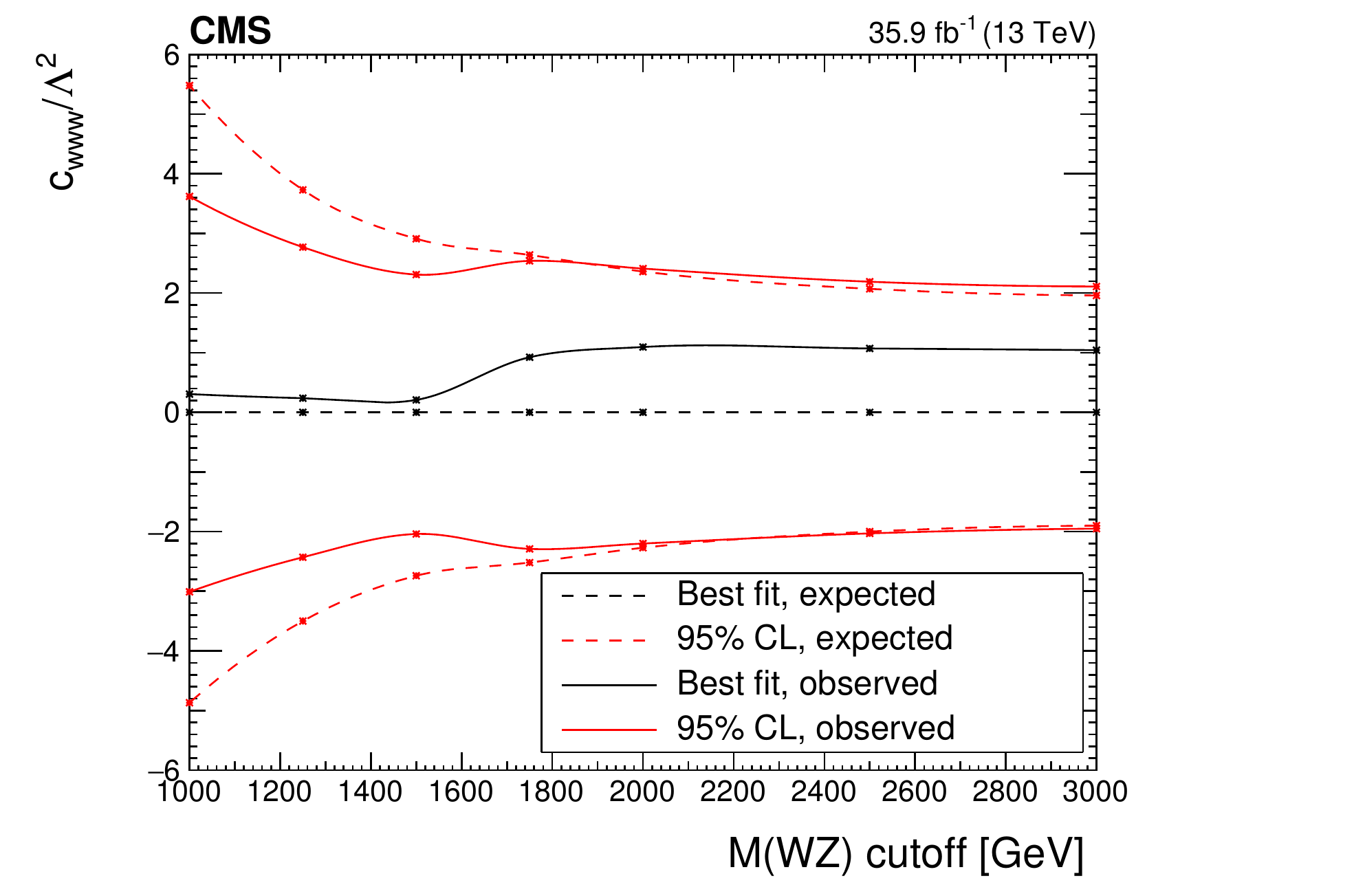}
\includegraphics[width=0.65\textwidth]{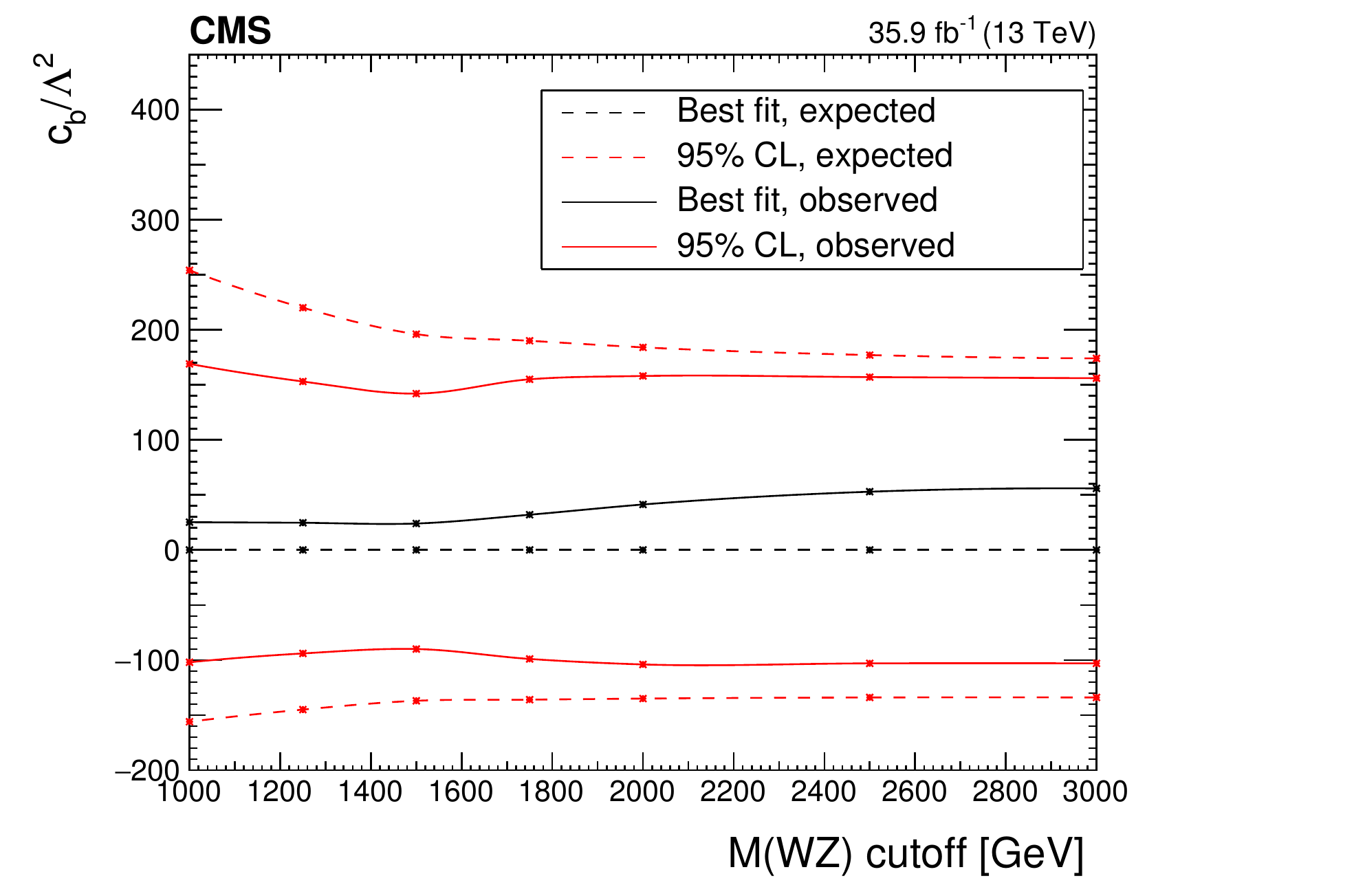}
\caption{Evolution of the expected and observed confidence intervals of the EFT anomalous coupling parameters in terms of the cutoff scale given by different restrictions in the \mwz variable. For each point and parameter, the confidence intervals are computed imposing the additional restriction of no anomalous coupling contribution over the given value of the \mwz cutoff. The last point is equivalent to no cutoff requirement being imposed. The parameters considered are: \cw (top), \cwww (middle) and \cb (bottom).}
\label{fig:acEVO}
\end{figure}

Restricting the effect of the anomalous couplings to a given range in the invariant mass of the diboson system can be used to impose unitarity in the aTGC models. While no direct computation of the invariant mass is possible in the leptonic decay of the \WZ\ channel, we use the \mwz variable as a reasonable substitute. We compute the confidence interval for each parameter based on multiple cutoff values of the \mwz value to obtain the results shown in Fig.~\ref{fig:acEVO}.

No anomalous effect has been observed, and the confidence regions obtained represent a significant improvement with respect to previous searches performed by the ATLAS~\cite{Aad:2016wpd} and CMS~\cite{Khachatryan:2016poo,Sirunyan:2017bey} Collaborations.

\section{Summary}
\label{sec:conclusions}

The production process \wzprod\ is studied in the trilepton final state at $\sqrts = 13\TeV$, using the full 2016 data set with a total integrated luminosity of \fulllumi collected with the CMS detector.

Fiducial results are obtained in each of the flavour categories (\eee, \eem, \emm, and \mmm) and in the combined category, and are extrapolated to the total \WZ\ production cross section for $60 < \mZ^{OSSF} < 120\GeV$.
The combined measurement yields a cross section of $\sigmatot(\Pp\Pp \to \WZ) = 48.09 \mathrm{ }^{+1.00}_{-0.96}\stat \mathrm{ }^{+0.44}_{-0.37}\thy \textrm{ }^{+2.39}_{-2.17}\syst \pm 1.39\lum$\unit{pb}, for a total uncertainty of $+2.98$ and $-2.78$\unit{pb}. The result is in good agreement with the~\textsc{MATRIX} next-to-next-to-leading-order (NNLO) prediction~\cite{theoWZ,Grazzini:2017mhc}, of $\sigma_{\mathrm{NNLO}}(\Pp\Pp \to \WZ) = 49.98^{+2.2\%}_{-2.0\%}$\unit{pb}. This result supersedes the result from the CMS Collaboration using data corresponding to a smaller integrated luminosity of 2.3\fbinv~\cite{Khachatryan:2016tgp}. A measurement in the fiducial region yields a value of $\sigma_{\text{fid}}(\Pp\Pp \to \WZ) = 257.5 \mathrm{ }^{+5.3}_{-5.0}\stat \mathrm{ }^{+2.3}_{-2.0}\thy \mathrm{ }^{+12.8}_{-11.6}\syst  \pm 7.4\lum$\unit{fb}, pointing to an excess over the \POWHEG next-to-leading-order cross section $\sigma_{\text{fid}}^{\POWHEG} = 227.6  \mathrm{ }^{+9.4}_{-8.0}$\unit{fb}. The cross sections are also measured independently for the two possible values of the \PW\ boson charge, yielding a ratio of $\awz = \sigmatot(\Pp\Pp \to \PW^+\cPZ)/\sigmatot(\Pp\Pp \to \PW^-\cPZ) = 1.48\pm 0.06$, which is compatible within uncertainties with the \POWHEG + \PYTHIA prediction of $1.43^{+0.06}_{-0.05}$. Similar results are obtained when splitting by flavour category. All the measurements of this paper are compatible with the SM when the appropriate order of theoretical calculations is considered.

Differential cross sections are measured as a function of the transverse momentum of the \cPZ\ boson, of the transverse momentum of the leading jet, and of an estimate of the mass of the \WZ\ system; results are compared with predictions from the \POWHEG and \MGvATNLO generators. Differential cross sections as a function of the transverse momentum of the leading jet are also measured for each sign of the \PW\ boson charge.
Confidence intervals for anomalous triple gauge boson couplings are extracted for each of the possible one- and two-dimensional combinations of the anomalous couplings parameters, using the \mwz variable in a maximum likelihood fit.
The confidence intervals obtained represent the most stringent results on the anomalous \PW\WZ\ triple gauge coupling to date.

\begin{acknowledgments}

\hyphenation{Bundes-ministerium Forschungs-gemeinschaft Forschungs-zentren Rachada-pisek} We congratulate our colleagues in the CERN accelerator departments for the excellent performance of the LHC and thank the technical and administrative staffs at CERN and at other CMS institutes for their contributions to the success of the CMS effort. In addition, we gratefully acknowledge the computing centres and personnel of the Worldwide LHC Computing Grid for delivering so effectively the computing infrastructure essential to our analyses. Finally, we acknowledge the enduring support for the construction and operation of the LHC and the CMS detector provided by the following funding agencies: the Austrian Federal Ministry of Science, Research and Economy and the Austrian Science Fund; the Belgian Fonds de la Recherche Scientifique, and Fonds voor Wetenschappelijk Onderzoek; the Brazilian Funding Agencies (CNPq, CAPES, FAPERJ, FAPERGS, and FAPESP); the Bulgarian Ministry of Education and Science; CERN; the Chinese Academy of Sciences, Ministry of Science and Technology, and National Natural Science Foundation of China; the Colombian Funding Agency (COLCIENCIAS); the Croatian Ministry of Science, Education and Sport, and the Croatian Science Foundation; the Research Promotion Foundation, Cyprus; the Secretariat for Higher Education, Science, Technology and Innovation, Ecuador; the Ministry of Education and Research, Estonian Research Council via IUT23-4 and IUT23-6 and European Regional Development Fund, Estonia; the Academy of Finland, Finnish Ministry of Education and Culture, and Helsinki Institute of Physics; the Institut National de Physique Nucl\'eaire et de Physique des Particules~/~CNRS, and Commissariat \`a l'\'Energie Atomique et aux \'Energies Alternatives~/~CEA, France; the Bundesministerium f\"ur Bildung und Forschung, Deutsche Forschungsgemeinschaft, and Helmholtz-Gemeinschaft Deutscher Forschungszentren, Germany; the General Secretariat for Research and Technology, Greece; the National Research, Development and Innovation Fund, Hungary; the Department of Atomic Energy and the Department of Science and Technology, India; the Institute for Studies in Theoretical Physics and Mathematics, Iran; the Science Foundation, Ireland; the Istituto Nazionale di Fisica Nucleare, Italy; the Ministry of Science, ICT and Future Planning, and National Research Foundation (NRF), Republic of Korea; the Lithuanian Academy of Sciences; the Ministry of Education, and University of Malaya (Malaysia); the Ministry of Science of Montenegro; the Mexican Funding Agencies (BUAP, CINVESTAV, CONACYT, LNS, SEP, and UASLP-FAI); the Ministry of Business, Innovation and Employment, New Zealand; the Pakistan Atomic Energy Commission; the Ministry of Science and Higher Education and the National Science Centre, Poland; the Funda\c{c}\~ao para a Ci\^encia e a Tecnologia, Portugal; JINR, Dubna; the Ministry of Education and Science of the Russian Federation, the Federal Agency of Atomic Energy of the Russian Federation, Russian Academy of Sciences, the Russian Foundation for Basic Research, and the National Research Center ``Kurchatov Institute"; the Ministry of Education, Science and Technological Development of Serbia; the Secretar\'{\i}a de Estado de Investigaci\'on, Desarrollo e Innovaci\'on, Programa Consolider-Ingenio 2010, Plan Estatal de Investigaci\'on Cient\'{\i}fica y T\'ecnica y de Innovaci\'on 2013-2016, Plan de Ciencia, Tecnolog\'{i}a e Innovaci\'on 2013-2017 del Principado de Asturias, and Fondo Europeo de Desarrollo Regional, Spain; the Ministry of Science, Technology and Research, Sri Lanka; the Swiss Funding Agencies (ETH Board, ETH Zurich, PSI, SNF, UniZH, Canton Zurich, and SER); the Ministry of Science and Technology, Taipei; the Thailand Center of Excellence in Physics, the Institute for the Promotion of Teaching Science and Technology of Thailand, Special Task Force for Activating Research and the National Science and Technology Development Agency of Thailand; the Scientific and Technical Research Council of Turkey, and Turkish Atomic Energy Authority; the National Academy of Sciences of Ukraine, and State Fund for Fundamental Researches, Ukraine; the Science and Technology Facilities Council, UK; the US Department of Energy, and the US National Science Foundation.

Individuals have received support from the Marie-Curie programme and the European Research Council and Horizon 2020 Grant, contract No. 675440 (European Union); the Leventis Foundation; the A. P. Sloan Foundation; the Alexander von Humboldt Foundation; the Belgian Federal Science Policy Office; the Fonds pour la Formation \`a la Recherche dans l'Industrie et dans l'Agriculture (FRIA-Belgium); the Agentschap voor Innovatie door Wetenschap en Technologie (IWT-Belgium); the F.R.S.-FNRS and FWO (Belgium) under the ``Excellence of Science - EOS" - be.h project n. 30820817; the Ministry of Education, Youth and Sports (MEYS) of the Czech Republic; the Lend\"ulet (``Momentum") Programme and the J\'anos Bolyai Research Scholarship of the Hungarian Academy of Sciences, the New National Excellence Program \'UNKP, the NKFIA research grants 123842, 123959, 124845, 124850 and 125105 (Hungary); the Council of Scientific and Industrial Research, India; the HOMING PLUS programme of the Foundation for Polish Science, cofinanced from European Union, Regional Development Fund, the Mobility Plus programme of the Ministry of Science and Higher Education, the National Science Center (Poland), contracts Harmonia 2014/14/M/ST2/00428, Opus 2014/13/B/ST2/02543, 2014/15/B/ST2/03998, and 2015/19/B/ST2/02861, Sonata-bis 2012/07/E/ST2/01406; the National Priorities Research Program by Qatar National Research Fund; the Programa de Excelencia Mar\'{i}a de Maeztu, and the Programa Severo Ochoa del Principado de Asturias; the Thalis and Aristeia programmes cofinanced by EU-ESF, and the Greek NSRF; the Rachadapisek Sompot Fund for Postdoctoral Fellowship, Chulalongkorn University, and the Chulalongkorn Academic into Its 2nd Century Project Advancement Project (Thailand); the Welch Foundation, contract C-1845; and the Weston Havens Foundation (USA).

\end{acknowledgments}

\bibliography{auto_generated}
\cleardoublepage \appendix\section{The CMS Collaboration \label{app:collab}}\begin{sloppypar}\hyphenpenalty=5000\widowpenalty=500\clubpenalty=5000\vskip\cmsinstskip
\textbf{Yerevan Physics Institute, Yerevan, Armenia}\\*[0pt]
A.M.~Sirunyan, A.~Tumasyan
\vskip\cmsinstskip
\textbf{Institut f\"{u}r Hochenergiephysik, Wien, Austria}\\*[0pt]
W.~Adam, F.~Ambrogi, E.~Asilar, T.~Bergauer, J.~Brandstetter, M.~Dragicevic, J.~Er\"{o}, A.~Escalante~Del~Valle, M.~Flechl, R.~Fr\"{u}hwirth\cmsAuthorMark{1}, V.M.~Ghete, J.~Hrubec, M.~Jeitler\cmsAuthorMark{1}, N.~Krammer, I.~Kr\"{a}tschmer, D.~Liko, T.~Madlener, I.~Mikulec, N.~Rad, H.~Rohringer, J.~Schieck\cmsAuthorMark{1}, R.~Sch\"{o}fbeck, M.~Spanring, D.~Spitzbart, A.~Taurok, W.~Waltenberger, J.~Wittmann, C.-E.~Wulz\cmsAuthorMark{1}, M.~Zarucki
% ...
\vskip\cmsinstskip
% ...
1:  Also at Vienna University of Technology, Vienna, Austria\\
% ...
\end{sloppypar}
\end{document}